\documentclass[11pt]{article}
\usepackage{amsmath, amsthm}
\usepackage{newpxtext}
\usepackage{newpxmath}
\usepackage{tikz}
\usepackage{float}
\usepackage{stackrel}

\usepackage{marginnote}

\usepackage{titlesec}
\usepackage[T1]{fontenc}
\usetikzlibrary{shapes.geometric,arrows}
\tikzset{
  basics/.style={minimum width=30mm, minimum height=7.5mm, text centered, draw=black},
  basicsone/.style={minimum width=15mm, minimum height=15mm, text centered, draw=black},
  startstop/.style={rectangle, rounded corners, basics},
  io/.style={trapezium, trapezium left angle=70, trapezium right angle=110, basics, fill=blue!30},
  process/.style={rectangle, basics},
  processone/.style={rectangle, basics},
  processtwo/.style={rectangle, basicsone},
  decision/.style={ellipse,basics},
  arrow/.style={thick,->,>=stealth},
  connected/.style={dashed,-},
}

\RequirePackage[colorlinks=true]{hyperref}
\hypersetup{
  linkcolor=[rgb]{0.3,0.3,0.6},
  citecolor=[rgb]{0.2, 0.6, 0.2},
  urlcolor=[rgb]{0.6, 0.2, 0.2}
}
\usepackage{mathtools}
\usepackage{comment}
\usepackage[margin=1in]{geometry}
\usetikzlibrary{decorations.pathreplacing,backgrounds}

\usepackage{algorithm}
\usepackage{algpseudocode}
\usepackage{accents}

\usepackage{bm}

\usepackage{fancybox}
\usepackage{color}
\usepackage{latexsym}

\usepackage{eepic}
\usepackage{epic}
\usepackage{epsf}
\usepackage{graphics}
\usepackage{dsfont}
\usepackage{anyfontsize}

\newtheorem{theorem}{Theorem}
\newtheorem*{theorem*}{Theorem}

\newtheorem{lemma}[theorem]{Lemma}

\newtheorem{fact}[theorem]{Fact}
\newtheorem{corollary}[theorem]{Corollary}
\newtheorem{claim}[theorem]{Claim}
\newtheorem{observation}[theorem]{Observation}
\newtheorem{proposition}[theorem]{Proposition}
\theoremstyle{definition}

\newtheorem{definition}[theorem]{Definition}
\newtheorem{remark}[theorem]{Remark}
\newenvironment{claimproof}[1]{\par\noindent{\em Proof of Claim.}~#1}

\newcommand{\B}{\mathfrak{B}}
\newcommand{\A}{\mathcal{A}}
\newcommand{\N}{\mathds{N}}

\newcommand{\M}{\mbox{\rm Mat}}

\newcommand{\U}{\mathfrak{U}}

\newcommand{\F}{\mathds{F}}

\newcommand{\Q}{\mathds{Q}}

\DeclareMathOperator{\ncrank}{\mathrm ncrank}

\DeclareMathOperator{\poly}{poly}
\DeclareMathOperator{\rank}{rank}

\DeclareMathOperator{\Gal}{Gal}

\DeclareMathOperator{\sing}{\rm{\textsc{Singular}}}
\DeclareMathOperator{\nsing}{\rm{\textsc{NSingular}}}
\DeclareMathOperator{\pcsing}{\rm{\textsc{PC-Singular}}}
\DeclareMathOperator{\rankincrement}{\rm\textsc{PC-Rank}}
\DeclareMathOperator{\pitsearch}{\rm PC-PIT}

\DeclareMathOperator{\diag}{diag}

\DeclareMathOperator{\cir}{circ}
\DeclareMathOperator{\pcrank}{pc-rank}
\DeclareMathOperator{\RANKINCREMENT}{\textsc{Rank-Increment}}

\newcommand{\Mat}{\mbox{\small\rm Mat}} 
\renewcommand{\P}{\mbox{\rm P}} 
\newcommand{\pc}{\mbox{\rm PC}} 
\renewcommand{\int}{\rm{\mbox{int}}}

\renewcommand{\ge}{\geqslant}
\renewcommand{\geq}{\geqslant}
\renewcommand{\le}{\leqslant}
\renewcommand{\leq}{\leqslant}

\DeclareFontFamily{OMX}{MnSymbolE}{}
\DeclareFontShape{OMX}{MnSymbolE}{m}{n}{
   <-6>  MnSymbolE5
   <6-7>  MnSymbolE6
   <7-8>  MnSymbolE7
   <8-9>  MnSymbolE8
   <9-10> MnSymbolE9
   <10-12> MnSymbolE10
   <12->   MnSymbolE12}{}
\DeclareSymbolFont{mnlargesymbols}{OMX}{MnSymbolE}{m}{n}
\SetSymbolFont{mnlargesymbols}{bold}{OMX}{MnSymbolE}{b}{n}
\DeclareMathDelimiter{\llangle}{\mathopen}{mnlargesymbols}{'164}{mnlargesymbols}{'164}
\DeclareMathDelimiter{\rrangle}{\mathclose}{mnlargesymbols}{'171}{mnlargesymbols}{'171}

\renewcommand{\angle}[1]{\langle #1 \rangle}
\newcommand{\ubar}[1]{\underaccent{\bar} #1}

\DeclareFontEncoding{LS1}{}{}
\DeclareFontSubstitution{LS1}{stix}{m}{n}
\DeclareSymbolFont{symbols2stix}{LS1}{stixfrak} {m} {n}
\DeclareMathSymbol{\lparenless}{\mathopen} {symbols2stix}{"32}
\DeclareMathSymbol{\rparengtr}{\mathclose}{symbols2stix}{"33}

\newcommand{\newbrak}[1]{{\lparenless} #1 {\rparengtr}}

\newcommand{\p}{\textit{\textsf{p}}} 

\title{Trading Determinism for Noncommutativity in Edmonds' Problem}

\author{
V. Arvind\thanks{Institute of Mathematical Sciences (HBNI), and Chennai Mathematical Institute, Chennai, India, \texttt{email: arvind@imsc.res.in.}}  
\and Abhranil Chatterjee\thanks{Indian Statistical Institute, Kolkata, \texttt{email: abhneil@gmail.com.} Research Supported by INSPIRE Faculty Fellowship provided by the Department of Science and Technology, Government of India.} 
\and Partha Mukhopadhyay\thanks{Chennai Mathematical Institute, Chennai, \texttt{email: partham@cmi.ac.in.}}
}

\date{}
\begin{document}
\maketitle

\begin{abstract}
 Let $X=X_1\sqcup X_2\sqcup\ldots\sqcup X_k$ be a partitioned set of variables such that the variables in each part $X_i$ are noncommuting but for any $i\neq j$, the variables $x\in X_i$ commute with the variables $x'\in X_j$. Given as input a square matrix $T$ whose entries are linear forms over $\Q\angle{X}$, we consider
 the problem of checking if $T$ is invertible or not over the universal skew field of fractions of the partially commutative polynomial ring $\Q\angle{X}$ \cite{KVV20}. In this paper, we design a deterministic polynomial-time algorithm for this problem for constant $k$. The special case $k=1$ is the noncommutative Edmonds' problem ($\nsing$) which has a 
 deterministic polynomial-time algorithm by recent results \cite{GGOW16, IQS18, HH21}.  
 
 En-route, we obtain the first deterministic polynomial-time algorithm for the equivalence testing problem of 
 $k$-tape \emph{weighted} automata %over the alphabet $X$ 
 (for constant $k$) resolving a long-standing open problem \cite{HK91, Worrell13}. Algebraically, the equivalence problem reduces to testing whether a partially commutative rational series over the partitioned set $X$ is zero or not \cite{Worrell13}. Decidability of this problem was established by Harju and Karhum\"{a}ki \cite{HK91}. Prior to this work, a \emph{randomized} polynomial-time algorithm for this problem was given by Worrell \cite{Worrell13} and, subsequently, a deterministic quasipolynomial-time algorithm was also developed \cite{ACDM21}.  \end{abstract}
\newpage
\tableofcontents
\newpage
\section{Introduction}\label{sec:intro}
Let $X=\{x_1, x_2, \ldots, x_n\}$ be a set of $n$ variables and $\F$ be a field. Consider the coefficient matrices $A_0, A_1, \ldots, A_n \in \M_s(\F)$, and define the $s \times s$ symbolic matrix $T$ as \[T = A_0 + A_1x_1+\ldots + A_nx_n.\]
In 1967, Edmonds introduced the problem of deciding whether $T$ is invertible over the rational function field $\F(x_1, x_2, \ldots, x_n)$~\cite{Edm67}, often referred to as the $\sing$ problem.
More generally, Edmonds was interested in computing the (commutative) rank of $T$ over the rational function field $\F(x_1, x_2, \ldots, x_n)$. The problem can be restated as computing the maximum rank of a matrix in the affine matrix space generated by the $\F$-linear span of the coefficient matrices $A_i, 1\le i\le n$. This 
was further studied by Lov\'{a}sz~\cite{Lov89}, in the context of graph matching and matroid-related problems. The $\sing$ problem, and more generally the rank computation problem, admits a simple randomized polynomial-time algorithm due to the Polynomial Identity Lemma~\cite{Sch80, Zip79, DL78}. However, the quest for an efficient \emph{deterministic} algorithm remains elusive. Eventually, Kabanets and Impagliazzo showed that any efficient deterministic algorithm for $\sing$ will imply a strong circuit lower bound, justifying the elusiveness over the years \cite{KI04}. Interestingly, the rank computation problem admits a deterministic PTAS algorithm~\cite{BJP18}.

%\textcolor{red}{In recent times, several groups of researchers have studied the complexity of this problem in the noncommutative setting.}
The rank computation problem is also well-studied in the noncommutative setting~\cite{Coh95, FR04}.
More precisely, $T$ is still a linear matrix but the variables $x_1, x_2, \ldots, x_n$ are noncommuting. The problem of testing whether $T$ is invertible ($\nsing$), or the rank computation question is naturally addressed over the noncommutative analog of the commutative function field, \emph{the free skew field} $\F\newbrak{X}=\F\newbrak{x_1, x_2, \ldots, x_n}$. %It takes some work to formally define the free skew field
The free skew field has been extensively studied in mathematics~\cite{Ami66, ami55, Coh71}.
%and the definition is somewhat involved.
%complicated. 
Intuitively, it suffices to state that
$\F\newbrak{X}$ is the smallest field over the noncommutative ring $\F\angle{X}$. %This is the noncommutative analog of the commutative function field $\F(\ubar{x})$. 

%Similar to the commutative setting, the noncommutative rank computation of $T$ has an equivalent definition involving the matrix space generated by the $\F$-linear span of the coefficient matrices. We say that the noncommutative rank of $T$ is $s-c$ if $c$ is the maximum integer such that there exists a \emph{$c$-shrunk subspace}\footnote{We say $U\leq \F^n$ is a $c$-shrunk subspace of a matrix space $\mathcal{B}$ if there exists $W\ \leq \F^n$ such that $\dim(W) \leq \dim(U) - c$ and for all $B\in \mathcal{B}$, the set $\{Bu:u\in U\}\leq W.$} of the coefficient matrix space.  %via matrix blow-up spaces, which we discuss in Section \ref{sec:overview}. 

Two independent breakthrough results showed that $\nsing$ is in $\P$ \cite{GGOW16, IQS18}. The algorithm of Garg, Gurvits, Oliveira, and Wigderson \cite{GGOW16} is analytic in nature and based on operator scaling which works over $\Q$. The algorithm of Ivanyos, Qiao, and Subrahmanyam \cite{IQS18} is purely algebraic, and it works over $\Q$ as well as fields of positive characteristic. Subsequently, a third algorithm based on convex optimization is also developed by Hamada and Hirai \cite{HH21}. Not only are these beautiful results, but also they have enriched the field of computational invariant theory greatly \cite{BFGOWW19, DM20}.  
%The noncommutative version of Edmonds’ problem asks to find the noncommutative rank of $T$ over the free skew field $\F\newbrak{x_1, \ldots, x_n}$.
%In the commutative setting, this problem is solvable in randomized polynomial time due to the Polynomial Identity Lemma. Derandomizing Edmonds' problem is a challenging open problem in theoretical computer science which also captures the PIT of ABPs. However, it admits a deterministic PTAS~\cite{BJP18}. Interestingly, the noncommutative counterpart of the problem admits a deterministic polynomial-time algorithm~\cite{GGOW16, IQS18}.

The main driving motivation for this work is to understand the trade-off between the role of noncommutativity and the complexity of Edmonds' problem. More precisely, let $X_{[k]} = X_1\sqcup X_2\sqcup\ldots\sqcup X_k$ be a partitioned set of variables such that the variables in each $X_i : 1\leq i\leq k$ are noncommuting and $|X_i|\leq n$. However, for each $i\neq j$, the variables in $X_i$ commute with the variables in $X_j$. Given a linear matrix $T$ with (affine)-linear form entries over $X_{[k]}$, the problem is to decide whether $T$ is invertible or not. Of course, in order to consider the invertibility of $T$, we need a skew field of the fractions of the partially commutative polynomial ring $\F\angle{X_{[k]}}$. A construction of such a skew field (which we call as $\U_{[k]}$) is known when the characteristic of $\F$ is zero \cite[Theorem 1.1]{KVV20}. Given the field $\U_{[k]}$, the definition of matrix rank is as usual, the maximum size of any invertible submatrix over $\U_{[k]}$. We define $\pcsing$ as the problem of checking whether such a linear matrix is invertible over $\U_{[k]}$ where $\pc$ stands for the partially commutative nature of the variables. The main result of this paper is the following theorem. 

\begin{theorem}\label{thm:main-theorem}
Given an $s\times s$ matrix $T$ whose entries are $\Q$-linear forms over the partially commutative set of variables $X_{[k]}$ (where $|X_i|\leq n$ for $1\leq i\leq k$), the rank of $T$ over $\U_{[k]}$ can be computed in deterministic $(ns)^{2^{O(k \log k)}}$ time. %$(ns)^{\eta{k^k}}$ time for a constant $\eta>0$. 
The bit complexity of the algorithm is also bounded by $(ns)^{2^{O(k \log k)}}$. % 
%$(ns)^{\eta{k^k}}$.
\end{theorem}

%\begin{remark}
 %   Moreover, if the rank of $T$ over $\U_{[k]}$ is $r$, our algorithm outputs a matrix substitution such that 
%\end{remark}

As a direct corollary of Theorem \ref{thm:main-theorem}, $\pcsing \in \P$  for $k=O(1)$. 
Notice that $\pcsing$ generalizes both $\nsing$ and $\sing$. For $k=1$ it is just $\nsing$ and the above theorem implies
$\nsing$ is in $\P$. Also, letting $|X_i|=1$ for each $i$, it captures the $\sing$ problem with $k$ as a running parameter. 

\begin{remark}\label{rmk:thm1-remark}
 We note two points regarding the choice of the field and the input parameters. 
\begin{enumerate}
\item Theorem \ref{thm:main-theorem} is stated over $\Q$ as the result of \cite{KVV20} works over characteristic zero fields, and we also want that the field arithmetic computation should be efficient. The other ingredients of the proof work also over sufficiently large fields of positive characteristic. 

\item For convenience (and w.l.o.g) throughout the paper we assume $s\geq n$ and express the run time, bit complexity, and the dimension of the matrices used as a function of $s$ and $k$ only. 
\end{enumerate}
\end{remark}
%{\textcolor{red}{Our key insight in obtaining the algorithm of Theorem~\ref{thm:main-theorem} is a new \emph{rank increment step} which we reduce to a polynomial identity testing problem for ABPs over partially commutative set of variables. This step is significantly different from the earlier algorithms for $\nsing$ \cite{GGOW16, IQS18} and makes the proof of Theorem \ref{thm:main-theorem} feasible.: confusing}  

It is to be noted that apart from $\nsing$, the \emph{deterministic polynomial-time} algorithm is known only for a few other special instances of $\sing$ problem defined over linear matrices. We refer the reader to Section \ref{sec:otherresults} for more details.   

%For example, a deterministic polynomial-time algorithm is known if the coefficient matrices of the symbolic matrix is of rank-one or rank-two skew-symmetric \cite{Lov89}. Raz and Wigderson have given a deterministic polynomial-time algorithm for another instance of $\sing$ problem originated in the context of graph rigidity \cite{RW19}. 
%Recently, Ivanyos, Mittal, and Qiao obtain a deterministic polynomial time algorithm where the coefficient matrices generate a matrix Lie algebra \cite{IMQ22}. 

\paragraph{Equivalence testing of multi-tape weighted automata }
En-route to the proof of Theorem \ref{thm:main-theorem}, we obtain the first deterministic polynomial-time algorithm for equivalence testing of $k$-tape weighted automata for $k=O(1)$ resolving a long standing open problem \cite{HK91, Worrell13}. Since the equivalence testing problem of multi-tape automata %originated in 
is closely related to the rich domain of trace monoids (or partially commutative monoids), we make a small detour to it.%provide some pointer before moving into our problem. 

%we resolve the complexity of a long standing problem %in the theory of rational series and algebraic automata theory \cite{BR11} which we discuss now. 

A trace is a set of strings over an alphabet where certain letters (variables) are allowed to commute and others are not. Historically, traces were introduced by Cartier and Foata to give a combinatorial proof of MacMahon's master theorem \cite{CF69}. The trace monoid or the partially commutative monoid is a monoid of traces. More formally, it is constructed by giving an independence relation on the set of commuting letters. This induces an equivalence relation and partitions the given trace into equivalences classes. The set of equivalence classes themselves form a monoid which is a quotient monoid. This is also called the trace monoid which is a foundational object in concurrency theory \cite{Diekert97, Maz95}. 
%Given a partitioned set of variables $X = X_1 \sqcup X_2 \sqcup \ldots \sqcup X_k$, we can define an independence relation on any two different set of commuting variables which induce an equivalence relation on the free monoid $X^*$. The quotient monoid is called a \emph{partially commutative monoid} or a \emph{trace monoid}. The trace monoid has been extensively studied and found numerous applications particularly to model concurrent computation \cite{Diekert97}. 
%\marginnote{\textcolor{red}{We had a line separation which I think was better.}}

%Now, we return to the context of equivalence testing of two multi-tape weighted automata which is our main interest. 
For us the alphabet is the partitioned set of  variables $X_{[k]} = X_1\sqcup X_2\sqcup\ldots\sqcup X_k$. The variables in $X_i$ are noncommuting but the variables in $X_i$ and $X_j$ for $i\neq j$ are mutually commuting. 
%\paragraph{Equivalence testing of algebraic automata}
%{\textcolor{red}{We can start from the equivalence of multitape automata and then introduce it in the algebraic setting.}}
%En-route to the proof of Theorem \ref{thm:main-theorem}, we resolve the complexity of a long standing problem %in the theory of rational series and algebraic automata theory \cite{BR11} which we discuss now. 
Given two $s\times s$ linear matrices $T_1, T_2$ over $X_{[k]}$%$=X_1\sqcup X_2\sqcup\ldots\sqcup X_k$ 
and vectors $u_1, u_2\in\F^{1\times s}$, $v_1,v_2\in\F^{s\times 1}$, the problem is to check whether 
the following infinite series are the same: 
\[
u_1 \left(\sum_{i\geq 0} T^{i}_1\right)v_1 \stackrel{?}{=} u_2 \left(\sum_{i\geq 0} T^{i}_2\right)v_2.
\]
Let $X^*_{[k]}$ denote the set of all monomials (or words) over the variables in $X_{[k]}$. Any monomial $m\in X^*_{[k]}$ can be obtained as some interleaving of monomials $m_i\in X_i^*, 1\le i\le k$. Conversely, given $m\in X^*_{[k]}$ we
can uniquely extract each $m_i\in X_i^*$ by dropping from monomial $m$ the variables in $X_{[k]}\setminus X_i$.
Essentially each $m_i$ is the restriction $m|_{X^{*}_i}$. Hence, two partially commutative monomials 
$m, m'\in X^*_{[k]}$ are the same if and only $m|_{X^{*}_i} = m'|_{X^{*}_i}$ for each $1\leq i\leq k$. 
This defines an equivalence relation $\sim $ over the set of monomials $X^{*}_{[k]}$. To see a simple example, consider $X_1=\{x_1, x_2\}$ and $X_2=\{x'_1, x'_2\}$. Then $x_1 x'_2 x_2 x'_1 \sim x_1 x_2 x'_2 x'_1$.  
%In the language of algebraic automata theory this is the problem of equivalence testing of $k$-tape \emph{weighted} automata given by their transition matrices $T_1$ and $T_2$.  
This is the algebraic formulation of the well-known $k$-tape weighted automata equivalence problem. See \cite[Section~3]{Worrell13} for a detailed discussion.
Equivalence testing of $k$-tape weighted automata was shown to be \emph{decidable} by Harju and Karhum\"{a}ki~\cite{HK91} using the theory of free groups. Indeed, a co-NP upper bound follows from their result as observed by Worrell \cite{Worrell13}. Improved complexity upper bounds for this problem remained elusive, until 
Worrell~\cite{Worrell13} obtained a \emph{randomized} polynomial-time algorithm for testing the equivalence of $k$-tape weighted automata for any constant $k$. Worrell's key insight was to reduce this problem to the polynomial identity testing of algebraic branching programs (ABPs) defined over the partially commutative set of variables $X_{[k]}$ (in other words, the linear forms on the edges of the ABP are in $\F\angle{X_{[k]}}$). Essentially, the reduction says that two infinite series are the same if and only if 
\[
u_1 \left(\sum_{i\leq s} T^{i}_1\right)v_1 = u_2 \left(\sum_{i\leq s} T^{i}_2\right)v_2.
\]
This is obtained by adapting such a result for $k=1$ case suitably for arbitrary $k$ \cite[Corollary~8.3]{Eilenberg74}. 
This is equivalent to the following identity testing problem:
\[
{u}\left(\sum_{i\leq s}T^i\right){v} \stackrel{?}{=} 0
\]

\[
\text{where,}\quad
{u}=\begin{pmatrix}
u_1 & u_2 
\end{pmatrix},~
{v}=\begin{pmatrix}
v_1\\
-v_2
\end{pmatrix}, ~
T=\begin{pmatrix}
T_1 & 0\\
0   & T_2
\end{pmatrix}. 
\]
Clearly, ${u}\left(\sum_{i\leq s}T^i\right){v}$ can be represented as an ABP of width $2s$ and degree $s$ defined over the variable set $X_{[k]}$. 
Then, Worrell developed a partially commutative analogue of the well-known Amitsur-Levitzki Theorem to solve the identity testing problem in randomized polynomial time \cite{AL50, Worrell13}. Building on Worrell's work, in \cite{ACDM21} a \emph{deterministic} quasipolynomial-time algorithm was obtained for any constant $k$. The key technical idea was a bootstrapping of the quasipolynomial-size hitting set for noncommutative ABPs \cite{FS13} to the partially commutative setting. However, the main open question of \cite{HK91, Worrell13} was to design a deterministic \emph{polynomial-time} test that remained elusive. In this paper, we fully resolve the problem by giving the first deterministic polynomial-time algorithm for any constant $k$.% via very different techniques. 
\begin{theorem}\label{thm:main-theorem-2}
Given an ABP of size $s$ whose edges are labeled by $\Q$-linear forms over the partially commutative set of variables $X_{[k]}$ (where $|X_i|\leq n$ for $1\leq i\leq k$), there is a deterministic $(ns)^{2^{O(k \log k)}}$ time algorithm %$(ns)^{\gamma{k^k}}$ time algorithm (for a constant $\gamma >0$) 
to check whether the ABP computes the zero polynomial. As a corollary, the equivalence testing of $k$-tape weighted automata can be solved in deterministic polynomial time for $k=O(1)$. The bit complexity of the algorithm is also bounded by $(ns)^{2^{O(k \log k)}}$. %$(ns)^{\gamma{k^k}}$. 
\end{theorem}
As already mentioned in Remark \ref{rmk:thm1-remark}, that for convenience we always take $s\geq n$. 
We provide more background and other results related to the equivalence testing problem of multi-tape weighted automata in Section \ref{sec:otherresults}.

%uses a  theorem  takes a different approach via Polynomial
%Identity Testing (PIT). In~\cite{Worrell13}, Worrell asked if the
%equivalence testing problem for $k$-tape weighted automata can be solved in \emph{deterministic} polynomial time, for constant $k$. 

%One may naturally wonder whether Theorem \ref{thm:main-theorem} can be obtained by any simple adaptation of techniques in \cite{IQS18} used to solve $\nsing$ problem. We believe that the answer would be "No". The techniques developed in \cite{IQS18} crucially uses the shrunk subspace criteria for the $\nsing$ problem. 

%In this paper, we propose a simpler algorithm for $\nsing$. The main algorithmic template of our algorithm is similar to \cite{IQS18}. However, there is an important difference in implementing one of the core steps that we explain in the next subsection.

%\begin{theorem}\label{thm:main-theorem}
%Given an $s\times s$ matrix $T$ whose entries are $\Q$-linear forms over the noncommuting variables $\{x_1, \ldots, x_n\}$, the noncommutative rank of $T$ over $\Q\newbrak{x_1, \ldots, x_n}$ can be computed in deterministic $\poly(s,n)$ time. As a special case, $\nsing\in\p$.
%\end{theorem}

%\begin{remark}\label{rmk:allfields}
%The result of \cite{IQS18} works over fields of positive characteristics using the rudiments of Galois theory. Similarly, our algorithm can also be extended over fields of positive characteristics. Since the key algorithmic ideas remain exactly the same, we prefer to describe the algorithm only over $\Q$ to minimize the use of the Galois theory machinery.  
%\end{remark}

\subsection{Proof Idea}\label{sec:proofidea}
When a ring $R$ is embeddable in a skew field $\mathfrak{F}$, the notions of rank and singularity of matrices over
$R$ are easier to work with. It is a remarkable fact that in the noncommutative world, even integral domains, in general, need not be embeddable in
a skew field! Cohn's text contains a detailed study of matrix rank over different rings \cite{Coh95}. We also refer the reader to the important paper of Malcev \cite{Mal37}. An $s\times s$ matrix $T$ over such a
ring $R$ is invertible if there is a matrix $T^{-1}$ over $\mathfrak{F}$ such that $TT^{-1}=T^{-1}T=I_s$.\footnote{The inverse if it exists will be unique and is hence denoted $T^{-1}$.}
An $s\times s$ matrix $T$ over this ring $R$ is invertible precisely when its rank is $s$. Likewise, 
the rank of an $s\times t$ matrix $M$ over such a ring $R$ is precisely the maximum $r$ such that $M$ has an $r\times r$ invertible submatrix. An example of this setting is the free noncommutative ring $R=\F\angle{X}$ which embeds in the free skew field $\F\newbrak{X}$. 

%If $X$ is a set of noncommuting variables and $T$ is a linear matrix defined over $\F\angle{X}$, the question whether $T$ is invertible %or not is addressed over the universal free skew field $\F\newbrak{X}$. 

For $S\subseteq [k]$, let $X_S$ be the set of variables in $X_i$ for $i\in S$. Now, if $X$ is a set of partially commutative variables $X=X_{[k]}$, singularity testing (or more generally the rank computation) of any linear matrix $T$ defined over $X_{[k]}$, the construction of a universal skew field containing $\F\angle{X_{[k]}}$ will be required. As already mentioned, such a construction is recently obtained \cite[Theorem 1.1]{KVV20} when $\F$ is characteristic zero. We will denote that universal skew field by $\U_{[k]}$. 
More generally, for a subset of indices $S\subseteq [k]$, we will denote by $\U_{S}$ the universal skew field containing the ring $\F\angle{X_{S}}$. 
This is the main reason that we state our results over fields of characteristic zero and for efficient computational purpose, we fix it to be $\Q$. 

We develop two recursive subroutines $\pitsearch$ and $\rankincrement$ which are the building blocks of our main results. The subroutine $\pitsearch$ takes as input an ABP whose edges are labeled by $\Q$-linear forms over the partially commutative variables $X_{[k]}$ and finds matrix assignments of the form \ref{eqn:tensorstruct1} to the variables in $X_1, X_2, \ldots, X_k$ such that the nonzeroness is preserved. For clarity, when the subroutine $\pitsearch$ handles ABPs over a $\ell$-partition set, we denote it by $\pitsearch_{\ell}$. For example, here we are interested in $\pitsearch_{k}$.  
%\ACnote{It can be mentioned here that the progress is on the parameter $k$.}

The subroutine $\rankincrement$ takes a linear matrix $T$ over $X_{[k]}$ as input and finds matrix assignments to the variables in $X$ of the form \ref{eqn:tensorstruct1} that attains the rank. More precisely, if the rank of $T$ in $\U_{[k]}$ is $r$ and the dimension of the matrices is $d$, then the rank of the scalar matrix obtained from $T$ after the substitution is $rd$. We use $\rankincrement_{\ell}$ to indicate that the subroutine is applied over a $\ell$-partition variable set. In essence, it turns out that the two subroutines $\pitsearch_k$ and $\rankincrement_k$ are interlinked. Indeed, $\pitsearch_{k}$ makes %works by making 
subroutine calls to $\rankincrement_{k-1}$ and, in turn, $\rankincrement_k$ makes %works by making 
subroutine calls to $\pitsearch_{k}$. 

As a warm-up, we first consider $\rankincrement$ subroutine for $k = 1$ case i.e.\ the $\nsing$ problem. This algorithm for $\nsing$, reduces the main algorithmic step (which is the rank increment step) to noncommutative ABP identity testing. It allows us to design a new algorithm for $\nsing$, presented in Section \ref{sect-idea}. It turns out that this connection to ABP identity testing can be lifted in the setting of partially commutative case and proved to be
a key conceptual component in the proofs of Theorem \ref{thm:main-theorem} and Theorem \ref{thm:main-theorem-2}. We do not know if other algorithms for $\nsing$, e.g.\  the algorithm in \cite{IQS18} which is based on the connection between singularity and the existence of shrunk subspaces \cite{FR04}, can be generalized to the partially commutative setting.\footnote{Neither do we know if the other approaches for $\nsing$ in \cite{GGOW16, HH21} are applicable in this setting.}   
%Over any field, the rank (or inner rank) of a matrix $T$ is the size of the largest submatrix which is invertible. 

A crucial notion that plays an algorithmic role in \cite{IQS18} and in our $\nsing$ algorithm is the blow-up rank \cite{DM17, IQS18}.  
Let $T$ be a linear matrix in noncommutative variables. Writing $T = A_0 + \sum_{i=1}^n A_i x_i$, where $A_0, A_1, \ldots, A_n$ are  coefficient matrices, the evaluation of $T$ at a matrix tuple $\ubar{M}=(M_1, M_2, \ldots, M_n)$ of dimension $d$, where each
$M_i$ has scalar entries is:
\[
T(\ubar{M})= A_0\otimes I_d + \sum_{i=1}^n A_i\otimes M_i.
\]
Define $T^{\{d\}}=\{T(\ubar{M})\mid$ each $M_i\in\F^{d\times d}\}$. Notice that $T^{\{d\}}$ contains $sd\times sd$ matrices.
Let $\rank(T^{\{d\}})$ be the maximum rank attained by a matrix in $T^{\{d\}}$. The regularity lemma \cite{IQS18, DM17} shows that $\rank(T^{\{d\}})$ is always a multiple of $d$. Moreover, $\ncrank(T)$ is the maximum $r$ such that for
some $d$ $\rank(T^{\{d\}})=rd$. If for a tuple $\ubar{M}$ of dimension $d$ the rank of $T(\ubar{M})\geq rd$, we say 
that $\ubar{M}$ is a witness of rank $r$.

Guided by the above notion of blow-up rank, the algorithm in \cite{IQS18} has two main steps applied iteratively: the rank increment step, and the rounding and blow-up control step. We briefly sketch their algorithm. Given a matrix $B$ in $T^{\{d\}}$ of rank $\geq rd$, 
the rank-increment step searches\footnote{%Using the notion 
Computing the limit point of a second Wong sequence \cite{IKQS15, IQS18}, %conceptually analogous to 
a non-trivial generalization of augmenting paths algorithm in the bipartite graph matching.} for a new matrix $B'$ in $T^{\{d'\}}$ (where $d' > d$) of rank $\geq rd' + 1$. If no such matrix exists, then $\ncrank(T) = r$ where $\ncrank(T)$ is the rank of $T$ in $\F\newbrak{X}$. Next, the rounding step is a constructive version of the regularity lemma to find another matrix $B''$ in $T^{\{d'\}}$ such that the rank of $B''$ is $r'd'$ where $r'$ is at least $r + 1$. A blow-up in the dimension of $\ubar{M}$ at each iteration incurs an exponential blow-up in the final dimension. They control the dimension increase by dropping rows and columns from the witness matrices along with repeated applications of the rounding step. Finally, it outputs a matrix $\hat{B}$ of rank $r'd''$ where $r' \geq r + 1$ and $d''\leq r'+1$. The rounding step crucially works with matrices from a division algebra (because nonzero matrices in a division algebra are of full rank).

Coming back to our $\nsing$ algorithm, the rounding and blow-up control step is very similar to that in \cite{IQS18}. As already mentioned, the main difference is the rank increment step which we reduce to PIT of noncommutative ABPs. As we show in Lemma \ref{lem:truncated-pcnew}, Lemma \ref{lem:nonzero-output}, and Corollary \ref{cor:immediate}, %for finding a witness of larger rank it suffices to compute a matrix tuple where a noncommutative ABP does not evaluate to zero. 
given a linear matrix $T$ and a rank-$r$ witness of dimension $d$, it essentially suffices to compute a nonzero matrix tuple for a noncommutative ABP of size $rd$ to find a witness of $T$ of noncommutative rank $r+1$.
This can be done with well-known identity testing algorithms \cite{RS05,AMS10}. 
 %This way one can compute a matrix tuple $\ubar{p'}$ of dimension $d_1$ such that the rank of $T(\ubar{p'}) > rd_1$. 
 %Interestingly, 
 This also avoids incurring any super-polynomial bit-complexity blow-up over $\Q$. 
%\ACnote{Can we be more specific here using the notations of the previous paragraph? Like given a $d\times d$ rank-$r$ witness, it suffices to compute a nonzero matrix tuple for a noncommutative ABP of size $rd$ to find a rank-$(r+1)$ witness.}

%\ACnote{Rank increment step of our algorithm can be implemented in NC unlike \cite{IQS18}. Should we mention that?}

 Armed with the intuition for the new algorithm for $\nsing$, we %are in a position to 
 now sketch the main ideas of %behind 
 the proofs of Theorem \ref{thm:main-theorem} and Theorem \ref{thm:main-theorem-2}. It is shown in \cite{KVV20} that a linear matrix $T$ over the partially commutative variable set $X_{[k]}$ is invertible (over the universal skew field $\U_{[k]}$)  if and only if 
 there exists matrix substitutions for the variables $x\in X_i : 1\leq i\leq k$ 
of the form 
\begin{equation}\label{eqn:tensorstruct1}
I_{d_1}\otimes I_{d_2}\otimes\cdots \otimes I_{d_{i-1}}\otimes M_x\otimes I_{d_{i+1}}\otimes\cdots\otimes I_{d_k}
\end{equation}
such that $T$ evaluates to an invertible matrix. Here, $M_x$ is a $d_i\times d_i$ matrix and $d_1, d_2, \ldots, d_k\in\N$. Notice that the structure of the matrices respect the partial commutativity. 
%However, the result in \cite{KVV20} does not provide any bound on the dimensions $d_1, d_2, \ldots, d_k$. 
The basic idea in our proof is to explicitly (and efficiently) find such matrices respecting partially commutative tensor product structures. Our algorithm also confirms that each dimension $d_i$ is at most $s+1$.\footnote{For $k=1$, the result in \cite[Theorem 1.8]{DM17} shows that for linear matrices of size $s$, $s-1$ dimension suffices.}

\paragraph{ABP identity testing over partially commutative variables}

We now give an overview of the $\pitsearch_k$ subroutine. In the noncommutative case ($k=1$), when $X$ is just a set of noncommuting variables, the PIT algorithm in \cite{RS05}, first homogenizes the ABP using standard techniques \cite[Lemma 2]{RS05} and then identity tests each homogenized component. Each homogenized ABP is processed layer by layer. An important feature of
homogeneous noncommutative ABPs is that every nonzero monomial $m$ has \emph{unique parsing}: more 
precisely, the only way the ABP can construct a monomial $m$ is from left to right, one variable at a time.
This allows the algorithm of \cite{RS05} to maintain at layer $i$ (of width $w$) at most $w$ monomials
of degree $i$ that have linearly independent coefficient vectors at that layer. 
%At layer $j$ when it has read a monomial prefix $m$, the algorithm knows that it has \emph{touched} all %monomials 
%with $m$ as prefix which can be created by the ABP in subsequent layers. 

This crucial unique parsing property does not hold for ABPs defined over partially commutative variables (for $k>1$).  
To handle this, we homogenize the input ABP $\A$ over the variable set $X_1$, treating the remaining variables as part of
the coefficients. More precisely, suppose the input ABP $\A$ is of width $w$,  degree $d$ and size $s$. Then it turns out each
$X_1$-homogenized component is an ABP whose edge labels are linear forms $\sum_i \alpha_i x_i$, with $x_i\in X_1$, such that the coefficients $\alpha_i$ are given by ABPs  of size $O(sd)=O(s^2)$ over variables $X_2, X_3, \ldots, X_k$ (Lemma \ref{lem:struct1}). For an 
$X_1$-homogenized ABP, inductively, assume that at the $j^{th}$ level, we have recorded the monomials  $m_1, m_2, \ldots, m_{w}\in X_1^j$ and the corresponding coefficient vectors are $v_1, v_2, \ldots, v_{w}$. The entries of the vectors $v_i$ are ABPs over $X_2, X_3, \ldots, X_k$ of size $O(s^2 j)$. The vector $v_i$ is the vector of coefficients of the monomial $m_i$ in the polynomials computed at each node of layer $j$. Additionally, we maintain the property that the vectors $v_1, v_2, \ldots, v_{w}$ are $\U_{[k]\setminus\{1\}}$-linearly independent and also $\U_{[k]\setminus\{1\}}$-spanning for the set of all vectors corresponding to all monomials in $X_1^j$ (spanning as a left $\U$-module).  
For the $(j+1)^{th}$ level, we need to now compute a similar set of $\U_{[k]\setminus\{1\}}$-linearly independent vectors from among the vectors corresponding to the monomials $\{m_i x_j : 1\leq i\leq w, x_j\in X_1\}$. Clearly the size of such a set is bounded by the width of the ABP. We will see that 
this is reducible to computing the rank of a matrix $M$ whose entries are ABPs over the variables in $X_{[k]\setminus \{1\}}$. 
%However, implementing Gaussian elimination for such matrices is not directly computationally efficient. 
It turns out that, we can linearize this rank problem by adapting a recent result \cite{ACCGM22} proved in the context of noncommutative ($k=1$) setting. 
That is, the rank computation of matrix $M$ is polynomial-time reducible to the rank computation of a 
\emph{linear} matrix $T$ over $X_2, X_3, \ldots, X_k$ of size $O(s^5)$.  This is the place where we recursively call $\rankincrement_{k-1}$ for linear 
matrices of size $O(s^5)$ over the variable set $X_{[k]\setminus \{1\}}$. 

At the end of this process we find a 
surviving monomial $m_1$  
over the variables in $X_1$ whose coefficient is nonzero.  
Given such a monomial, we can use a standard idea (by now) to produce assignments $\{M_x\}_{x\in X_1}$ which makes the polynomial evaluate to a nonzero matrix of polynomials over $X_2, X_3, \ldots, X_k$ \cite{AMS10}. The dimension of the matrices $\{M_x\}_{x\in X_1}$ is bounded by $\deg(f)+1$ (recall that $f$ is the polynomial computed by the ABP), and the entries are over $\{0,1\}$. 

One can then recover the ABP $\mathcal{A}_{m_1}$ over $X_2, X_3, \ldots, X_k$ which is the coefficient of $m_1$. The size of the ABP will be $O(sd^2)=O(s^3)$, however the degree is still bounded by $\deg(f)$. 
%This is the place, the procedure 
It now recursively invokes 
$\pitsearch_{k-1}$ over $\mathcal{A}_{m_1}$ to compute the matrix assignments for the variables in $X_{[k]\setminus \{1\}}$ such that $\mathcal{A}_{m_1}$ evaluates to a nonzero matrix. 

%and repeat the process to find a surviving monomial $m_2$ over $X_2$ and compute the matrices for it. Repeating this process for $k$-times one can recover the matrix substitutions for all the sets $X_1, X_2, \ldots, X_k$ that makes the (nonzero) input polynomial evaluate to a nonzero scalar matrix. 
Since the dimension of the matrices is only a function of $\deg(f)$, it is always bounded by $s+1$ for each $X_i$. Finally, the matrix substitution for the variables $x\in X_i$ will be identified with the form~\ref{eqn:tensorstruct1}. 
%\begin{equation}
%I_{d_1}\otimes I_{d_2}\otimes\cdots \otimes I_{d_{i-1}}\otimes M_x\otimes I_{d_{i+1}}\otimes\cdots\otimes I_{d_k}
%\end{equation}
To summarize, the upshot is that the $\pitsearch_k$ problem over $X_{[k]}$ is deterministic $\poly(s)$-time reducible to $O(s^4)$ instances of $\rankincrement_{k-1}$ for linear matrices over $X_2, X_3, \ldots, X_k$ of size $O(s^5)$ and at most $s$ recursive calls to $\pitsearch_{k-1}$ for ABPs of size $O(s^3)$ defined over $X_{[k]\setminus \{1\}}$ (taking all homogenized components into account).
%. Thus to prove Theorem \ref{thm:main-theorem-2}, it suffices to solve $\rankincrement_k$ problem for linear matrices ($k=O(1)$) in deterministic polynomial time.

\paragraph{Computing linear matrix rank over partially commutative variables}

Now we discuss the construction of the subroutine $\rankincrement_k$. 
Given a linear matrix $T$ of size $s$ over the partially commutative variables 
$X_{[k]}$, let $\pcrank(T)$ denote its rank over the universal skew field $\U_{[k]}$. This is the size of the largest invertible submatrix (over $\U_{[k]}$) of $T$. The subroutine $\rankincrement_k$ finds the matrix assignments of the form \ref{eqn:tensorstruct1} to the variables such the rank of the new scalar matrix becomes a multiple of $\pcrank(T)$. 
%More precisely, the matrix assignment for the variables in $x\in X_i$ is of the form 
%\begin{equation}\label{eqn:witness}
%I_{d_1}\otimes I_{d_2}\otimes\cdots \otimes I_{d_{i-1}}\otimes M_x\otimes I_{d_{i+1}}\otimes\cdots\otimes I_{d_k}. 
%\end{equation}
More precisely, the rank of the scalar matrix after the matrix 
assignments is $d'\cdot \pcrank(T)$ where $d'=d_1 d_2 \cdots d_k$. 
%The result in \cite{KVV20} confirms that such matrix assignments exist. 

%By an induction on $k$, assume that we have already constructed the subroutines $\rankincrement_{\ell}$ for $\ell\leq k-1$. The base case of the induction is $k=1$ which is the $\nsing$ problem. At this point 
Let us define the notion of the witness for $\pcrank$. A set of matrix tuples of the form \ref{eqn:tensorstruct1} is a rank $r$ witness for $T$ if after the substitution, the rank of the scalar matrix is at least $rd'$. 
Now to construct the subroutine $\rankincrement_{k}$, the main idea is to do an induction over $r$. Namely, given a witness for $\pcrank$ $r$ (which we call as the matrix tuple $\ubar{M}$), we would like to construct another witness for $\pcrank$ $r+1$ in deterministic polynomial time unless $r$ is already the $\pcrank(T)$. Note that to construct a witness for rank $r=1$, it suffices to assign values to the variables such that any nonzero linear form in $T$ becomes nonzero, which is clearly trivial. 

Let the input matrix $T$ (of size $s$) be of the following form:
\[
T(X_1, \ldots, X_k) = A_0 + \sum_{j=1}^k\sum_{x \in X_j} A_{x}x.
\]

Given a $\pcrank$ witness $r$ for $T$ of the form 
\ref{eqn:tensorstruct1}, the rank of 
\begin{equation}\label{eqn:six-witness}
T''_{1}=A_0\otimes I_{d'} + \sum_{j=1}^k \sum_{x\in X_{j}} A_x\otimes (I_{d_1}\otimes\cdots I_{d_{j-1}}\otimes M_x\otimes I_{d_{j+1}}\otimes\cdots\otimes I_{d_k})  
\end{equation}
is at least $rd'$ where $T''_{1}$ is the evaluation of $T$ on the witness tuple. Additionally, assume that 
for $1\leq j\leq k$, the dimension $d_j\leq s^3$. We call $\ubar{d}=(d_1, d_2, \ldots, d_k)$ as the shape of the tensor product. 

Let $T_{d'}(Z)$ denote the matrix obtained from $T$ by replacing the variable $x\in X_i$ by the matrix $I_{d_1}\otimes \cdots\otimes I_{d_{i-1}}\otimes Z_x\otimes I_{d_{i+1}}\otimes\cdots\otimes I_{d_k}$ where the dimension of the generic matrix $Z_x$ is $d_i$. 

By generic, we mean that the entries of $Z_x$ are indeterminate variables $z_{x, \ell_1, \ell_2} : 1\leq \ell_1, \ell_2 \leq d_i$. Furthermore, the variables in $Z_i=\{Z_x\}_{x\in X_i}$ are noncommuting but variables across $Z_i$ and $Z_j$ are commuting for $i\neq j$. 
Equivalently, one can view each $Z_{i}$ as the set of variables $\{z_{x, \ell_1, \ell_2}\}_{x\in X_i, 1\leq \ell_1,\ell_2\leq d_{i}}$. 
Thus we have a new set of partially commutative variables over $Z=(Z_1, \ldots, Z_k)$ but with equal number of partitions. It is important to note that $\pcrank(T_{d'}(Z))=d'\cdot \pcrank(T)$ (Corollary \ref{cor:pc-blow-up}). We require a similar observation for our $\nsing$ algorithm (Lemma \ref{lem:scaling-blowup}).  
%By Corollary \ref{cor:pc-blow-up}, $\pcrank(T_{d'}(Z))=d'\cdot \pcrank(T)$. The notation $Z$ is used for $(Z_1, \ldots, Z_k)$ where $Z_{\ell}=\{Z_x\}_{x\in X_{\ell}}$. Equivalently, one can view each $Z_{\ell}$ as the set of variables $\{z_{\ell, i, j}\}_{1\leq i,j\leq d_{\ell}}$.    

In $T_{d'}(Z)$ 
replace the matrices $I_{d_1}\otimes \cdots\otimes I_{d_{i-1}}\otimes Z_x\otimes I_{d_{i+1}}\otimes\cdots\otimes I_{d_k}$ corresponding to $x\in X_i$ by 
\[
I_{d_1}\otimes \cdots\otimes I_{d_{i-1}}\otimes (Z_x + M_x)\otimes I_{d_{i+1}}\otimes\cdots\otimes I_{d_k}
\]
and obtain the matrix $T_{d'}(Z+\ubar{M})$. Note that the scalar part of the matrix is $T''_1$.  
In other words, 
%${T}_{d'_1}$ to be the following linear matrix. 
%\[
%{T}_{d'_1}(Z+M) = T''_{1} +\sum_{j'=1}^k \sum_{x\in X_{j'}} (A_x\otimes I_{d'_1}) ~x + \sum_{x\in X_{1}} A_x\otimes Z_x  
%\]
\begin{equation}\label{eqn:new6-intro}
{T}_{d'}(Z+\ubar{M}) = T''_{1} +\sum_{i=1}^k \sum_{x\in X_{i}} A_x\otimes I_{d_1}\otimes \cdots\otimes I_{d_{i-1}}\otimes Z_x\otimes I_{d_{i+1}}\otimes\cdots\otimes I_{d_k}.  
\end{equation}
By the simple property that the rank of a linear matrix is invariant under shifting of the variables by scalars, we get that $\pcrank({T}_{d'}(Z+\ubar{M}))=\pcrank({T}_{d'}(Z))$.  
%where $\{Z_x : x\in X_{1}\}$ are generic $d'_1\times d'_1$ matrices with noncommutative entries. Collectively, $Z_{1}$ is the set of all variables in $\{Z_x : x\in X_{1}\}$. 
%Note that $T_{d'_1}$ is the matrix obtained from $T$ by first shifting the variables $x\in X_1$ by $M_x\otimes I_{\widetilde{d}}$ and then replacing $x$ by $Z_x$. Similarly the variables $x\in X_{j'}$ is shifted by $I_{d_1}\otimes \widetilde{M}_x$ for $2\leq j'\leq k$.  

%Let us define, 
%\begin{equation}\label{eqn:new2}
%\widehat{T} =  T''_{1} +\sum_{j'=1}^k \sum_{x\in X_{j'}} (A_x\otimes I_{d'_1}) ~x. 
%\end{equation}
%By Lemma \ref{lem:scaling-blowup-pc}, we know that $d'_1\cdot \pcrank(\widehat{T})=\pcrank({T}_{d'_1})$. 
Applying Gaussian elimination, we can transform $T_{d'}(Z+\ubar{M})$ to the following shape:  
\begin{equation}\label{eqn:new7-proofidea}
{T}_{d'}(Z+\ubar{M}) \rightarrow
\left(
\begin{array}{c|c}
I_{rd'} - L & 0 \\
\hline
0 & C - B(I_{rd'} - L)^{-1}A
\end{array}
\right). 
\end{equation}
%Suppose for each $x\in X_j$, we have a $d\times d$ matrix substitution $p_x$ as a witness of rank $r$. We now substitute each $x\in X_j$ by a $d\times d$ generic matrix $Z_x = \left(z^{\{x\}}_{\hat{i}\hat{j}}\right)$.
%We now check whether $\ncrank(T) > r$ or not over $\DR_{j+1}\newbrak{X_j}$ by checking whether $\ncrank(T_d) > rd$ or not over $\DR_{j+1}\newbrak{\ubar{Z}}$. Observe that,
%\begin{observation}
%Evaluating $T$ at $(p_1, \ldots, p_n)$ is equivalent to evaluating $T^{\{d\}}$ by substituting each $z_{i,j,k}$ by $p_{i,j,k}$.    
%\end{observation}
%\[
%{T}_{d'}(Z+M) =
%U\left(
%\begin{array}{c|c}
%I_{rd'} - L & A \\
%\hline
%B & C
%\end{array}
%\right)V. 
%\]
%Furthermore,
%\begin{equation}\label{eqn:new7}
%{T}_{d'}(Z+M) =
%U U'\left(
%\begin{array}{c|c}
%I_{rd'} - L & 0 \\
%\hline
%0 & C - B(I_{rd'} - L)^{-1}A
%\end{array}
%\right) V'V. 
%\end{equation}

%\[
%\text{Here,~}
%U'= \left(
%\begin{array}{c|c}
%I_{rd'}  & 0 \\
%\hline
%B (I_{rd'} - L)^{-1} & I_{(s-r)d'}
%\end{array}
%\right)
%,
%\quad\quad 
%V'= \left(
%\begin{array}{c|c}
%I_{rd'}  & (I_{rd'}-L)^{-1}A \\
%\hline
%0 & I_{(s-r)d'}
%\end{array}
%\right).
%\]
  
The matrices $L, A, B, C$ are linear matrices over the variables in $Z_{1}, Z_{2}, \ldots, Z_k$. The $(\ell_1,\ell_2)^{th}$ entry of $C - B(I_{rd'} - L)^{-1}A$ is given by $S_{\ell_1, \ell_2}=C_{\ell_1\ell_2} - B_{\ell_1}(I_{rd'} - L)^{-1}A_{\ell_2}$ where $B_{\ell_1}$ is the $\ell_1^{th}$ row vector of $B$ and $A_{\ell_2}$ is the $\ell_2^{th}$ column vector of $A$. Now we notice a simple fact that shows $\pcrank(T)>r$ if and only if $S_{\ell_1,\ell_2}\neq 0$ for at least one pair $(\ell_1,\ell_2)$ (Lemma \ref{lemma:reduce-to-pc-PIT}). This is the partially commutative version of Lemma \ref{lemma:reduce-to-PIT} that we prove in the context of $\nsing$ problem. 

Notice that $S_{\ell_1, \ell_2}$ has the following series expansion
\[
S_{\ell_1 \ell_2}=C_{\ell_1\ell_2} - B_{\ell_1}\left(\sum_{i\geq 0} L^i\right)A_{\ell_2}. 
\]
The refined goal is to find a nonzero for the series which allows us to construct a witness of $\pcrank(T)\geq r+1$. 
If the series is defined over the free noncommuting variables, a standard result \cite[Corollary 8.3]{Eilenberg74} shows that the infinite series is nonzero if and only if the polynomial 
\[
S^{\leq rd'}_{\ell_1 \ell_2}=C_{\ell_1\ell_2} - B_{\ell_1}\left(\sum_{i\leq rd'} L^i\right)A_{\ell_2}\neq 0. 
\]
%A self contained proof is given in Lemma \ref{lem:trucated}. 
The same result can be extended to the partially commutative case to obtain a similar statement. In \cite[Proposition 5]{Worrell13}, Worrell proves this statement using Ore domains. A self contained proof is given in Lemma \ref{lem:truncated-pcnew}. 
 The important (and simple) observation is that the polynomial $S^{\leq rd'}_{\ell_1 \ell_2}$ can be represented by a partially commutative ABP of width $\leq rd'$ and degree $rd'+2$ over the variable set $Z_1, Z_2, \ldots, Z_k$. 

Hence, we can apply $\pitsearch_k$ subroutine on the ABP computing the partially commutative polynomial $S^{\leq rd'}_{\ell_1 \ell_2}$. 
%As already outlined, the $\pitsearch_k$ subroutine reduces the problem to $\rankincrement_{\ell}$ ($\ell\leq k-1$) computation for linear matrices over $\bigcup_{i\in S} X_i$ where $S\subseteq \{2,3,\ldots,k\}$ and $|S|=\ell$, and finds a nonzero of the polynomial $S^{\leq rd'}_{\ell_1 \ell_2}$. 
Additionally, we observe that a suitable scaling of the nonzero of the ABP will be a nonzero for the infinite series $S_{\ell_1 \ell_2}$ also. This is by the combined effect of applying Theorem \ref{thm:pittorank} and Lemma \ref{lem:bound-zero-finding}. As a result, we obtain a matrix tuple that witness the $\pcrank(T)>r$. Now we need a rounding operation that should produce a witness for $\pcrank(T)\geq r+1$ and also a blow-up control procedure that controls the dimension of the matrices.
This step is somewhat similar in spirit to the rounding and blow-up control steps for the $\nsing$ algorithm, but requires additional conceptual ideas. More precisely, given a linear matrix $T$ over $X_{[k]}$ of size $s$ and matrix tuple of shape $(d_1, \ldots, d_k)$ such that the rank of the image $> r d_1 d_2\cdots d_k$,  
our idea is to update the matrix substitution such that the rank of the image of the new substitution $\geq (r+1)d_1 d_2\cdots d_k$. Indeed, assuming that %each $d_i$ is a distinct prime number (or even if 
the $d_i$ are pairwise relatively prime, the rounding step turns out to be essentially like the noncommutative case. However, this assumption makes the blow-up control step harder. Even if we start with a substitution of shape $(d_1, \ldots, d_k)$ where each $d_i$ is prime, it might fail for $d_i - 1$. To overcome this, our idea is to relax the dimension upper bound of the witness matrix. Instead of reducing $d_i$ one at a time, we allow it to drop to the next (suitable) prime number
less than $d_i$. A theorem about the distribution of primes in small intervals helps us find such a prime close enough to $d_i$ \cite{LS12}.

\subsection{Other Related Results}\label{sec:otherresults}
Among the specific instances of $\sing$ problem, a deterministic polynomial-time algorithm is known if the coefficient matrices of the symbolic matrix is of rank-one or rank-two skew-symmetric \cite{Lov89}. Raz and Wigderson have given a deterministic polynomial-time algorithm for another instance of $\sing$ problem originated in the context of graph rigidity \cite{RW19}. Another result by Ivanyos, and Qiao gives deterministic polynomial-time algorithm for a special case of $\sing$ problem related to symmetrization or skew-symmetrization problem \cite{IQ19}. 
Recently, Ivanyos, Mittal, and Qiao obtain a deterministic polynomial-time algorithm where the coefficient matrices generate a matrix Lie algebra \cite{IMQ22}. 

Equivalence testing of multitape automata is a foundational algorithmic question and has a long history. 
One-way multitape automata were introduced in the seminal paper of Rabin and Scott \cite{RS59}. The equivalence testing problem of multitape nondeterministic automata is undecidable \cite{Griff68}. Here the 
equivalence means the words accepted as sets and the question is to decide whether two sets are the same.  
The problem was shown to be decidable for $2$-tape \emph{deterministic} automata independently by
Bird \cite{Bird73} and Valiant \cite{Val74}. Subsequently, an exponential upper bound was obtained for it \cite{Beeri76}. 
Eventually, for two-tape deterministic automata, a polynomial-time algorithm was given in \cite{FG82}. 
As already mentioned, using the theory of free groups, Harju and Karhum\"{a}ki \cite{HK91} established the decidability of \emph{multiplicity equivalence} of multitape nondeterministic automata. More generally, they prove that the weighted equivalence testing of multitape automata is decidable. One of their open questions was to give an efficient algorithm for the weighted equivalence testing problem when the number of tapes is any constant. Worrell's result giving a randomized polynomial-time algorithm is the first major progress in this direction \cite{Worrell13}, followed by the quasipolynomial deterministic bound given in \cite{ACDM21}. A relatively recent result analyzes the combinatorial method of Bird \cite{Bird73} more carefully, and it shows a polynomial-time algorithm for the equivalence problem for $k$-tape \emph{deterministic} automata (where the coefficients are only $0-1$) when $k=O(1)$ \cite{GS20}. Our paper completes this line of investigation by obtaining the first deterministic polynomial-time equivalence test for weighted $k$-tape automata for $k=O(1)$, thereby improving on the previous algorithmic results.

%that subsumes all the algorithms mentioned so far.            

\subsection*{Organization.} In Section \ref{sec:prelim}, we collect background results from algebraic complexity theory and cyclic division algebras. We give the algorithm for noncommutative singularity testing in Section \ref{sect-idea}. The main results (Theorem \ref{thm:main-theorem} and Theorem \ref{thm:main-theorem-2}) are proved in Section \ref{sec:rank-general-case}. We state a few question for further research in Section \ref{sec:discussion}.

\section{Background and Notation}\label{sec:prelim} 

%\subsection*{Basic notation.}
Throughout the paper, we use $\F, F, K$ to denote fields. ${\M_m(\F)}$ (or ${\M_m(F)}, {\M_m(K)}$) will denote $m$-dimensional matrix algebras over $\F$ (resp.\ $F$ or $K$) where $m$ will be clear from the context. Similarly, ${\M_m(\F)}^n$ (resp. ${\M_m(F)}^n, {\M_m(K)}^n$) will denote the set of $n$ tuples over ${\M_m(\F)}$ (resp. ${\M_m(F)}, {\M_m(K)}$).  
$D$ is used to denote a division algebra. We use $X$ to denote a set of variables. 
Sometimes, we use $\ubar{p}, \ubar{q}, \ubar{M}, \ubar{N}$ to denote matrix tuples in suitable matrix algebras. The free noncommutative ring or partially commutative ring of polynomials over a field $\F$ is denoted by $\F\angle{X}$ where 
$X$ is clear from the context. The notation $A\otimes B$ denotes the usual tensor product of the matrices $A$ and $B$. We use $[k]$ to denote the set $\{1,2,\ldots,k\}$. Let $X=X_1 \sqcup X_2 \sqcup \cdots\sqcup X_k$. For $S\subseteq [k]$, let $X_S$ be the set of variables in $\bigsqcup_{i\in S} X_i$. In particular, if $X$ is a set of partially commutative variables, it is denoted by $X_{[k]}$.  

%For a series (or polynomial) $S$, the coefficient of a monomial (word) in $S$ is denoted by $[m]S$. The formal power series over $\M_m(\F)$ is denoted by $\M_m(\F)\dangle{\ubar{x}}$.  
%We use $\Supp(S)$ to denote the \emph{support} of the series $S$: $ \Supp(S)=\{m\mid [m]S\ne 0\}$.
%Noncommutative monomials are also words. 

%$\F\angle{\ubar{x}}$ is the $\F$-algebra of noncommutative polynomials and $R_{\F}\newbrak{x}$ is the set of rational expressions. For a rational expression $r$, let $\dom(r)$ be the set of matrix tuple (of any dimension) where $r$ is defined. 

\subsection{Algebraic Complexity} 

%We first recall the definition of an algebraic branching program.

\begin{definition}[Algebraic Branching Program]\label{abpdefn}
An \emph{algebraic branching program} (ABP) is a layered directed
acyclic graph. The vertex set is partitioned into layers
$0,1,\ldots,d$, with directed edges only between adjacent layers ($i$
to $i+1$). There is a \emph{source} vertex of in-degree $0$ in the layer
$0$, and one out-degree $0$ \emph{sink} vertex in layer $d$. Each edge
is labeled by an affine $\F$-linear form. The polynomial computed by
the ABP is the sum over all source-to-sink directed paths of the
ordered product of affine forms labeling the path edges. 
\end{definition}

The \emph{size} of the ABP is defined as the total number of nodes and the \emph{width} is the maximum number of nodes in a
layer. The ABP model can compute commutative or noncommutative polynomials (depending on the variable set $X$). ABPs of width $w$ can also be seen as iterated matrix multiplication $ \ubar{c}\cdot M_1 M_2 \cdots M_{\ell} \cdot\ubar{b} $, where $\ubar{c}, \ubar{b}$ are $1\times w$ and $w \times 1$ vectors respectively and each $M_i$ is a $w \times w$ matrix, whose entries are affine linear forms over $\ubar{x}$. 

Similarly, the ABP model can be used to compute polynomials over a set $X_{[k]}$ of partially commutative variables. The only difference is that the linear forms are over $\F\angle{X_{[k]}}$ and two monomials $m,m'\in X^{*}$ are same under the equivalence relation $\sim$ as described in Section \ref{sec:intro}. 

\begin{definition}[Linear Pencil for Noncommutative Polynomials]\label{def:linear-pencil}
A noncommutative polynomial $g\in\F\angle{X}$ is said to have a size $s$ \emph{linear pencil} $L$ if $L$ is an $s\times s$ invertible linear matrix over $X$ such that $g$ is computed in the $(1,s)^{th}$ entry of $L^{-1}$.
\end{definition}

We can now generalize Definition \ref{def:linear-pencil} for partially commutative polynomials also where $X = X_{[k]}$.

\begin{definition}[Linear Pencil for Partially Commutative Polynomials]\label{def:linear-pencil-pc}
A partially commutative polynomial $g\in\Q\angle{X_{[k]}}$ is said to have a size $s$ \emph{linear pencil} $L$ if $L$ is an $s\times s$ invertible linear matrix over $X_{[k]}$ such that $g$ is computed in the $(1,s)^{th}$ entry of $L^{-1}$.
\end{definition}

Since we will be using the result in \cite{KVV20} throughout the paper, whenever we talk about invertibility over the partially commutative setting, the field is always fixed to be $\Q$. 

Given an ABP that computes a polynomial $f$ at the the $(1,w)^{th}$ entry of the matrix product $M_1 M_2 \cdots M_d$ where each $M_i$ is of size $w\times w$, it is well-known that the polynomial can be computed at the upper right corner of the inverse of a linear matrix $L_f$ of small size. This was explicitly stated in \cite[Equation 6.4]{HW15} in the context of noncommutative variables. However, we can immediately see that the construction also holds for a partially commutative set of variables. We give the formal statement.   

\begin{proposition}\label{prop:abp-pencil}
An ABP of size $s$ (width $w$, and depth $d$) computing a polynomial $f$ over the partially commutative variables $X_{[k]}=\bigsqcup_{i=1}^k X_i$ has the following linear pencil of size bounded by $2s$:  
\[
L_f = 
\begin{bmatrix}
I_w &-M_1\\
& I_w & -M_2\\
&&\ddots &\ddots\\
&&&I_w &-M_d\\
&&&&I_w
\end{bmatrix}.
\]
The polynomial $f$ is computed at the upper right corner of $L^{-1}_f$. 
\end{proposition}

We also record the following simple observation that talks about the partial evaluation of a polynomial defined over $X_{[k]}$. This is field independent. 

\begin{observation}\label{obs:partial-eval}
Let $f\in \F\angle{X_{[k]}}$ be a partially commutative polynomial. For each $i\in [k]$ and $x\in X_i$, let $M_x$ be a $d_i\times d_i$ matrix. Consider the following matrices:
\begin{enumerate}
\item Substitute each $x\in X_1$ by $M_x$ in $f$ and obtain a $d_1\times d_1$ matrix $M_1\in \Mat_{d_1}(\F\angle{X_{[k]\setminus \{1\}}})$. Similarly, define a $(d_1  d_2\cdots d_i)\times (d_1 d_2\cdots d_i)$ matrix $M_i\in \Mat_{d_1d_2\cdots d_i}(\F\angle{X_{[k]\setminus [i]}})$ by substituting each $x\in X_i$ by $M_x$ in the $(d_1d_2\cdots d_{i-1})\times (d_1d_2\cdots d_{i-1})$ matrix $M_{i-1}\in \Mat_{d_1d_2\cdots d_{i-1}}(\F\angle{X_{[k]\setminus [i-1]}})$. Let $M_k$ be the final matrix.

\item Let $M^*$ be the matrix evaluation of $f(X_{[k]})$ substituting for each $i\in [k]$, each $x\in X_i$ by
\[
I_{d_1}\otimes \cdots I_{d_{i-1}}\otimes M_x\otimes I_{d_{i+1}} \cdots I_{d_k}.
\] 
\end{enumerate}
Then, it computes the same matrix i.e.\ $M_k = M^*$.
\end{observation}

\begin{proof}
Consider a monomial $m$ in $f(X_{[k]})$. We can write $m=m_1 m_2 \cdots m_k$ where $m_i\in X^*_i$. Let $N_{i,m}=\prod_{x\in m_i} M_x$ be the $d_i\times d_i$ matrix. Now, $M_1 = \sum_{m} N_{1,m}\otimes m_2\otimes \cdots \otimes m_k$ from the definition. Therefore, 
$M_k = \sum_m N_{1,m}\otimes N_{2,m}\otimes \cdots\otimes N_{k,m}$. Clearly from the definition of each $N_{i,m}$ the contribution of each $m$ in $M^*$ is also $N_{1,m}\otimes N_{2,m}\otimes \cdots\otimes N_{k,m}$. 
\end{proof}

%This construction is well-known and also used in \cite{HW15}.
%if their restrictions over each $X^{*}_i$ are identical. In other words, $m\sim m'$ if and only if for each $1\leq i\leq k$, the restriction words $m|_{X^{*}_i}=m'|_{X^{*}_i}$. 

%Here $X^{*}_i$ is the set of monomials (or words) over the variable set $X_i$.

%\marginnote{Do we need to rewrite the sentence about monomial equivalence $\sim$? I did some
%rewrite of this...} 
%We also consider commutative set-multilinear ABPs and read-once oblivious ABPs (ROABPs). For the set-multilinear case, the
%(commutative) variable set is partitioned as $Y = Y_1\sqcup
%Y_2\sqcup\cdots \sqcup Y_d$ where for each $j\in [d]$, $Y_j =
%\{y_{ij}\}_{i=1}^n$. An ABP $B$ is a homogeneous set-multilinear if each edge in the $j^{th}$ layer of the ABP is labelled by linear forms over $Y_j$. For ROABP, a different variable is used for each layer, and the edge labels are univariate polynomials. Therefore, an ROABP of $d$ layers can be represented as $ \boldsymbol{c} \cdot M_1(v_1) M_2(v_2) \cdots M_{v_d}(d) \cdot\boldsymbol{b}$. We say that the ROABP respects the variable order $v_1 < v_2 < \cdots < v_d$.

%Given a homogeneous degree $d$ noncommutative polynomial $f$, its set-multilinearization $\sm(f)$ is the corresponding set-multilinear polynomial obtained by replacing $x_i$ in the $j^{th}$ position (in a monomial) by $y_{i,j}$ in every monomial. Clearly, $f\equiv 0$ if and only if $\sm(f)\equiv 0$.

%\vspace{1mm}
%\subsubsection*{Identity testing results}
%\begin{description}
\subsubsection{Identity testing results}\label{sec:pit-results} 
%\end{description}

For noncommutative ABPs, Raz and Shpilka obtained a deterministic polynomial-time algorithm for identity testing \cite{RS05}. 

\begin{theorem}[Raz-Shpilka \cite{RS05}]\label{razshpilka}
Given as input a noncommutative ABP of width $w$ and $d$ many layers computing a polynomial $f \in \F \angle{X}$, there is a deterministic $\poly(w,d,n)$ time algorithm to test whether or not $f \equiv 0$. 
\end{theorem} 
In fact, the following corollary is standard by now. This was first formally observed in \cite{AMS10} using a minor adaptation of \cite{RS05}. 

\begin{corollary}\label{cor:rs-ams-adapted}
Given a noncommutative ABP of width $w$ and $d$ many layers computing a nonzero polynomial $f \in \F \angle{X}$, there is a deterministic $\poly(w,d,n)$ time algorithm which outputs a nonzero monomial $m$ in $f$. If $\F=\Q$, the bit complexity of the algorithm is $\poly(w,d,n,b)$ where $b$ is the maximum bit complexity of any coefficient in the input ABP.  
\end{corollary}

Essentially, the algorithm of Raz and Shpilka maintains basis vectors (indexed by at most $w$ monomials) in each layer of the ABP using simple linear algebraic computations. The entries of the basis vectors are the coefficients of the indexing monomials in different nodes of that layer of the ABP. 

Given such a monomial $m=x_{i_1} x_{i_2} \ldots x_{i_d}$, \cite{AMS10} introduced a simple trick to produce a matrix tuple in 
$\M_{d+1}(\F)^n$ on which $f$ evaluates to nonzero. To see that consider a $d+1$ state deterministic finite automaton $\mathcal{A}$ that accepts only the string $x_{i_1} x_{i_2} \ldots x_{i_d}$ over the alphabet $\{x_1, x_2, \ldots, x_n\}$. The transition matrix tuple 
$(M_{x_1}, \ldots, M_{x_n})$ of $\mathcal{A}$ have the property that $f(M_{x_1}, \ldots, M_{x_n})\neq 0$. More precisely, the automaton $\mathcal{A}$ is the following.

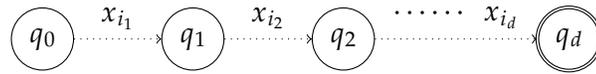
\begin{figure}[H]
\begin{center}
\begin{tikzpicture}
\node(pseudo) at (-1,0){};
\node(0) at (0,0)[shape=circle,draw]        {$q_0$};
\node(1) at (2,0)[shape=circle,draw]        {$q_1$};
\node(2) at (4,0)[shape=circle,draw]        {$q_2$};
\node(3) at (7,0)[shape=circle,draw,double] {$q_{d}$};
\path [->]
  (0)      edge                 node [above]  {$x_{i_1}$}     (1)
  (1)      edge                 node [above]  {$x_{i_2}$}     (2)
  (2)   [dotted]   edge                 node [above]  {$\cdots \cdots ~~x_{i_d}$}     (3);
  %(2)      edge [bend left=30]  node [below]  {a}     (0)
  %(0)      edge [loop above]    node [above]  {$\xi_1$}     ()
  %(1)      edge [loop above]    node [above]  {$\xi_2$}     ()
%(2)      edge [loop above]    node [above]  {$\xi_3$}     ()
 % (3)      edge [loop above]    node [above]  {$\xi_{k +1}$}   ()
  %(pseudo) edge                                       (0);
\end{tikzpicture}
\end{center}
\end{figure}

The transition matrices $M_{x_j} : 1\leq j\leq n$ are $(d+1)$ dimensional $(0,1)$-matrices with the property that $M_{x_j}(\ell,\ell+1)=1$ if and only if $x_j$ is the edge label between $q_{\ell}$ and $q_{\ell+1}$ for $0\leq \ell\leq d-1$.  
This we record as a corollary. 

\begin{corollary}\label{cor:abp-witness}
Given a noncommutative ABP of width $w$ and $d$ layers computing a nonzero polynomial $f\in\F\angle{X}$, there is a deterministic polynomial-time algorithm that can output a matrix tuple $(M_1, M_2, \ldots, M_n)$ of dimension at most $d+1$ such that $f(M_1, M_2,\ldots, M_n)\neq 0$. 
\end{corollary}

%\subsubsection{A structural result}\label{sec:structural} 
\subsubsection{Homogenization}\label{sec:structural} 
A noncommutative (or commutative ABP) over variable set $X$ can be easily homogenized using standard ideas. 
The standard reference is the survey by Shpilka and Yehudayoff \cite[Chapter 2]{SY10}. This is also explained in \cite[Lemma 2]{RS05}. We observe that the same homogenization extends to partially commutative ABPs defined over the variable set $X_{[k]}$ in the following sense. This result is field independent. 

%Recall that each set $X_i$ the variables are noncommuting and accross different sets $X_i, X_j ~(i\neq j)$ %they are commuting. 

\begin{lemma}\label{lem:struct1}
Let $f\in \F\angle{X_{[k]}}$ be a partially commutative polynomial of degree $d$ computed by an ABP of size $s$. Then for any $1\leq j\leq k$, we can efficiently homogenize the ABP over the variable set $X_j$, and the coefficients are also computed by ABPs over $\F\angle{X_{{[k]}\setminus\{j\}}}$ of size $O(sd)$.   
\end{lemma}

\begin{proof}
W.l.o.g, we describe the homogenization w.r.t.\ the variable set $X_1$. The construction is standard and we provide a sketch. Every node  $v$ is replaced by a set of nodes $(v,0), (v,1), \ldots, (v,d)$ where the node $(v,j)$ computes the $j^{th}$ homogenized component of the polynomial computed at the node $v$. Let $v\rightarrow u$ be an edge in the ABP labeled by $L'+L$ where 
$L$ is a linear form over $X_1$ and $L'$ is an affine linear form over $X_2\sqcup \cdots \sqcup X_k$. Then we connect $(v,i)$ to $(u,i)$ with a label $L'$ for $0\leq i\leq d$. 
Similarly, we connect $(v,i)$ to $(u,i+1)$ with a label $L$. 

The next step is to get rid of edges that labeled with affine linear forms over $\F\angle{X_2, \ldots, X_k}$. 
This process is repeated layer by layer starting from the source vertex at the left most layer. 
Suppose that there is an edge between $(v,i)\rightarrow (u,i)$ labeled with $L'$ over $\F\angle{X_2, \ldots, X_k}$. 
For an edge $(w,i-1)\rightarrow (v,i)$ already labeled with an ABP $g$ will be changed to $(w,i-1)\rightarrow (u,i)$ with the label $g\cdot L'$. If there is already an edge between $(w,i-1)$ to $(u,i)$ with a label $g'$ which is an ABP, we update the edge label $(w,i-1)\rightarrow (u,i)$ by $g\cdot L' + g'$. We repeat this process until we get rid of all the edges carrying affine linear forms over $\F\angle{X_{{[k]}\setminus\{1\}}}$. Clearly each of the ABPs on the edges are of size $O(sd)$.   
\end{proof}

\subsection{Cyclic Division Algebras}\label{sec:cyclic}
A division algebra $D$ is an associative algebra over a (commutative) field $\F$ such that all 
nonzero elements in $D$ are units (they have a multiplicative inverse). In the context of this 
paper, we are interested in finite-dimensional division algebras. Specifically, we focus on cyclic division algebras and their construction \cite[Chapter 5]{Lam01}. 

We describe the construction over $\F=\Q$. 
Let $F=\Q(z)$, where $z$ is a commuting indeterminate. Let $\omega$ be an $\ell^{th}$ primitive root of unity. To
be specific, let $\omega= e^{2\pi i/\ell}$. Let
$K=F(\omega)=\Q(\omega,z)$ be the cyclic Galois extension of $F$ obtained by
adjoining $\omega$. The elements of $K$ are polynomials in $\omega$ (of
degree at most $\ell-1$) with coefficients from $F$.

Define $\sigma:K\to K$ by letting $\sigma(\omega)=\omega^k$ for some $k$
relatively prime to $\ell$ and stipulating that $\sigma(a)=a$ for all
$a\in F$. Then $\sigma$ is an automorphism of $K$ with $F$ as fixed
field and it generates the Galois group $\Gal(K/F)$.

The division algebra $D=(K/F,\sigma,z)$ is defined using a new
indeterminate $x$ as the $\ell$-dimensional vector space:
\[
D = K\oplus Kx\oplus \cdots \oplus Kx^{\ell-1},
\]
where the (noncommutative) multiplication for $D$ is defined by
$x^\ell = z$ and $xb = \sigma(b)x$ for all $b\in K$. Then $D$ is a
division algebra of dimension $\ell^2$ over $F$
\cite[Theorem 14.9]{Lam01}. 
\begin{definition}\label{def:index}
The \emph{index} of $D$ is defined to be the square root of the dimension of $D$ over $F$. In our example, $D$ is of index $\ell$.
\end{definition}

%The Galois automorphism $\sigma$ naturally extends from $D\rightarrow D$ by fixing $x$. In other words $\sigma(x)=x$. 
The elements of $D$ has matrix representation 
in $K^{\ell\times \ell}$ from its action on the basis 
$\mathcal{X}=\{1,x,\ldots,x^{\ell-1}\}$. I.e., for $a\in D$ and $x^j\in\mathcal{X}$, the $j^{th}$ row of the matrix representation is obtained by writing $x^{j} a$ in the $\mathcal{X}$-basis. 
%Its elements have matrix representations %in
%$K^{\ell \times \ell}$ (the regular matrix representation defined by
%multiplication from the left) given below:

For example, the matrix representation $M(x)$ of $x$ is:

\[
        M(x)[i,j] = \begin{dcases}
                        1 & \text{ if } j=i+1, i\le \ell-1 \\
                        z & \text{ if } i=\ell, j=1\\
                        0 & \text{ otherwise.}
                    \end{dcases}
\]

$$
M(x)=\begin{bmatrix}
    0       & 1 & 0 & \cdots & 0 \\
     0       & 0 & 1 &\cdots  &  0 \\
      \vdots & \vdots &\ddots &\ddots  &  \vdots \\
      0       & 0 &\cdots                  & 0 &1 \\
      z       & 0 & \cdots & 0 & 0
\end{bmatrix}.
$$

For each $b\in K$ its matrix representation $M(b)$ is:

\[
        M(b)[i,j] = \begin{dcases}
                        b & \text{ if } i=j=1 \\
                        \sigma^{i-1}(b) & \text{ if } i=j, i\ge 2\\
                        0 & \text{ otherwise.}
                    \end{dcases}
\]

\[M(b) = 
\begin{bmatrix}
b & 0 & 0 & 0 & 0 & 0  \\
0 & \sigma(b) & 0 & 0 & 0 & 0 \\
0 & 0 & \sigma^2(b) & 0 & 0 & 0 \\
0 & 0 & 0 & \ddots & 0 & 0 \\
0 & 0 & 0 & 0 & \sigma^{\ell-2}(b) & 0 \\
0 & 0 & 0 & 0 & 0 & \sigma^{\ell-1}(b)
\end{bmatrix}
\]
        
\begin{remark}
We note that $M(x)$ has a ``circulant'' matrix structure and $M(b)$ is
a diagonal matrix. For a vector $v\in K^\ell$, it is convenient to
write $\cir(v_1,v_2,\ldots,v_\ell)$ for the $\ell\times \ell$ matrix
with $(i,i+1)^{th}$ entry $v_i$ for $i\le \ell-1$, $(\ell,1)^{th}$
entry as $v_{\ell}$ and remaining entries zero. Thus, we have
$M(x)=\cir(1,1,\ldots,1,z)$.  Similarly, we write
$\diag(v_1,v_2,\ldots,v_\ell)$ for the diagonal matrix with entries
$v_i$.
\end{remark}

\begin{fact}
  The $F$-algebra generated by $M(x)$ and $M(b), b\in K$ is an
  isomorphic copy of the cyclic division algebra in the matrix algebra
  $\M_{\ell}(K)$.
\end{fact}

\begin{proposition}\label{circ-in-D}
  For all $b\in K$, $\cir(b,\sigma(b),\ldots,z\sigma^{\ell-1}(b)) = M(b)\cdot M(x)$.
\end{proposition}

Define $C_{i,j}= M(\omega^{j-1}) \cdot M(x^{i-1})$ for $1\leq i,j\leq \ell$. Observe that, $\B=\{C_{ij}, i,j \in [\ell]\}$ be a $F$-generating set for the division algebra $D$. The following proposition is a standard fact. 

%\begin{definition}\label{def:center}
%The center $\mathfrak{C}$ of the division algebra is the set of elements 
%$a\in D$ such that $a$ commutes with every %element in $D$. 
%\end{definition}
%It is easy to observe that $\mathfrak{C}$ is a field. Another standard fact is the following. 

\begin{proposition}\cite[Section 14(14.13)]{Lam01}\label{full-space}
Then $K$ linear span of $\B$ is the entire matrix algebra 
$\M_{\ell}(K)$. 
\end{proposition}

\subsection{Partially Commutative Rational Series}\label{sec:ncratseries}
In the following lemma, we prove that the zero testing of a series defined over partially commutative variables can be reduced to the zero testing of a polynomial of low degree. This extends such a result known for $k=1$ \cite[Corollary 8.3, Page 145]{Eilenberg74} (Also, see \cite[Example 8.2, Page 23]{DK21}) 
to the partially commutative setting where  $X=X_{[k]}$. The proof is linear algebraic and we crucially use the fact that the partially commutative ring $\F\angle{X_{[k]}}$ is embedded in the universal skew field $\U_{[k]}$ \cite{KVV20} as mentioned in Section \ref{sec:intro} (a formal statement regarding the construction 
of $\U_{[k]}$ is given in Theorem \ref{thm:gen-skew-field}). In \cite[Proposition 5]{Worrell13}, Worrell has proved the same result using Ore domains. %\marginnote{We say $\le s$ here but $<s$ in the
%previous lemma...} 

\begin{lemma}\label{lem:truncated-pcnew}
Consider the universal skew field $\U_{[k]}$ over $\F\angle{X_{[k]}}$. Let $L\in \U_{[k]}^{s\times s}$ be a linear matrix over $X_{[k]}$, ${u}, {v}$ are $1\times s$ and $s\times 1$ dimensional vectors whose entries are linear forms over $X_{[k]}$. Then ${u} \left(\sum_{i\geq 0} L^i\right){v}=0$ if and only if ${u} \left(\sum_{i\leq s} L^i\right){v}=0$.  
\end{lemma}

\begin{proof}
If ${u} \left(\sum_{i\geq 0} L^i\right){v}=0$ then clearly ${u} \left(\sum_{i\leq s} L^i\right){v}=0$ since different homogeneous components will not mix together. 

To see the other direction, we first note that ${u} \left(\sum_{i\leq s} L^i\right){v}=0$
implies ${u}L^i{v}=0, 0\le i\le s$ as each term in the sum is a different homogeneous part. Now consider the $s+1$ many vectors $v_i={u}\cdot L^i$ for $0\leq i\leq s$. Since each $v_i$ is in the left $\U_{[k]}$-module $U^s$, and $\U_{[k]}$ is a (skew) field, they cannot all be $\U_{[k]}$-linearly independent. That means there are $\lambda_0, \ldots, \lambda_s$ in $\U_{[k]}$, not all zero, such that the \emph{left} linear combination $\sum_{j=0}^s \lambda_j v_j=\sum_{j=0}^s \lambda_j {u} L^j=0$. Let $t$ be the largest index such that $\lambda_t$ is nonzero. 
Then we can write ${u}\cdot L^{{t}}=-\sum_{j=0}^{t-1}{{\lambda_t}}^{-1}\lambda_j {u}\cdot L^j$. Multiplying both sides on the right by $L^{s+1-t}$, we obtain 
\[
{u}\cdot L^{{s+1}}=-\sum_{j=0}^{t-1}{{\lambda_t}}^{-1}\lambda_j {u}\cdot L^{j+s+1-t}.
\]
But this will imply that ${u}\cdot L^{s+1}\cdot {v}=0$ since ${u}\cdot L^{j+s+1-t}{v}=0$ for $j\leq t-1$. Now, assuming inductively that ${u}L^i{v}=0$
for some $i \ge s+1$, we can similarly prove that ${u}L^{i+1}{v}=0$. It follows that 
the entire series is zero.   
\end{proof}

\subsection{Equivalent Notions of Matrix Rank}\label{sec:rank-equivalence}
We first recall the definition of the noncommutative rank of a linear matrix in noncommutative 
variables, the computationally useful notion of its blow-up rank, and their equivalence.\footnote{We note that Cohn's text \cite{Coh95} has a detailed discussion of matrix rank over general noncommutative rings.}

\begin{definition}\label{defn:ncrank}
The noncommutative rank ($\ncrank$) of an $s\times s$ linear matrix $T$ over the noncommuting variables $x_1, x_2, \ldots, x_n$ is equal to the size of the largest invertible (square) submatrix of $T$.  
\end{definition}

Let $T$ be an $s\times s$ matrix whose entries are affine linear forms over $\{x_1,x_2, \ldots,x_n\}$. We can write $T = A_0 + \sum_{i=1}^n A_i x_i$ where $A_0, A_1, \ldots, A_n$ are the coefficient matrices.   
Given matrix $T$, for $d\in\mathds{N}$ we define the set of "blow-up" matrices
\[
T^{\{d\}}\ =\ \{T(\ubar{M}) \mid \ubar{M} \in {\M_d(\F)}^n\},
\]  
where $T(\ubar{M})=A_0\otimes I_d + \sum_{i=1}^n A_i\otimes M_i$. Then
we define the \emph{blow-up rank} of $T$ at $d$ as 
$\rank(T^{\{d\}})= \max_{\ubar{M}} \{\rank(T(\ubar{M}))\}.$  The regularity 
lemma \cite{IQS17, DM17, IQS18} shows that $\rank(T^{\{d\}})$ is always a multiple $bd$
of $d$. Thus, we can define the \emph{blow-up rank} of $T$ as $b$, which the largest
positive integer such that for some $d$ we have $\rank(T^{\{d\}})=bd$. The regularity
lemma also implies that the blow-up rank of $T$ is precisely $\ncrank(T)$.

\begin{fact}\label{fact:ncblow-up}
For a linear matrix $T$ in noncommutative variables $\ncrank(T)$ is its
blow-up rank.
%If $\ncrank(T)=r$, the lemma shows that for some $d$ there is a matrix tuple 
%$\ubar{p}\in\M_d(\F)^n$ such that $\rank(T(\ubar{p}))=rd$.
\end{fact}

%\marginnote{We are not using the term blow-up rank in the definition?}

%\marginnote{Modified the text regarding blow-up rank.}

The blow-up rank is algorithmically useful \cite{IQS17,DM17,IQS18}.  In Section \ref{sec:construct-regular}, 
we will discuss this aspect further. Let us now consider a set $X=X_{[k]}$ of partially commutative variables and $T$ be a linear matrix with linear forms over $\F\angle{X_{[k]}}$. The main result of \cite{KVV20} is stated in the following theorem. 

\begin{theorem}\cite[Theorem 1.1]{KVV20}\label{thm:gen-skew-field}
For arbitrary $k\in \N$, the ring $\F\angle{X_{[k]}}$ can be embedded in a universal skew field of fractions $\U_{[k]}$.  \end{theorem}

As a consequence of the above theorem and some properties of noncommutative rings
\cite{Coh95}, we can define the rank of matrices over $\F\angle{X_{[k]}}$ for a partially
commutative variable set $X_{[k]}$ as follows.

\begin{definition}\label{defn:pcrank}
The partially commutative rank ($\pcrank$) of an $s\times s$ linear matrix $T$ over the 
partially commutative variable set $X_{[k]}$ is equal to the size of the largest invertible (over $\U_{[k]}$) 
square submatrix of $T$. 
\end{definition}

The following crucial result is also shown in \cite{KVV20}. 

\begin{proposition}\cite[Proposition 3.8]{KVV20}\label{prop:pc-folded}
A matrix $T$ is invertible over $\U_{[k]}$ if and only if there exists matrix substitutions for 
the variables $x\in X_i : 1\leq i\leq k$  of the form 
\begin{equation}\label{eqn:1prelim}
I_{d_1}\otimes I_{d_2}\otimes\cdots \otimes I_{d_{i-1}}\otimes M_x\otimes I_{d_{i+1}}\otimes\cdots\otimes I_{d_k}
\end{equation}
such that $T$ evaluates to an invertible matrix. Here $M_x$ is a $d_i\times d_i$ matrix and $d_1, \ldots, d_k\in\N$.    
\end{proposition}

The above proposition in fact gives us the right analogue of blow-up rank for the partially
 commutative ring $\F\angle{X_{[k]}}$, and is crucial for our algorithms.
Since we will be using partially commutative matrix substitutions of the above kind for variables
in $X_{[k]}$ for rank computations in this paper, we introduce
the following useful definition. 

\begin{definition}\label{def:type-fold-tensor}
We call the matrix substitution of the form given in the expression \ref{eqn:1prelim} as a type-$i$ $k$-fold tensor product. Also $\ubar{d}=(d_1,d_2,\ldots,d_k)$ is the shape of the tensor. 
\end{definition}

Thus, for a linear matrix $T$ over $X_{[k]}$ we seek type-$i$ matrix
substitutions for variables in $X_i$ for each $i$, which are $(d_1,d_2,\ldots,d_k)$ shape tensor products
for a suitable choice of the dimensions $d_i, 1\le i\le k$.

\section{An Algorithm for NSINGULAR based on NC-PIT}\label{sect-idea}

The key ideas for the proofs of Theorems \ref{thm:main-theorem} and \ref{thm:main-theorem-2} come from the design of a somewhat simpler algorithm for $\nsing$ (which is the case for $k=1$) that we discuss in this section. As explained earlier, the algorithm in \cite{IQS18} has two main steps: rank increment, rounding and blow-up control. In the simpler algorithm, rounding and blow-up control is essentially the same as
in \cite{IQS18}. But the rank increment step is quite different. It is based on an efficient reduction to the noncommutative ABP identity testing. This connection extends to the partially commutative setting and plays a crucial role in the proofs of Theorems \ref{thm:main-theorem} and \ref{thm:main-theorem-2}. Motivated by Fact \ref{fact:ncblow-up}, we give the following definition. We fix the field to be $\Q$.

\begin{definition}[Witness of $\ncrank$ $r$]\label{defn:nc-rank-witness}
Let $A_0, A_1, \ldots, A_n \in \M_s(\Q)$ and $T = A_0 + \sum_{i=1}^n A_ix_i$. 
%We say $(\alpha_1, \ldots, \alpha_n) \in \F^n$ is a witness of commutative rank $r$ of $T$ if $\rank(T(\alpha_1, \ldots, \alpha_n))\geq r$. 
We say that $\ubar{p}=(p_1, \ldots, p_n) \in \M_d(\Q)^n$ for some $d$ is a witness of noncommutative rank (at least) $r$ of $T$, if $\rank(T(\ubar{p}))\geq rd$.
\end{definition}

\subsection{Constructive Regularity Lemma}\label{sec:construct-regular}

Suppose that for a linear matrix $T$, we already have a matrix tuple $\ubar{q}$ over $\M_d(\Q)$, a witness of rank $r$ of $T$ such that $\rank(T(\ubar{q}))>rd$. Then the constructive regularity lemma offers
a simple and general procedure to get a $d\times d$ witness of rank $r+1$ for $T$ \cite{IQS18}. 
We present essentially the same proof as described in \cite{IQS18}. But for clarity and for setting the context of the main results in the next section, we use the explicit cyclic division algebra construction described in Section~\ref{sec:cyclic}. Following Section~\ref{sec:cyclic}, the field $F=\Q(z)$ and $K=F(\omega)$.    
%The proof uses the construction of a central division algebra described in \cite[Lemma~5.7]{IQS18}. 

%\marginnote{Please see the rewrite sentence "...clarity...". Is it okay?}

\begin{lemma}\label{lem:const-reg}\cite{IQS18}
For any $s\times s$ matrix $T = A_0 + \sum_{i=1}^n A_i x_i$, and a matrix tuple $\ubar{q}=(q_1,\ldots, q_n) \in \M_d(\Q)^n$ such that $\rank(T(\ubar{q})) > rd$, there exists a deterministic $\poly(n,s,d)$-time algorithm that returns another
matrix substitution $\ubar{q'}=(q'_1,\ldots, q'_n)\in \M_d(\Q)^n$ such that $\rank(T(\ubar{q'})) \geq (r+1)d$.
\end{lemma}

\begin{proof}

Let $D = (K/F,\sigma,z)$ be the cyclic division algebra described in Section~\ref{sec:cyclic}. Recall that $\B = \{C_{i,j} : i,j\in[d]\}$ is a $F$-generating set of $D$. 

\begin{enumerate}
\item By Proposition \ref{full-space}, we can express $q_k = \sum_{i,j} \lambda_{i,j,k}C_{i,j}$ where $\lambda_{i,j,k}$,  $1\leq k\leq n$ are unknown variables which take values in $K$. A linear algebraic computation yields the values $\lambda^{0}_{i,j,k}$ where $1\leq i,j\leq \ell$, and $1\leq k\leq n$ for the unknowns in $K$.\\

\item Now the goal is to compute a $d\times d$ tuple $\ubar{q''}=(q''_1,\ldots, q''_n)$ such that $q''_k= \sum_{i,j} \mu^{0}_{i,j,k}C_{i,j}$ where $\mu^{0}_{i,j,k}\in \Q$ and $\rank(T(\ubar{q''}))\geq (r+1)d$. 
We briefly describe the procedure outlined in~\cite{IQS18}. Write $\tilde{q}_1= \mu_{1,1,1} C_{1,1,1} + \sum_{(i,j)\neq (1,1)} \lambda^0_{i,j,1}C_{i,j}$ where $\mu_{1,1,1}$ is a variable. There will be a sub-matrix of size $>rd$ whose minor is non-zero, under the current substitution $(\tilde{q}_1, q_2, \ldots, q_n)$. Since the determinant of that sub-matrix is a univariate polynomial in $\mu_{1,1,1}$ and degree $\poly(r,d)$, we can easily fix the value of $\mu_{1,1,1}$ from $\Q$ such that the minor remains nonzero. Repeating the procedure, we can compute a tuple $\ubar{q''}$.      
Since $\ubar{q''}$ is a tuple over the division algebra, $\rank(T(\ubar{q''}))\geq (r+1)d$. 
\end{enumerate}
%\begin{remark}\label{rmk:rounding}
The last line of the above proof is easy to see. The matrix $T(\ubar{q''})$ can be viewed as a $s\times s$ block-matrix of $d$-dimensional blocks, and each such block is an element in $D$.   
%There will be a minor in $T(\ubar{q'})$ of rank $>rd$. 
Since Gaussian elimination is supported over division algebras, up to elementary row and column operations, we can transform $T(\ubar{q''})$ as:
\[
\left(
\begin{array}{c|c}
I & 0 \\
\hline
0 & 0
\end{array}
\right)
\]
where $I$ is an identity matrix which has at least $r+1$ blocks of identity matrices
$I_d$ on its diagonal. 
%\[
%I=\begin{bmatrix}
%I_{d}\\
%&I_{d}\\
%&&\ddots\\
%&&&&I_d
%\end{bmatrix}
%\]
%\end{remark}
%and the number of $I_d$ blocks is at least $r+1$. 
Hence $\rank(T(\ubar{q''}))\geq (r+1)d$. 
%\end{remark}
From the tuple $\ubar{q''}$, we can easily obtain the desired tuple $\ubar{q'}$ as follows. We can think $\omega$ and $z$ as fresh commutative parameters $t_1, t_2$. Clearly, after the substitution the determinant of that $(r+1)d$ dimensional submatrix is a bivariate polynomial in $t_1, t_2$ of degree $\leq (r+1)d$. We can set the variables from a set of size $O(rd)$ in $\Q$ such that the submatrix remains invertible. Replacing the variables $t_1, t_2$ by such values over $\Q$, we get the tuple $\ubar{q'}$ defined over $\Q$.    
\end{proof}
%\subsection{The Plan of the Algorithm}\label{sec:algo-plan}
%We first give a simple template.

%\begin{description}
%\item[Algorithm Template]\

%\textbf{Input:} $T = A_0 + \sum_{i=1}^n A_ix_i$ where $A_0, A_1, \ldots, A_n \in \M_s(\F)$.

%\textbf{Output:} The noncommutative rank of $T$.

%\vspace{0.2cm}
%It is an incremental algorithm that 
%The algorithm gradually 
%constructs a witness at every stage. Suppose we already have a \emph{witness of rank $r$} for $T$. 
%\vspace{0.1cm}
%\begin{enumerate}
%\item Is $r$ the maximum rank?

%\item If yes, output $r$ to be the noncommutative rank of $T$.

%\item Otherwise, find a witness of rank at least $r+1$ and go to Step~1.
%\end{enumerate}
%\end{description}
%We now discuss each step in detail. 

\subsection{Rank Increment Step}\label{sec:rankincrement}
This is quite different from the rank increment step in \cite{IQS18}. More importantly, this turns out to be 
readily extendable to the partially commutative case in the proof of Theorem \ref{thm:main-theorem}. The increment step gradually constructs a witness at every stage. Given a \emph{witness of rank $r$} for $T$, the algorithm checks if $r$ is the maximum possible rank. If not, it produces a witness of rank at least $r+1$ by solving an instance of ABP identity testing and iterates. At a high level, it has a conceptual similarity with the idea used in \cite{BBJP19} in approximating commutative rank.
%\marginnote{Please see the second sentence rewrite...}

For an $s\times s$ linear matrix $T(\ubar{x}) = A_0 + \sum_{i=1}^n A_ix_i$ and $d\in\N$, define 
\[T_d(Z) = A_0\otimes I_d + \sum_{i=1}^n A_i\otimes Z_i\] 
where $Z_i = (z^{(i)}_{jk})_{1\leq j,k \leq d}$ is a $d\times d$ generic matrix with noncommutative indeterminates. In other words, $Z=(Z_1, Z_2, \ldots, Z_n)$ is the substitution used for the variables $x_1,x_2,\ldots,x_n$ in $T$. Now $T_d(Z)$ is a linear matrix of dimension $sd$ over the variables $\{z^{(i)}_{jk}\}_{1\leq j,k \leq d, 1\leq i\leq n}$. 
\begin{remark}\label{rmk:matrix-shift}
%Since we substitute the variables $x_1,\ldots,x_n$ by the $d\times d$ matrices $Z_1,\ldots,Z_n$, one can 
It is immediate to see that any $d\times d$ matrix shift $T_d(Z_1 + p_1, Z_2 + p_2,\ldots, Z_n + p_n)$ is indeed a scalar shift for the variables 
$\{z^{(i)}_{jk}\}_{1\leq j,k \leq d, 1\leq i\leq n}$ in the matrix $T_d$.
\end{remark}
%Intuitively, we would like to consider any $d\times d$ matrix shift of $T$ as a scalar substitution of $T_d$.

%For any integer $d'\in\mathbb{N}$, we define the bijective map $\iota_{d'}: \M_{d'}(K)^{nd^2} \to \M_{dd'}(K)^n$ as follows:
%\[
%\ubar{p}=(p^{(1)}_{11}, \ldots, p^{(1)}_{dd}, \ldots, p^{(n)}_{11}, \ldots, p^{(n)}_{dd})
%\mapsto \ubar{q}=(q_1, \ldots, q_n)
%\]
%where each $q_i = (p^{(i)}_{jk})_{1\leq j,k\leq d}$, i.e. we think of $q_i$ as $d\times d$ block matrix where the $(j,k)^{th}$ block is $p^{(i)}_{jk}$. 
%The inverse image is also well-defined.
%Notice that $T_{d}(\ubar{p})=A_0\otimes I_{dd'} + \sum_{i=1}^n A_i\otimes q_i = T(\ubar{q})$ where $\ubar{q}=\iota_{d'}(\ubar{p})$. 
%Notice that, for any matrix tuple $(q_1, \ldots, q_n)\in \M^n_{dd'}(\F)$, $T(\ubar{q}) = T_d(\iota^{-1}_{d'}(\ubar{q}))$.

\begin{lemma}\label{lem:scaling-blowup}
$\ncrank(T_{d}) = d \cdot \ncrank(T).$
\end{lemma}

\begin{proof}
Let $\ncrank(T) = r$. Then, for every sufficiently large $d''$, the maximum rank obtained by evaluating $T$ over all  $d''\times d''$ matrix tuples is $rd''$. Let $d'' = dd'$ be a multiple of $d$ and let $\ubar{q}=(q_1, \ldots, q_n)$ be a matrix tuple such that $\rank(T(\ubar{q})) = rdd'$. Let 
\[
\ubar{p}=(p^{(1)}_{11}, \ldots, p^{(1)}_{dd}, \ldots, p^{(n)}_{11}, \ldots, p^{(n)}_{dd})
\]
be the matrix tuple such that each $q_i = (p^{(i)}_{jk})_{1\leq j,k\leq d}$. That is, we think of $q_i$ as the $d\times d$ block matrix where the $(j,k)^{th}$ block is $p^{(i)}_{jk}$. Notice that $T_{d}(\ubar{p})=A_0\otimes I_{dd'} + \sum_{i=1}^n A_i\otimes q_i = T(\ubar{q})$, with the matrix $q_i$ substituted for the variable $x_i$ in $T$. Therefore, $\rank(T(\ubar{q})) = \rank(T_d(\ubar{p}))$ and $\ncrank(T_d)\geq rd$.
%Suppose each $q_i = (p_{ijk})$, i.e. we think of $q_i$ as $d\times d$ block matrix where the $j,k$-th block is $p_{ijk}$. Consider evaluating $T_d$ by substituting $p_{ijk}$ for each $z_{ijk}$ variable. So we have a $d'\times d'$ substitution for the $z_{ijk}$ variables such that the rank is $rdd'$. 

%To show that $\ncrank(T_d)\leq rd$, assume to the contrary that $\ncrank(T_d)> rd$. Suppose, $\ncrank(T_d) = rd+k$ for some positive integer $k$. Therefore, for some $d'\times d'$ matrix tuple substitution $\ubar{p}\in\M_{d'}(K)^{nd^2}$, $\rank(T_d(\ubar{p})) = (rd+k)d' = rdd' + kd'$. From the definition, $\rank(T_d(\ubar{p})) = \rank(T(\iota_{d'}(\ubar{p}))) > rdd'$.
%Now define a $dd'\times dd'$ tuple $(q_1, \ldots, q_n)$ such that $q_i = (p_{ijk})$, i.e. we think of $q_i$ as $d\times d$ block matrix where the $j,k$-th block is $p_{ijk}$. Therefore, $\rank(T(\ubar{q})) = rdd' + kd'$. 
%However, $\ncrank(T) = r$ and for any $(q_1, \ldots, q_n)\in \M_{dd'}(K)^n$, $\rank(T(\ubar{q}))\leq rdd'$, hence a contradiction. Therefore, $\ncrank(T_d)\leq rd$.
For the other direction, as $\ncrank(T) = r$, we can write $T = P Q$ where $P, Q$ are $s\times r$ and $r\times s$ matrices respectively with linear entries \cite{Coh95}. 
%This follows from the equivalence of inner rank and $\ncrank$ \cite{Coh95}. 
We can now define an $sd\times rd$ matrix $P'(Z)$ by substituting each $x_i$ by $Z_i$ in the matrix $P(\ubar{x})$. Similarly, we can define a $rd\times sd$ matrix $Q'(Z)$ from $Q(\ubar{x})$. Notice that, $T_d = P' Q'$. Therefore, $\ncrank(T_d)\leq rd$. %\footnote{The same conclusion can be easily derived from the size of the largest submatrix (which is $r$) also.}. 
Hence, the lemma follows. 
\end{proof}

\subsubsection{A noncommutative ABP identity testing reduction step}\label{sec:block-reduction}
Suppose, now, that we have computed a witness of noncommutative rank $r$ of $T$, namely $\ubar{p}=(p_1, \ldots, p_n) \in \M_d(\Q)^n$ (by construction, we will ensure that $d\leq r+1$). %If $\ncrank(T) > r$, we now construct a witness of rank $r+1$.
We will now describe how to check whether $\ncrank(T) > r$ or not. Observe that
%\begin{observation}
%Evaluating $T$ at $(p_1, \ldots, p_n)$ is equivalent to evaluating $T^{\{d\}}$ by substituting each $z_{i,j,k}$ by $p_{i,j,k}$.    
%\end{observation}
\[
T_d(Z_1 + p_1, \ldots, Z_n + p_n) =
U\left(
\begin{array}{c|c}
I_{rd} - L & A \\
\hline
B & C
\end{array}
\right)V
\]
for invertible transformations $U, V$ in $\M_{rd}(\Q)$. In fact, applying further invertible transformations $U', V'$, we can write 
\[
T_d(Z_1 + p_1, \ldots, Z_n + p_n) =
U U'\left(
\begin{array}{c|c}
I_{rd} - L & 0 \\
\hline
0 & C - B(I_{rd} - L)^{-1}A
\end{array}
\right) V'V. 
\]

\[
\text{Here,~}
U'= \left(
\begin{array}{c|c}
I_{rd}  & 0 \\
\hline
B (I_{rd} - L)^{-1} & I_{(s-r)d}
\end{array}
\right)
,
\quad\quad 
V'= \left(
\begin{array}{c|c}
I_{rd}  & (I_{rd}-L)^{-1}A \\
\hline
0 & I_{(s-r)d}
\end{array}
\right).
\]

Let $\widetilde{T_d}=C - B(I_{rd} - L)^{-1}A$. The entries in $C, B, L, A$ are linear forms over the variables $X$. 
Notice that the $(i,j)^{th}$ entry of $\widetilde{T_d}$ is given by $\widetilde{(T_d)}_{ij} = C_{ij} - B_{i}(I_{rd} - L)^{-1}A_{j}$ where $B_i$ is the $i^{th}$ row vector of $B$ and $A_j$ is the $j^{th}$ column vector of $A$. 

\begin{lemma}\label{lemma:reduce-to-PIT}
$\ncrank(T) > r$ if and only if $\widetilde{(T_d)}_{ij}\neq 0$ for some choice of $i,j$.
\end{lemma}

\begin{proof}
 %Multiplying further by invertible matrices, we can write,
%\[
%T_d({Z}+\ubar{p}) =
%U\left(
%\begin{array}{c|c}
%I_{rd} - L & 0 \\
%\hline
%0 & C - B(I_{rd} - L)^{-1}A
%\end{array}
%\right) V. 
%\]
Let $\ncrank(T) > r$. Then by Lemma \ref{lem:scaling-blowup}, $\ncrank(T_d)>rd$. The noncommutative rank of a linear matrix is invariant under a scalar shift \footnote{Suppose a linear matrix $L$ achieves the maximum rank at matrix substitution $\ubar{q}$ of some dimension $d$. Then, for any scalar shift $(\alpha_1,\ldots,\alpha_n)$, the linear matrix $L(\ubar{x}+\ubar{\alpha})$ achieves the same rank at the matrix substitution $\ubar{q}-\ubar{\alpha}\otimes I_d$.\label{footnote}}, 
hence $\ncrank(T_d(Z_1+p_1, \ldots, Z_n + p_n)) = \ncrank (T_d) > rd$. However, if $C - B(I_{rd} - L)^{-1}A$ is a zero matrix, this is impossible. 

Conversely if $\widetilde{(T_d)}_{ij} = C_{ij} - B_{i}(I_{rd} - L)^{-1}A_{j}$ is nonzero for some indices $i,j$, we can find matrix substitutions $\tilde{p}^{(k)}_{\ell_1\ell_2}$ of dimension $d'$ for the variables $\{{z}^{(k)}_{\ell_1\ell_2}\}_{1\leq \ell_1, \ell_2\leq d, 1\leq k\leq n}$, such that the rank of $T_d(Z_1+p_1, \ldots, Z_n+p_n)$ on that substitution is more than $rdd'$. Therefore, $\ncrank(T_d(Z_1+p_1, \ldots, Z_n+p_n)) > rd$.
%More precisely, let $\tilde{\ubar{p}}=\left(\tilde{p}^{\{k\}}_{\ell_1,\ell_2}\right)_{1\leq \ell_1, \ell_2\leq d, 1\leq k\leq n}$. 
%$\ubar{q}$ from sufficiently large dimension $dd'$ such that 
%Let $\ubar{q}=\iota_{d'}(\tilde{\ubar{p}})$ and $\ubar{p}\otimes I_{d'}=(p_1\otimes I_{d'}, \ldots, p_n\otimes I_{d'})$, then the rank of $T_d(\ubar{q}+\ubar{p}\otimes I_{d'})$ is more than $rdd'$. 
%This is same as saying the rank of $T_d(Z)$ is more than $rdd'$ when the variables $z^{(k)}_{\ell_1,\ell_2}$ is substituted by $\tilde{p}^{k}_{\ell_1,\ell_2} + p_k(\ell_1,\ell_2) \otimes I_{d'}$ for $1\leq k\leq n, 1\leq \ell_1, \ell_2\leq d$.
Hence $\ncrank(T_d)>rd$. 
By Lemma \ref{lem:scaling-blowup}, we get that $\ncrank(T)>r$. 
\end{proof}

Now, applying Lemma \ref{lem:truncated-pcnew} for $k=1$ we note that the infinite series $\widetilde{(T_d)}_{ij}\neq 0$ if and only if the truncated polynomial 

\begin{equation}\label{eqn:zero}
\widetilde{P}_{ij}=C_{ij} - B_i\left(\sum_{k\leq rd} L^k\right)A_j\neq 0. 
\end{equation}

To see that Lemma \ref{lem:truncated-pcnew} is applicable above, notice that the $C_{ij}$ is a linear form and $B_i\left(\sum_{k\leq rd} L^k\right)A_j$ generates terms of degree at least $2$. 
Next, we apply Corollary \ref{cor:rs-ams-adapted} and Corollary \ref{cor:abp-witness} to output a matrix tuple efficiently on which $\widetilde{(T_d)}_{ij}$ evaluates to nonzero and $I_{rd}-L$ evaluates to a full rank matrix. 

\begin{lemma}\label{lem:nonzero-output}
There is a deterministic $\poly(n,r,d)$-time algorithm that can output a matrix tuple $\ubar{q}$ of dimension at most $d'=2rd$ for the ${Z}$ variables such that $I_{rdd'}-L(\ubar{q})$ is invertible and $\widetilde{(T_d)}_{ij}(\ubar{q})\neq 0$. 
%for some choice of $i,j$. Moreover, if $\widetilde{(T_d)}_{i,j}\neq 0$ for some $i,j$, it outputs a witness of its non-zeroness. 
\end{lemma}

\begin{proof}
 Notice that $\widetilde{P}_{ij}$ is an ABP of size $\poly(r,d)$ and the number of layers is at most $rd+1$. Applying Corollary \ref{cor:abp-witness}, we get a matrix tuple of dimension at most $rd+2$ such that $\widetilde{P}_{ij}$ evaluates on it to nonzero. By simple padding, we can get a matrix tuple $\ubar{q'}$ of dimension $d'=2rd$ such that $\widetilde{P}_{ij}(\ubar{q'})\neq 0$. Since $\ubar{q'}$ is a substitution for the $Z$ variables 
 $\{z^{(k)}_{\ell_1\ell_2}\}$ where $1\leq k\leq n, 1\leq \ell_1, \ell_2\leq d$, we write $\ubar{q'}=(q'^{(1)}_{11}, \ldots, q'^{(1)}_{dd}, \ldots, q'^{(n)}_{11}, \ldots, q'^{(n)}_{dd})$ for more clarity. Here each $q'^{(k)}_{\ell_1\ell_2}$ is a $d'$ dimensional matrix.  
 
 Consider a commutative variable $t$ and the scaled matrix tuple $t\ubar{q'}$. It is easy to see that the infinite series $C_{ij} - B_i (I_{rdd'}-L(t\ubar{q'}))^{-1} A_j$ is nonzero since the $k^{th}$ homogeneous part $t^k B_i L^k(\ubar{q'}) A_j$ will not mix with other homogeneous components.   
 
 However this also has a rational representation $\widetilde{(T_d)}_{ij}(t\ubar{q'})=\gamma_1(t)/\gamma_2(t)$ where $t$-degrees of the polynomials $\gamma_1(t), \gamma_2(t)$ are bounded by $rdd'$. Moreover,  
 $I_{rdd'}-L(t\ubar{q'})$ is an invertible matrix and the degree of $\det(I_{rdd'}-L(t\ubar{q'}))$ is bounded by $rdd'$ over the variable $t$. Simply by varying the variable $t$ over a suitable large set $\Gamma$ of size $O(rd)$, we can fix a value for $t=t_0$ such that $\widetilde{(T_d)}_{ij}(t_0\ubar{q'})\neq 0$ and $I_{rdd'}-L(t_0\ubar{q'})$ is of rank $rdd'$. Define $\ubar{q}=t_0\ubar{q'}$.  
 %The proof follows by using Schutzenberger's theorem and the algorithm of Raz and Shpilka. RANDOM paper
\end{proof}

Following is an immediate corollary. 
\begin{corollary}\label{cor:immediate}
Suppose Lemma~\ref{lem:nonzero-output} outputs a matrix tuple $\ubar{q}$. We can compute another matrix tuple $\ubar{p'}$ of dimension $dd'$ which is a witness of $\ncrank(T)>r$. 
\end{corollary}
\begin{proof}
Define the matrix tuple 
$\ubar{q''}=(q''^{(1)}_{11}, \ldots, q''^{(1)}_{dd}, \ldots, q''^{(n)}_{11}, \ldots, q''^{(n)}_{dd})$ where $q''^{(k)}_{\ell_1\ell_2}=q^{(k)}_{\ell_1\ell_2} + p^{(k)}_{\ell_1\ell_2}\otimes I_{d'}$ is a $d'$ dimensional matrix tuple for $1\leq k\leq n, 1\leq \ell_1,\ell_2\leq d$.

Lemma \ref{lem:nonzero-output} shows that the rank of $T_d$ evaluated on the matrix tuple $\ubar{q''}$ is more than $rdd'$. This is same as saying that $T_d(Z)$ is of rank more than $rdd'$ when the variable $z^{k}_{\ell_1,\ell_2} : 1\leq k\leq n, 1\leq \ell_1, \ell_2\leq d$ is substituted by $q''^{(k)}_{\ell_1\ell_2}$. 
%\[
%\ubar{q'}+\ubar{p}\otimes I_{2r}=(q'^{(1)}_{11} + (p_1)_{11}\otimes I_{2r}, \ldots, q'^{(1)}_{dd} + (p_1)_{dd}\otimes I_{2r}, \ldots, q'^{(n)}_{11} + (p_n)_{11}\otimes I_{2r}, \ldots, 
%q'^{(n)}_{dd} + (p_n)_{dd} + \otimes I_{2r})
%\] 
%is more than $rdd'$.  
%(\ubar{q}+I_{2r}\otimes \ubar{p})>rdd'$. 
Hence $\ncrank(T_d)>rd$. By Lemma \ref{lem:scaling-blowup}, we know that $\ncrank(T)>r$.
Moreover, we obtain a matrix tuple $\ubar{p'}=({p'_1}, {p'_2}, \ldots, {p'_n})$ which is a witness of $\ncrank(T)>r$, where ${p'_k}=\left(q''^{(k)}_{\ell_1\ell_2}\right)_{1\leq \ell_1,\ell_2\leq d} : 1\leq k\leq n$. Notice that $\ubar{p'}$ is the substitution for the $\ubar{x}$ variables. 
%Moreover, the matrix tuple $\ubar{p'}=({p'_1}, {p'_2}, \ldots, {p'_n})$ for the $\ubar{x}$ variables, where ${p'_k}=\left(q''^{(k)}_{\ell_1\ell_2}\right)_{1\leq \ell_1,\ell_2\leq d}$ ($1\leq k\leq n$), a witness of $\ncrank(T)>r$. 
\end{proof}
%Equivalently, the matrices $(Z_1, Z_2, \ldots, Z_n)$ will be substituted by the $n$-tuple $\tilde{\ubar{q}}$ where 

%\textcolor{red}{Let $\iota_{d'}(\ubar{q'})=\tilde{\ubar{q}}$. 
%blow up each $z_{i,j,k} : 1\leq i\leq n, 1\leq j,k\leq d$ 
%variable by the matrix $q_{i,j,k}$ to construct a matrix tuple $\tilde{\ubar{{q}}}$ of dimension $dd'$. 
%It is easy to see that, $\ubar{q^{''}}=\tilde{\ubar{{q}}} + \ubar{p}\otimes I_{d'}$ is a substitution for the $\ubar{x}$ variables to witness that $\ncrank(T)>r$. }
%We can now construct $(p'_1, \ldots, p'_n)\in (\F^{d'\times d'})^n$ such that $\rank(T(\ubar{p'})) > rd'$. 

\subsubsection{Rounding and blow-up Control}\label{sec:roundblowup}

Next, we apply Lemma \ref{lem:const-reg} which gives a rounding procedure to get a matrix tuple of dimension $d_1=dd'$ to witness that $\ncrank(T)=r'$ where $r' \geq r+1$. Call that new matrix tuple as $\ubar{p''}$. 

However, we cannot afford to have such a dimension blow-up for the witness matrix tuple in every step of the iteration as it incurs an exponential blow-up in the dimension of the final witness. To control that, we use a simple trick from \cite{IQS18} which we describe for the sake of completeness.

\begin{lemma}\label{lem:blow-up-control}
Consider an $s\times s$ linear matrix $T$ and a matrix tuple $\ubar{p''}$ in $\M_{d_1}(\Q)^n$ such that $\ubar{p''}$ is a witness of rank $r'$ of $T$. We can efficiently compute another matrix tuple $\widehat{\ubar{p}}$ of dimension at most $r'+1$ (over $\Q$) such that $\widehat{\ubar{p}}$ is also a witness of rank $r'$ of $T$. 
\end{lemma}

\begin{proof}
%Let $d_1=dd'$. 
Consider a sub-matrix $A$ in $T(\ubar{p''})$ such that $\rank(A)$ is at least $r'd_1$. From each matrix in the tuple $\ubar{p''}$, remove the last row and the column to get another tuple $\widetilde{{\ubar{p}}}$. We claim that the corresponding sub-matrix $A'$ in $T(\widetilde{{\ubar{p}}})$ is of rank $>(r'-1)(d_1-1)$ as long as $d_1>r'+1$. Otherwise, $\rank(A)\leq \rank(A') + 2r'\leq (r'-1)(d_1-1)+2r' = r'd_1 - d_1 + r'+1 < r'd_1$. 
%Applying the procedure repeatedly, we can control the blow-up in the dimension. 
Now we can use the constructive regularity lemma (Lemma~\ref{lem:const-reg}) on the tuple $\widetilde{{\ubar{p}}}$ to obtain another witness of dimension $d_1-1$ which is a witness of rank $r'$ of $T$. Applying the procedure repeatedly, we can control the blow-up in the dimension within $r'+1$ and get the witness tuple $\widehat{\ubar{p}}$. 
\end{proof}

\subsection{The Algorithm for NSINGULAR}\label{sec:final-algorithm}
We formally state the main steps of the algorithm. 

\begin{description}
\item[Algorithm for $\nsing$]\

\textbf{Input:} $T = A_0 + \sum_{i=1}^n A_ix_i$ where $A_0, A_1, \ldots, A_n \in \M_s(\Q)$.

\textbf{Output:} The noncommutative rank of $T$ and a set of matrix assignments that witness $\ncrank(T)$.\\

%It is an incremental algorithm that constructs a witness at every stage. 
The algorithm gradually increases the rank and finds a witness for it. Suppose at any intermediate stage, we already have a matrix tuple $\ubar{p}$ in $\M_d(\Q)^n$, a \emph{witness of rank $r$} of $T$. 

\begin{enumerate}
\item (Is $r$ the maximum rank?) Use Theorem \ref{razshpilka} to check whether the polynomial $\widetilde{P}_{ij}\neq 0$ (as defined in Equation \ref{eqn:zero}) for some choice of $i,j$.
\item If no such choice for $i,j$ can be found, then STOP and output $r$ to be the noncommutative rank of $T$.
\item (Otherwise, construct a witness of rank $r+1$ and repeat Step~1) We implement the following steps to construct a rank $(r+1)$-witness:
\begin{enumerate}
\item~[Rank increment step] Apply Corollary \ref{cor:immediate}
%Section \ref{sec:rankincrement} 
to find a $d_1\times d_1$ matrix substitution $\ubar{p'}=(p'_1,\ldots, p'_n)$ such that $\rank(T(\ubar{p'})) > rd_1$ where $d_1=2rd^2$. 
%(from Step~1).

\item~[Rounding using the regularity lemma] Apply Lemma \ref{lem:const-reg} to find another $d_1\times d_1$ matrix substitution $(p''_1, \ldots, p''_n)$ such that the rank of $T$ evaluated at $(p''_1, \ldots, p''_n)$ is $r'd_1$ where $r'\geq r+1$. %[Using the constructive version of regularity lemma as discussed in Section \ref{sec:construct-regular} and in Section \ref{sec:roundblowup}].

\item ~[Reducing the witness size] Apply Lemma \ref{lem:blow-up-control} to find a matrix substitution $\widehat{\ubar{p}}=(\widehat{p}_1, \ldots, \widehat{p}_n)$ of dimension $d' \leq r'+1$, such that the rank of $T$ evaluated at $\widehat{\ubar{p}}$ is $\geq r'd'$. 
%[Using the blow-up control operation described in Section \ref{sec:roundblowup}].
\end{enumerate}
\end{enumerate}
\end{description}

Next we analyze the performance of the algorithm. 
%\begin{description}
%\item[Analysis] 
%\end{description}
\paragraph{Analysis}
Since the noncommutative rank of $T$ is at most $s$, the algorithm iterates at most $s$ steps. Lemma~\ref{lem:truncated-pcnew} (for $k=1$), Theorem \ref{razshpilka}, and Lemma \ref{lem:nonzero-output} guarantee that Step 1 and Step 3(a) can be done in $\poly(n,r,d)$ steps. Step 3(b) and 3(c) require straightforward linear algebraic computations discussed in Section \ref{sec:roundblowup} which can be performed in $\poly(n,d,r)$ time. Since $d\leq s+1$ throughout the process, the run time is bounded by $\poly(n,s)$. 

We now explain the simple analysis of the bit complexity of the algorithm (since $\F=\Q$). 
Suppose the witness of rank $r$ computed by the algorithm has bit complexity $b$. Notice that in the rank increment step the matrix constructed in Corollary \ref{cor:abp-witness} has only $0,1$ entries and the parameter $t_0$ is of size $\poly(s,d)$. Hence, the bit complexity after step 3(a) can change to $O(b + \log(sd))$ at most. Step 3(b) is a simple linear algebraic step that can incur an additive term of 
$\poly(s,d)$ to the bit complexity. Thus, the bit complexity of the witness of rank $r+1$ is bounded
by $b+\poly(s,d)$. Since the bit complexity for the first step is bounded by the input coefficients,
it follows that the overall bit complexity of the algorithm is polynomial in $s$ and the input size. 

%\marginnote{I rewrote the bit complexity analysis a 'bit'... please take a look}

\begin{remark}\label{rmk:nsing-prime}
%We can also fix the final dimension of the witness matrices a prime number between $s+1$ and $2(s+1)$ by truncating the blowup-control at this stage. We use this in the construction of $\rankincrement$ subroutine in Section \ref{sec:rank-general-case}. Also, 
The algorithm of $\nsing$ can be adapted over fields of positive characteristic by extending the division algebra construction over such fields \cite[Section 15.4]{Pie82}. However, since our main motivation is to prove Theorem \ref{thm:main-theorem} and Theorem \ref{thm:main-theorem-2}, we prefer to state the algorithm for $\nsing$ over $\F=\Q$.  
\end{remark}

\section{Proofs of the Main Theorems}\label{sec:rank-general-case}
The goal of this section is to present the proofs of Theorems \ref{thm:main-theorem} and \ref{thm:main-theorem-2} by designing the subroutines $\pitsearch_k$ and $\rankincrement_k$. %overall plan will be to design a recursive algorithm whose recursion depth is bounded by $k$ and at depth $1\leq j\leq k$ the algorithm finds the assignments for the variables in $X_j$ such that the $\pcrank$ is always preserved.    
%to have $k$ rounds and in round $j\in\{1,2,\ldots,k\}$, the variables in the set $X_j$ will be substituted by suitable matrices such that rank of the input matrix $T$ is always preserved. 

The subroutine $\pitsearch_k$ takes as input an ABP $\mathcal{A}$ of size $s$ computing a polynomial $f\in\F\angle{X_{[k]}}$. It finds substitution matrices of the form \ref{eqn:tensorstruct1} for the variables $x\in X_{[k]}$ such that $f$ evaluates to a nonzero matrix if $f$ is a nonzero polynomial. 
%output matrix $f(\{M_x\}_{x\in X_1}, X_2, \ldots, X_k)$ is nonzero if $f$ is nonzero. 
Moreover, the dimension of the substitution matrices is a polynomial function of the input size. 
%Notice that the entries of the matrix $f(\{M_x\}_{x\in X_1}, X_2, \ldots, X_k)$ are polynomials over $X_2, \ldots, X_k$. 

The subroutine $\rankincrement_k$ takes as input a linear matrix $T$ of size $s$ over the set of variables $X_{[k]}$ and finds matrix assignments of the form \ref{eqn:tensorstruct1} and dimension $d$, to the variables such that the rank of the final 
matrix is $d\cdot \pcrank(T)$. Moreover, the dimension $d$ is a polynomial of the input size. 
%and the matrices respect the partial commutativity structure. 

These two recursive subroutines intertwine, giving the proofs of Theorem \ref{thm:main-theorem} and Theorem \ref{thm:main-theorem-2}. Recall that for a set $S\subseteq [k]$, $X_S$ refers to $\bigsqcup_{i\in S} X_i$. When we use $\rankincrement$ (resp. $\pitsearch$) subroutine on $X_S$ with $|S|=\ell$, we refer it as $\rankincrement_{\ell}$ (resp. $\pitsearch_{\ell}$). Also, recall from Definition \ref{def:type-fold-tensor} that the substitution matrices of the form \ref{eqn:tensorstruct1} are type-$i$ $k$-fold tensors of shape $\ubar{d}=(d_1, d_2, \ldots, d_k)$.  
%In several places of our algorithms,  we will be describing a polynomial-time reduction from $\rankincrement_k$ to $\rankincrement_{k-1}$.     

%For any $j$, the input matrix can be viewed over the variables in $X_j$ and the coefficients are over $\F\angle{X_1\sqcup X_2\sqcup\ldots\sqcup X_{j-1}\sqcup X_{j+1}\ldots\sqcup X_k}$. By Theorem \ref{thm:gen-skew-field}, the ring $\F\angle{X_1\sqcup X_2\sqcup\ldots\sqcup X_{j-1}\sqcup X_{j+1}\ldots\sqcup X_k}$ can also be embedded inside the universal skew field of fractions $\U$.\footnote{Because $\F\angle{X_1\sqcup X_2\sqcup\ldots\sqcup X_{j-1}\sqcup X_{j+1}\ldots\sqcup X_k}$ is a subring of $\F\angle{X}$
%which is embeddable in $\U$.} The subroutine $\rankincrement$ takes as an input a linear matrix over the variables $X$, and output assignments to the variables such that the maximum rank is achieved. The subroutine $\pitsearch$ takes as an input an ABP over $X$ and find nonzero assignments to the variables in any set (w.l.o.g, say $X_1$) viewing the coefficients of the polynomial over $\U$. These two subroutines intertwine to give the proof of Theorem \ref{thm:main-theorem} and Theorem \ref{thm:main-theorem-2}.    

%\marginnote{Shouldn't we use $\U_k$ to denote the universal skew field? Because we also need
%to reason about $\U_{k-1}$...}

\subsection{Identity testing of partially commutative ABPs}\label{sec:pc-pit}
In this section, we describe the subroutine $\pitsearch_k$. Basically, given a partially commutative ABP as input with edges labeled by linear forms over 
$\Q\angle{X}$ where $X=X_{[k]}$, we develop a deterministic algorithm for identity testing of such ABPs.

We need to generalize the following result shown in \cite{ACGMR23} for the noncommutative case. 
%We recall the formal
%definition of a linear pencils for elements in the free noncommutative ring $\F\angle{X}$. 
%\begin{definition}[Linear Pencil]\label{def:linear-pencil}
%A noncommutative polynomial $g\in\F\angle{X}$ is said to have a size $s$ \emph{linear pencil} $L$ if $L$ is an $s\times s$ invertible linear matrix over $X$ such that $g$ is computed in the $(1,s)^{th}$ entry of $L^{-1}$.
%\end{definition}
Suppose $M$ is a matrix over $\F\angle{X}$, for noncommutative $X$, such that each $M_{ij}$ is given as 
input by a linear pencil (See, the definition \ref{def:linear-pencil}). Then we can efficiently reduces rank computation of $M$ to the 
rank computation of a (noncommutative) linear matrix over $X$.

\begin{lemma}{\rm\cite[Lemma 23]{ACGMR23}}\label{lemma-ncrank-connection}
Let $X=\{x_1, \ldots, x_n\}$ be a set of noncommutative variables. Let  $M \in \F\angle{X}^{m\times m}$ be a matrix where each $(i,j)^{th}$ entry $M_{ij}\in\F\angle{X}$ is given as input by a size $s$ linear pencil $L_{ij}$. Then, there is a polynomial-time algorithm that computes a linear matrix $L$ of size $m^2s + m$ such that,
\[
\ncrank(L) = m^2s + \ncrank(M).
\]
\end{lemma}
%\marginnote{Better to replace "Then, one can construct..." with "Then, there is a
%polynomial-time algorithm that computes...", and next lemma also?}

%Although the lemma is stated for the case when $X$ is a set of noncommutative variables, 

%We can now generalize Definition \ref{def:linear-pencil} for partially commutative polynomials also where $X = X_{[k]}$.

%\begin{definition}[Linear Pencil of Partially Commutative Polynomials]
%A partially commutative polynomial $g\in\F\angle{X_{[k]}}$ is said to have a size $s$ \emph{linear pencil} $L$ if $L$ is an $s\times s$ invertible linear matrix over $X_{[k]}$ such that $g$ is computed in the $(1,s)^{th}$ entry of $L^{-1}$.
%\end{definition}

It is easy to see by inspection that the proof of the above lemma \cite{ACGMR23} holds even when 
$X_{[k]}$ is a set of partially commutative variables. More precisely, we have the following generalization, proved in the appendix. 

%\footnote{In general, whenever we can embed $\F\angle{X}$ in a skew field, such a result can be proved.}

\begin{lemma}\label{lemma-pcrank-connection}
Let $X=X_{[k]}$ be a set of partially commutative variables. Let  $M \in \Q\angle{X_{[k]}}^{m\times m}$ be a matrix where each $M_{ij}$ is given by a linear
pencil $L_{ij}$ of size $s$.\footnote{See Definition \ref{def:linear-pencil-pc}.} Then, there is a polynomial-time algorithm that computes a linear matrix $L$ of size $m^2s + m$ such that,
\[
\pcrank(L) = m^2s + \pcrank(M).
\]
\end{lemma}

The actual application of this lemma in the next theorem is as follows: Suppose $M$ is an input matrix 
whose entries are ABPs defined over the set of partially commutative variables $X_{[k]}$. By Proposition \ref{prop:abp-pencil}, size $s$ ABPs have linear pencils of size $O(s)$ and, moreover, the linear pencils can be computed in time $\poly(s)$. %This proposition is stated for $X$ noncommutative, but the same construction also works for $X_{[k]}$ partially commutative.
As a result, the rank computation problem for such a matrix $M$ can be reduced in $\poly(s)$ time to rank computation of a linear matrix over $X_{[k]}$. 

\begin{theorem}\label{thm:pittorank}
Given an input ABP $\mathcal{A}$ of size $s$, width $w$, 
computing a polynomial $f\in\Q\angle{X_{[k]}}$ of degree $d$, 
the subroutine $\pitsearch_k(\mathcal{A}, s, w, d, X_{[k]})$ reduces
the identity testing problem for $f$ to at most 
%$O(ns^3)$ 
%$s^{c^k}$ 
$O(ds^3)$ instances of  $\rankincrement_{k-1}$ problem for linear matrices of size $O(s^5)$ and at most $d$ recursive calls of $\pitsearch_{k-1}$ for an ABP of size $O(sd^2)$, width $O(sd)$, computing a polynomial of degree $\leq d$ in $\Q\angle{X_{[k]\setminus\{1\}}}$ in deterministic $\poly(s)$ time. 
%(for $\ell\leq k-1$) for linear matrices (of size at most $s^{c^k}$) 
%over $X_2\sqcup\ldots\sqcup X_k$ 
%in deterministic time $\poly(s^{c^k})$, where $c>0$ is a constant.\footnote{For convenience, we will assume that $s$ is at least $|X|$. 
%and it bounds the number of nodes in the ABP and also the sizes of the coefficients from $\F$ labeling the edges.
%} 
Moreover, it finds assignments to the variables in 
$X_j : 1\leq j\leq k$ which are of the form $I_{d_1}\otimes \cdots\otimes I_{d_{j-1}}\otimes M_x\otimes I_{d_{j+1}}\otimes \cdots \otimes I_{d_k}$ such that $f$ evaluates to a nonzero matrix if $f$ is originally a nonzero polynomial. The dimensions $d_1, d_2, \ldots, d_k$ are at most $d+1$. 
%The intermediate bit complexity of the reduction is also $\poly(s^{c^k})$ bounded.  
%$M_{x}$ to the variables $x\in X_1$ which makes the matrix $f(\{M_x\}_{x\in X_1}, X_2, \ldots, X_k)$ nonzero if $f$ is originally a nonzero polynomial.      
\end{theorem}

\begin{proof}
Firstly, we explain how $\pitsearch_{k}$ finds the substitution matrices for the variables in $X_1$. 
We view the edge labels as affine linear forms over the variables in $X_1$ and the coefficients are over the ring $\Q\angle{X_{[k]\setminus\{1\}}}$ inside $\U_{[k]\setminus \{1\}}$ by Theorem \ref{thm:gen-skew-field}. 
%Note that  $\U_{[k]\setminus \{1\}}$ is a subfield of $\U_{[k]}$. 

As discussed in Section \ref{sec:prelim}, the Raz-Shpilka algorithm \cite{RS05}, which is for a noncommutative set of variables $X$, is linear algebraic: We can assume the ABP is layered and the width is $w$ at each layer. 
For each monomial $m$ of degree $j$, there is a corresponding $w$-dimensional vector $v_m\in\F^w$ of $m$'s 
coefficients at the $w$ nodes in layer $j$. Now, the idea is to maintain a set of at most $w$ many monomials $m_1,m_2,\ldots,m_{w}$ such that their corresponding vectors $v_{m_i}$ are linearly independent and
their $\Q$-linear span includes all such coefficient vectors $v_m$. Then, the Raz-Shpilka algorithm proceeds
to layer $j+1$ with some linear algebraic computation.

%The idea is to maintain at layer $j$, a maximally linearly independent  spanning set of vectors (indexed by at %most $w$ monomials) in each layer of the ABP with straightforward linear algebraic computation. 
%Here $w$ is the width of the ABP which is bounded by $s$. 
%The entries of such vectors are the coefficients of the indexing monomials in different nodes of the ABP along %the width. So these vectors are $w$ dimensional. 

We will broadly use the same approach for the partially commutative case. Applying the procedure discussed in the proof of Lemma \ref{lem:struct1}, we first homogenize the ABP with respect to the variables in $X_1$. It suffices to solve the identity testing problem for such an $X_1$-homogenized ABP. It is easy to check that the edges of this homogenized ABP are labeled by linear forms $\sum_{i=1}^n \alpha_ix_i$ in variables $x_i\in X_1$, where the $\alpha_i$ are polynomials in $\Q\angle{X_{[k]\setminus\{1\}}}$. Moreover, each $\alpha_i$ is given by an ABP of size $O(sd)=O(s^2)$ by Lemma \ref{lem:struct1}.

Inductively, at the $j^{th}$ level, suppose the monomials computed are $m_1, m_2,\ldots, m_{w'}$ in $X_1^j$, where $w'\leq w$. Let the corresponding coefficient vectors be $v_1, v_2, \ldots, v_{w'}$ over the ring $\Q\angle{X_{[k]\setminus\{1\}}}$. Again by Lemma \ref{lem:struct1}, entries of the $v_i$ are given by ABPs over $X_{[k]\setminus\{1\}}$ of size $O(s^2j)$. Moreover, the vectors $v_1, v_2, \ldots, v_{w'}$ are $\U_{[k]\setminus \{1\}}$-spanning set for the coefficient vectors of monomials at layer $j$ (to be precise, as a left $\U_{[k]\setminus \{1\}}$-module). 

Now, for the $(j+1)^{th}$ level, we need to compute at most $w$ many $\U_{[k]\setminus \{1\}}$-linearly independent vectors from the at most $nw$ many coefficient vectors of the $\{m_i x_j : 1\leq i\leq w', x_j\in X_1\}$. Clearly, this is the problem of computing the rank of these at most $nw$ coefficient vectors whose entries are ABPs over the variables in $X_{[k]\setminus\{1\}}$. This is because, given a set of $w$-dimensional $\U_{[k]\setminus \{1\}}$-linearly independent vectors $v'_1, \ldots, v'_{\ell'}$ and another vector $v$, the rank of this matrix with $\ell'+1$ columns is $\ell'$  precisely if $v$ is in the $\U_{[k]\setminus \{1\}}$-span of $v'_1, \ldots, v'_{\ell'}$. The columns of the matrix are $v'_1, \ldots, v'_{\ell'}, v$ and we can make it a square matrix by padding with zero columns. Applying Lemma \ref{lemma-pcrank-connection}, we can reduce it to the $\rankincrement_{k-1}$ problem for linear matrices of size $O(s^5)$ over the variable set 
$X_{[k]\setminus\{1\}}$.
%{\textcolor{red}{We should change this $\rankincrement_{k-1}$ notation. $\rankincrement_{k-1}$ and $\rankincrement_{k}$ are the same subroutines, only the arguments are different. We should mention that instead of a different subroutine}} 
Equivalently, we need to compute the rank of these linear matrices 
over the skew field $U_{[k]\setminus \{1\}}$, which has $k-1$ parts in the set of partially commutative variables.

At the end, the $\pitsearch_k$ algorithm will compute a monomial $m$ over $X_1$ and its coefficient, which 
is an ABP over the remaining variables $X_{[k]\setminus\{1\}}$. If $f\ne 0$, then given such a monomial $m$, as discussed in Section \ref{sec:pit-results}, we can efficiently find scalar matrix substitutions 
$\{M_x\}_{x\in X_1}$ for the $X_1$-variables such that the polynomial $f$ remains nonzero. We can even ensure that the entries of each $M_x$ is in $\{0,1\}$ and $\dim(M_x)\le d+1\leq s+1$ as explained in Section \ref{sec:pit-results}.

The $\pitsearch_k$ procedure described above computes a nonzero monomial $m\in X_1^d$ for 
some $d\leq s$ whose coefficient is a nonzero ABP $\mathcal{A}_m$ in $\Q\angle{X_{[k]\setminus\{1\}}}$.
By Lemma~\ref{lem:struct1}, the size of $\mathcal{A}_m$ is $O(sd^2)=O(s^3)$, width $O(sd)$ and computes a polynomial of degree $\leq d$. 
%and more specifically it is bounded by $O(s^3)$. 
Hence we can recursively apply $\pitsearch_{k-1}(\mathcal{A}_m, O(s^3), O(sd), d, X_{[k]\setminus\{1\}})$. 
%to compute matrix substitutions for variables in $X_2$ preserving nonzero property. Repeated application proves the theorem. A routine careful inspection shows that the intermediate $\rankincrement_{\ell} : \ell\leq k-1$ subroutines are applied over linear matrices of size at most $s^{c^k}$ for some constant $c>0$. Note that the blow up happens due to the increase of the width of the ABP but the degree remains the same.  

The $\pitsearch_{k-1}$ subroutine outputs 
the substitution matrices for the variables $x\in X_j : 2\leq j\leq k$ which are of tensor product structure $I_{d_2}\otimes \cdots\otimes I_{d_{j-1}}\otimes M_x\otimes I_{d_{j+1}}\otimes \cdots \otimes I_{d_k}$ and the dimensions $d_2, \ldots, d_k$ are at most $d+1$. Combining with the substitution matrices for 
$X_1$, the final structure of the matrix substitutions for $x\in X_j$ is of the form 
$I_{d_1}\otimes I_{d_2}\otimes \cdots\otimes I_{d_{j-1}}\otimes M_x\otimes I_{d_{j+1}}\otimes \cdots \otimes I_{d_k}$. This follows from Observation \ref{obs:partial-eval}. Now the theorem follows by considering the procedure above for every $X_1$-homogenized ABPs.   \qedhere
\end{proof}
\begin{remark}\label{rem:pc-pit-dimension}
    By Theorem~\ref{thm:pittorank}, each $d_j \leq d+1$. However,
    we can relax the bound for each $d_j$ and choose any larger value. This can be easily done 
    using a standard idea of padding sufficient number of zero rows and columns to the matrix construction shown in 
    Subsection~\ref{sec:pit-results}. We will require this in Subsection~\ref{pc-rounding}, where we need to ensure 
    that $d_j, 1\le j\le k$ are distinct prime numbers bounded by $\poly(s)$.  
\end{remark}
\subsubsection{Matrix substitution witnessing nonzero of a series}
In the design of the subroutine $\rankincrement$, we need to find nonzero of a series over partially commutative variables.
To that end, using Theorem \ref{thm:pittorank} we prove the following lemma.

\begin{lemma}\label{lem:bound-zero-finding}
%Let $X=\bigcup_{j=1}^k X_k$ be a set of partially commutative variables and 
Let $S={b} (I - L)^{-1} {a}$ be a series over the partially commutative variable set $X_{[k]}$. The dimension of $I, L$ are $s\times s$, ${b}, {a}$ are $1\times s$ and $s\times 1$ dimensional vectors respectively. The entries in ${b}, {a}, L$ are linear forms over $X_{[k]}$. Then, there is a deterministic polynomial time algorithm, with access to subroutine $\pitsearch_k$ for linear matrices,
that computes matrix substitutions on which $S$ evaluates to a nonzero matrix  
%$$\{M_x\}_{x\in X_1}$ of dimension $\poly(s)$  such that 
%$S(\{M_x\}_{x\in X_1}, X_2, \ldots, X_k)\neq 0$ 
if $S\neq 0$. 
\end{lemma}

\begin{proof}
%Let $X=\sqcup_{j=1}^k X_k$ be a set of partially commutative variables and $S=c - B (I - L)^{-1} A$ be a series. The dimensions of $I, L$ are $s\times s$, of $B, A$ are $1\times s$ and $s\times 1$ dimensional 
%vectors respectively. The element $c$ and the entries in $B, L, A$ are linear forms over $X$.
%Consider the problem of checking if $S$ is nonzero, and if so to find a matrix substitution
%$\{M_x\mid x\in X_1\}$ for the variables in $X_1$ such that the resulting series remains nonzero.
%We now show that this problem is deterministic $\poly(s)$ time reducible to $\pitsearch$.   
For $k=1$, in Section \ref{sect-idea} we showed that finding a nonzero of the series $S$ reduces to finding a nonzero of its $s$-term truncation $P_S = {b} \left(\sum_{k\leq s} L^k\right) {a}$ (using Lemma \ref{lem:truncated-pcnew}), and a scaling trick. In this section, we extend the approach for $k>1$.
%to obtain the desired reduction
%to $\pitsearch$. 
%Since the terms occurring in $B(I-L)^{-1}A$ are of degrees at least two, if $S=0$ then $c=0$ and $B(I-L)^{-1}A=0$. 
Lemma \ref{lem:truncated-pcnew} implies that $S=0$ if and only if $P_S={b} \left(\sum_{k\leq s} L^k\right) {a}=0$. 

Apply $\pitsearch_k$ on $P_S$ and using Theorem \ref{thm:pittorank} compute substitution matrices which has tensor product structure. For the convenience of notation, let $\ubar{M_1}, \ubar{M_2}, \ldots, \ubar{M_k}$ be the tuples of the matrices for the variables in $X_1, X_2, \ldots, X_k$ respectively. Let $t$ be a commutative variable and by $t \ubar{M_j}$, we mean that each matrix in the tuple is scaled by the factor $t$. Notice that $S(t \ubar{M_1}, \ldots, t \ubar{M_k})$ is nonzero 
since $P_S(t \ubar{M_1}, \ldots, t \ubar{M_k})\neq 0$ and different $t$ degrees homogenized components will not mix together. Let $d_1, d_2, \ldots, d_k$ be the dimension of the matrices in different components as promised by Theorem \ref{thm:pittorank}, and let $d=d_1 d_2 \cdots d_k$. Thus $(I-L)(t \ubar{M_1}, \ldots, t \ubar{M_k})$ evaluates to a matrix of dimension $sd$ over the variable $t$. Similarly, ${b}(t \ubar{M_1}, \ldots, t \ubar{M_k})$ and ${a}(t \ubar{M_1}, \ldots, t \ubar{M_k})$ are $s$ dimensional vectors of matrices of dimension $d$. We want a value for the parameter $t$ that makes $\det[(I-L)(t \ubar{M_1}, \ldots, t \ubar{M_k})]\neq 0$ and $S={b} (I - L)^{-1} {a}((t \ubar{M_1}, \ldots, t \ubar{M_k})))\neq 0$. Hence, it suffices to avoid the roots of the univariate polynomials in $t$ originating from the determinant computation and the entries of ${b} (I - L)^{-1} {a}[t \ubar{M_1}, \ldots, t \ubar{M_k}]$. %Since $d\leq s^{c^k}$ by Theorem \ref{thm:pittorank}, 
Since $d = s^{O(k)}$ by Theorem \ref{thm:pittorank}, we can find a suitable value of $t$ from a $\poly(s^k)$ size finite subset of $\Q$.
\end{proof}

\subsection{The procedure for PC-RANK}\label{sec:proc}

We are now ready to design the subroutine $\rankincrement_k$. We can write the input linear matrix $T$ of size $s$ as:
\begin{equation}\label{eqn:4}
T(X_1, \ldots, X_k) = A_0 + \sum_{j=1}^k\sum_{x \in X_j} A_{x}x.
\end{equation}
%The plan is to design a recursive algorithm with recursion depth bounded by $k$. At the $j^{th}$ level 
%of the recursion the algorithm finds matrix substitutions for the variables in $X_j$ for $1\leq j\leq k$.
%preserving $\pcrank(T)$. {\textcolor{red}{It is misleading to say "preserving $\pcrank(T)$".}}
%Inductively, we assume that at the end of iteration $j$, for each $x\in X_1\sqcup \ldots \sqcup X_j$, we have a witness of the following form: for each $j' \leq j$, each $x \in X_{j'}$ is substituted by 

%When the algorithm terminates, 
The algorithm computes the matrix substitution for each $x\in X_j$ ($1\leq j\leq k$) of the form
\begin{equation}\label{eqn:5}
x\leftarrow I_{d_1}\otimes \cdots I_{d_{j - 1}}\otimes M_x\otimes I_{d_{j + 1}}\otimes \cdots \otimes I_{d_k},
\end{equation}
where matrix $M_x$ is $d_j$-dimensional %and $d_{j}\leq 2^j s^3$. Moreover, we ensure that each $d_j$ is a distinct prime number. 
and the rank of the resulting scalar matrix will be $(d_1 d_2\cdots d_k)\cdot\pcrank(T)$. Recall from the definition \ref{def:type-fold-tensor}, that the substitutions of the form \ref{eqn:5} is a type-$j$ $k$-fold tensor. Consider the following definition of witness of $\pcrank$. 
%For the convenience, we introduce the following definition. 
%\begin{definition}
%We call the matrix substitution of the form given in the expression \ref{eqn:5} as a type-$j$ $k$-fold tensor product. Also $\ubar{d}=(d_1,d_2,\ldots,d_k)$ is the shape of the tensor. 
%\end{definition}
%Let $\pcrank(T)=r$. By Proposition \ref{prop:pc-folded} we know that there \emph{exist} matrix substitutions such that variables in $X_i$ are 
%substituted by type-$i$ $k$-fold tensor products for each $i$, of some shape $\ubar{d}$, such that 
%there is a size $r$ submatrix of $T$ that evaluates to an invertible matrix. 
%This motivates the following definition. 

\begin{definition}[Witness of $\pcrank$ $r$]\label{defn:pc-rank-witness}
Let $T(X_1, \ldots, X_k)$ be the given linear matrix of the form $T = A_0 + \sum_{j=1}^k \sum_{x\in X_j} A_x x$ such that $A_0, A_x\in \Mat_s(\Q)$ for $x\in X_{[k]}$. %$x\in \bigsqcup_{j=1}^k X_j$.  
%We say $(\alpha_1, \ldots, \alpha_n) \in \F^n$ is a witness of commutative rank $r$ of $T$ if $\rank(T(\alpha_1, \ldots, \alpha_n))\geq r$. 
We say that a matrix substitution of shape $\ubar{d}=(d_1, \ldots, d_k)$ that assigns type-$j$ $k$-fold tensor products for variables in $X_j$ ($1\leq j\leq k$), is a witness of $\pcrank(T)\geq r$ if 
$T$ evaluates to a scalar matrix of rank at least $rd_1 d_2\cdots d_k$ after the substitution.  
\end{definition}

Now, we describe a rank increment procedure that computes new matrix assignments to the variables in $X_i$ ($1\leq i\leq k$) that witness the $\pcrank(T(X_{[k]}))$ is at least $r+1$, if such a rank increment is possible. 
\begin{comment}
Suppose, for the linear matrix $T$ we have found substitutions $\{M_x\}_{x\in X_{1}}$ of dimension 
$d_1$ such that the resulting matrix $T'_1$, after substitution has rank at least $rd_1$.
%\begin{comment}
More precisely, the rank of 
\begin{equation}\label{eqn:new1}
T'_{1}=A_0\otimes I_{d_1} + \sum_{x\in X_{1}}(A_x\otimes M_x) + \sum_{j=2}^k \sum_{x\in X_{j}} (A_x\otimes I_{d_1}) x 
\end{equation}
%
has rank at least $rd_1$ over the skew field $\U_{[k]\setminus\{1\}}$. As $T'_1$ is an instance of $\rankincrement_{k-1}$ we can invoke the procedure for it to compute substitutions $\{\widetilde{M}_x : x\in X_{j} : 2\leq j\leq k\}$ in $T'_{1}$ that witness that $\pcrank(T'_1)\ge rd_1$. Suppose $(d_2,d_3,\ldots,d_k)$ is the shape of this
$(k-1)$-fold tensor product substitution $\widetilde{M}_x$. Then, together with the substitution for $X_1$ variables
we get a $k$-fold tensor product substitution for $T$ such that the 
scalar matrix will have rank at least $rd_1d_2\cdots d_k$.
More precisely, we have the following matrix
\begin{equation}\label{eqn:six}
T''_{1}=A_0\otimes I_{d_1\widetilde{d}} + \sum_{x\in X_{1}}A_x\otimes (M_x\otimes I_{\widetilde{d}}) + \sum_{j=2}^k \sum_{x\in X_{j}} A_x\otimes (I_{d_1}\otimes\widetilde{M}_x) 
\end{equation}
is of rank at least $rd_1\widetilde{d}$, where $\widetilde{d}=\prod_{j=2}^k d_{j}$. 
\end{comment}
To do that, we need the following lemma and corollary as preparatory results.  
%\end{comment}
%\marginnote{I think we should mention explicitly the partially evaluated matrices like in the earlier version.}
%To adapt the algorithm in Section \ref{sect-idea} for finding substitutions for $X_{1}$ variables that 
%increase the rank further, we first prove a lemma which is analogous to Lemma \ref{lem:scaling-blowup}. 
%. Of course to do that, as expected from Section \ref{sect-idea}, we need to use 
%$\pitsearch$ to employ the identity testing problem on a partially commutative ABP over $X_{1}, \ldots, X_k$. The ABP instance will be obtained by truncating an infinite series.     
%This should follow the rounding and blow-up control steps to bring down the dimension of the witness matrices for rank $r+1$ suitably. However, as we show in this section that implementing the intuitive ideas are much harder than in Section \ref{sect-idea}.   
%We also note down the following lemma which is analogous to Lemma \ref{lem:scaling-blowup}. 

\begin{lemma}\label{lem:scaling-blowup-pc}
Let $T=A_0 + \sum_{i=1}^k \sum_{x\in X_i} A_x x$ be an $s\times s$ linear matrix over variables $X_{[k]}$. 
For $d\in \N$ define  $T_d=A_0\otimes I_d + \sum_{x\in X_1} (A_x \otimes Z_x) + \sum_{i=2}^k \sum_{x\in X_i} (A_x \otimes I_d) x$ where 
$\{Z_x=(z_{x,i,j})_{1\leq i,j\leq d}\}_{x\in X_1}$ be a set of generic matrices of noncommutative variables which are commuting with $X_2, \ldots, X_k$.   
Then, $\pcrank(T_d) = d \cdot \pcrank(T)$. 
\end{lemma}

\begin{proof}
Write $T_d=A_0\otimes I_d + \sum_{x,i,j} A_{x,i,j} ~z_{x,i,j} + \sum_{i=2}^k \sum_{x\in X_i} (A_x \otimes I_d) ~x$ 
where each $\{A_{x,i,j} : x\in X_1, 1\leq i,j\leq d\}$ is an $sd\times sd$ matrix.  

Let $\pcrank(T) = r$. Then, $T$ has a submatrix $M$ of size $r$ invertible over the skew field $\U_{[k]}$
%${Z_x\}_{x\in X_1}\bigcup X_{[k]\setminus\{1\}}}$
(by Theorem \ref{thm:gen-skew-field}). By Proposition \ref{prop:pc-folded}, there are matrix substitutions for the variables $x\in X_i : 1\leq i\leq k$ of the form $I_{d'_{1}}\otimes I_{d'_{2}}\otimes\cdots \otimes I_{d'_{i-1}}\otimes p_{{x}}\otimes I_{d'_{i+1}}\otimes\cdots\otimes I_{d'_k}$ 
such that $M$ evaluates to an invertible scalar matrix. Here, $p_{{x}}$ is a $d'_i\times d'_i$ matrix and $d'_{1}, \ldots, d'_k\in\N$. Also, w.l.o.g, we can assume $d'_i : 1\leq i\leq k$ to be multiple of $d$. In particular, let $d''_1=d'_1/d$.  

For each $x\in X_1$, let us write the matrix $p_{{x}}$ as a matrix of blocks of dimension $d''_1\times d''_1$. So the $(i,j)^{th}$ block in $[d]\times [d]$ is a matrix $q_{x,i,j}$. 
Now, it is not hard to see that $M$ corresponds to a submatrix of size $rd$ in $T_d$ which becomes invertible by the substitutions
\[
z_{x,i,j}\leftarrow q_{x,i,j}\otimes I_{d'_{2}}\otimes\cdots \otimes I_{d'_{i-1}}\otimes I_{d'_i}\otimes I_{d'_{i+1}}\otimes\cdots\otimes I_{d'_k},
\]
for $x\in X_1$ and 
\[
x\leftarrow I_{d''_1}\otimes I_{d'_{2}}\otimes\cdots \otimes I_{d'_{i-1}}\otimes p_{{x}}\otimes I_{d'_{i+1}}\otimes\cdots\otimes I_{d'_k},
\]
for $x\in X_i : 2\leq i\leq k$. Hence $\pcrank(T_d)\geq rd$. 

For the other direction, as $\pcrank(T) = r$, we can write 
\[
T=U\cdot \left(
\begin{array}{c|c}
I_{r}  & 0 \\
\hline
0 & 0
\end{array}
\right)\cdot V, 
\]
for invertible transformations $U,V$ over the skew field $\U_{[k]}$. 
Hence, $\pcrank(T_d)\leq rd$. This proves the lemma.  
\end{proof}

We apply Lemma \ref{lem:scaling-blowup-pc} repeatedly to prove the following corollary. 

\begin{corollary}\label{cor:pc-blow-up}
Let $T=A_0 + \sum_{i=1}^k \sum_{x\in X_i} A_x x$ be an $s\times s$ linear matrix over the partially commutative set of variables $X_{[k]}$. %X=\bigsqcup_{i=1}^k X_i$. 
Let $d_1, d_2, \ldots, d_k\in \N$, and define  $T_{d_1, d_2, \ldots, d_k}=A_0\otimes I_{d_1 d_2\cdots d_k} +\sum_{i=1}^k \sum_{x\in X_i} A_x \otimes I_{d_1}\otimes \cdots \otimes I_{d_{i-1}}\otimes Z_x\otimes I_{d_{i+1}}\otimes\cdots\otimes I_{d_k}$ where
the dimension of the generic noncommutative matrices $Z_x$ for the variables  $x\in X_i$ is $d_i$, and the variables in $\{Z_x\}_{x\in X_i}$ and $\{Z_x\}_{x\in X_j}$ are mutually commuting for $i\neq j$. 
Then, $\pcrank(T_{d_1, d_2, \ldots, d_k}) = d_1 d_2\cdots d_k \cdot \pcrank(T)$. 
\end{corollary}

\begin{proof}
For clarity we explain the proof up to stage two where we handle the variables in $X_1$ and $X_2$. Then a simple induction on $k$ gives the general result. 

For $d_1\in \N$, define 
$T_{d_1}=A_0\otimes I_{d_1} + \sum_{x\in X_1} A_x \otimes Z_x + \sum_{i=2}^k \sum_{x\in X_i} (A_x \otimes I_{d_1}) x$ where $Z_x$ is a $d_1$ dimensional generic matrix. Then by Lemma \ref{lem:scaling-blowup-pc}, we know that $\pcrank(T_{d_1})=d_1\cdot \pcrank(T)$. Let $A'_0=A_0\otimes I_{d_1} + \sum_{x\in X_1} A_x\otimes Z_x$. Also, for each $x\in \bigsqcup_{i=2}^k X_i$, we use $A'_x$ to denote the matrix $A_x\otimes I_{d_1}$. Thus, 
\[
T_{d_1}=A'_0 + \sum_{i=2}^k \sum_{x\in X_i} A'_x x. 
\]
Now replace the variables $x\in X_2$ by generic matrices $Z_x$ of dimension $d_2$ to get the matrix
\[
T_{d_1,d_2}=A'_0 \otimes I_{d_2} + \sum_{i=3}^k \sum_{x\in X_i} (A'_x\otimes I_{d_2}) x + \sum_{x\in X_2} A'_x\otimes Z_x. 
\]
Applying the Lemma \ref{lem:scaling-blowup-pc} again, we know that $\pcrank(T_{d_1,d_2})=d_2\cdot \pcrank(T_{d_1})=d_1 d_2 \cdot \pcrank(T)$. Note that to get $T_{d_1,d_2}$ from $T$, we need to substitute the variables $x\in X_1$ by matrices of the form $Z_x\otimes I_{d_2}$. Similarly, the matrices for $x\in X_2$ are given by $I_{d_1}\otimes Z_x$. Repeating the process $k$ times we get the desired result. 
\end{proof}

\subsubsection{Rank increment step}\label{sec:rankincrement-pc}

We now return to the construction of the subroutine $\rankincrement_k$. 
%let $\ubar{M}$ denote the entire matrix tuple assignment for all the variables in $X_{[k]}$.\footnote{Clearly for the base case ($r=1$), %we need to make a linear form nonzero after the evaluation, which is easy.
%we can easily make a linear form nonzero after the evaluation.
%} 
The main idea is that, given an input linear matrix $T$ over $X_{[k]}$, we do an induction on the rank parameter $r$. 
Clearly for the base case ($r=1$), 
we can easily make a linear form nonzero after the evaluation. 
Suppose that we have already computed a rank $r$ witness $\ubar{M}$, which is a type-$j$ $k$-fold tensor 
product matrix assignments for the variables in $X_j$ ($1\leq j\leq k$) such that:
\begin{itemize}
\item $\rank(T(\ubar{M}))$ is at least $rd'$ where $(d_1,d_2,\ldots,d_k)$ is the shape of the tensor and $d'=d_1 d_2\cdots d_k$. 
%\item For $1\le j\le k$ and  each variable $x\in X_j$, the matrix assignment is of the form shown in the Expression \ref{eqn:5}. 
\item Moreover, for each $1\leq j\leq k$, $d_j\leq s^3$ and $d_1, \ldots, d_k$ are distinct prime numbers.  
\end{itemize}

%Define $d'=d_1 d_2\cdots d_k$. 
Let $T_{d'}(Z)$ denote the matrix obtained from $T$ by replacing the variables $x\in \bigsqcup_{i=1}^k X_i$ by the matrices $I_{d_1}\otimes \cdots\otimes I_{d_{i-1}}\otimes Z_x\otimes I_{d_{i+1}}\otimes\cdots\otimes I_{d_k}$ where the dimension of the generic matrix $Z_x$ is $d_i$. By Corollary \ref{cor:pc-blow-up}, $\pcrank(T_{d'}(Z))=d'\cdot \pcrank(T)$. Let $Z$ denote the tuple $(Z_1, \ldots, Z_k)$ where $Z_{\ell}=\{Z_x\}_{x\in X_{\ell}}$ for each $\ell$. Equivalently, if we regard each matrix $Z_{\ell}$ as the set of variables $\{z_{x, {i'}, {j'}}\}_{1\leq i',j'\leq d_{\ell};x\in X_{\ell}}$, then $Z$ is essentially the new set of partially commutative variables. That is, in each $Z_{\ell}$ the variables are noncommuting and variables across different sets $Z_{\ell_1}, Z_{\ell_2}$,  for $\ell_1\neq \ell_2$, are mutually commuting.    

Next, in $T_{d'}(Z)$ replace each matrix $Z_x\otimes I_{d_2}\otimes \cdots\otimes I_{d_k}$ for $x\in X_1$ by 
\[
(Z_x + M_x) \otimes I_{d_2}\otimes \cdots\otimes I_{d_k}.
\] 
Similarly, the matrices $I_{d_1}\otimes \cdots\otimes I_{d_{i-1}}\otimes Z_x\otimes I_{d_{i+1}}\otimes\cdots\otimes I_{d_k}$ corresponding to %$x\in \bigcup_{i=2}^k X_i$ 
$x\in X_{[k]\setminus \{1\}}$ are replaced by 
\[
I_{d_1}\otimes \cdots\otimes I_{d_{i-1}}\otimes (Z_x + M_x)\otimes I_{d_{i+1}}\otimes\cdots\otimes I_{d_k}.
\]
For the simplicity, we write the matrix obtained as $T_{d'}(Z+\ubar{M})$. 

Notice that, 
%${T}_{d'_1}$ to be the following linear matrix. 
%\[
%{T}_{d'_1}(Z+M) = T''_{1} +\sum_{j'=1}^k \sum_{x\in X_{j'}} (A_x\otimes I_{d'_1}) ~x + \sum_{x\in X_{1}} A_x\otimes Z_x  
%\]
\begin{equation}\label{eqn:new6}
{T}_{d'}(Z+\ubar{M}) = T''_{1} +\sum_{i=1}^k \sum_{x\in X_{i}} A_x\otimes I_{d_1}\otimes \cdots\otimes I_{d_{i-1}}\otimes Z_x\otimes I_{d_{i+1}}\otimes\cdots\otimes I_{d_k},
\end{equation}
recalling the discussion in Section~\ref{sec:proofidea} (see Equation~\ref{eqn:new6-intro}).

Since the rank of a linear matrix is invariant under shifting of the variables by scalars (See, footnote \ref{footnote}), we get that $\pcrank({T}_{d'}(Z+\ubar{M}))=\pcrank({T}_{d'}(Z))$.  
%where $\{Z_x : x\in X_{1}\}$ are generic $d'_1\times d'_1$ matrices with noncommutative entries. Collectively, $Z_{1}$ is the set of all variables in $\{Z_x : x\in X_{1}\}$. 
%Note that $T_{d'_1}$ is the matrix obtained from $T$ by first shifting the variables $x\in X_1$ by $M_x\otimes I_{\widetilde{d}}$ and then replacing $x$ by $Z_x$. Similarly the variables $x\in X_{j'}$ is shifted by $I_{d_1}\otimes \widetilde{M}_x$ for $2\leq j'\leq k$.  

%Let us define, 
%\begin{equation}\label{eqn:new2}
%\widehat{T} =  T''_{1} +\sum_{j'=1}^k \sum_{x\in X_{j'}} (A_x\otimes I_{d'_1}) ~x. 
%\end{equation}
%By Lemma \ref{lem:scaling-blowup-pc}, we know that $d'_1\cdot \pcrank(\widehat{T})=\pcrank({T}_{d'_1})$. 
For invertible transformations $U, V$ over $\Q$, we can write 
%Suppose for each $x\in X_j$, we have a $d\times d$ matrix substitution $p_x$ as a witness of rank $r$. We now substitute each $x\in X_j$ by a $d\times d$ generic matrix $Z_x = \left(z^{\{x\}}_{\hat{i}\hat{j}}\right)$.
%We now check whether $\ncrank(T) > r$ or not over $\DR_{j+1}\newbrak{X_j}$ by checking whether $\ncrank(T_d) > rd$ or not over $\DR_{j+1}\newbrak{\ubar{Z}}$. Observe that,
%\begin{observation}
%Evaluating $T$ at $(p_1, \ldots, p_n)$ is equivalent to evaluating $T^{\{d\}}$ by substituting each $z_{i,j,k}$ by $p_{i,j,k}$.    
%\end{observation}
\[
{T}_{d'}(Z+\ubar{M}) =
U\left(
\begin{array}{c|c}
I_{rd'} - L & A \\
\hline
B & C
\end{array}
\right)V. 
\]
Furthermore,
\begin{equation}\label{eqn:new7}
{T}_{d'}(Z+\ubar{M}) =
U U'\left(
\begin{array}{c|c}
I_{rd'} - L & 0 \\
\hline
0 & C - B(I_{rd'} - L)^{-1}A
\end{array}
\right) V'V. 
\end{equation}

\[
\text{Here,~}
U'= \left(
\begin{array}{c|c}
I_{rd'}  & 0 \\
\hline
B (I_{rd'} - L)^{-1} & I_{(s-r)d'}
\end{array}
\right)
,
\quad\quad 
V'= \left(
\begin{array}{c|c}
I_{rd'}  & (I_{rd'}-L)^{-1}A \\
\hline
0 & I_{(s-r)d'}
\end{array}
\right).
\]
  
Notice that $L, A, B, C$ are linear matrices over the variables in $Z_{1}, Z_{2}, \ldots, Z_k$. The $(\ell_1,\ell_2)^{th}$ entry of $C - B(I_{rd'} - L)^{-1}A$ is given by $C_{\ell_1\ell_2} - B_{\ell_1}(I_{rd'} - L)^{-1}A_{\ell_2}$ where $B_{\ell_1}$ is the $\ell_1^{th}$ row vector of $B$ and $A_{\ell_2}$ is the $\ell_2^{th}$ column vector of $A$. We prove the following lemma which is the partially commutative version of Lemma \ref{lemma:reduce-to-PIT}.

\begin{lemma}\label{lemma:reduce-to-pc-PIT}
$\pcrank({T}) > r$ if and only if $S_{\ell_1\ell_2}=C_{\ell_1\ell_2} - B_{\ell_1}(I_{rd'} - L)^{-1}A_{\ell_2}\neq 0$ for some choice of $\ell_1,\ell_2$.
\end{lemma}

\begin{proof}
 Suppose $\pcrank({T}) > r$. Then, by Corollary \ref{cor:pc-blow-up}, $\pcrank({T}_{d'}(Z+\ubar{M}))=\pcrank({T}_{d'}(Z))>r d'$. 
 %The noncommutative rank of a linear matrix is invariant under a scalar shift \footnote{Consider a linear matrix $L$ that achieves the maximum rank for a matrix substitution $\ubar{q}$ of some dimension $d$. Then, for any scalar shift $(\alpha_1,\ldots,\alpha_n)$, the shifted linear matrix $L(\ubar{x}+\ubar{\alpha})$ achieves the same rank on the matrix tuple $\ubar{q}-\ubar{\alpha}\otimes I_d$.}, 
%hence $\ncrank(T_d(Z_1+p_1, \ldots, Z_n + p_n)) = \ncrank (T_d) > rd$. 
However, if $C - B(I_{rd'} - L)^{-1}A$ is a zero matrix, this is impossible. 

Conversely, if $({T}_{d'}(Z+\ubar{M}))_{\ell_1\ell_2} = C_{\ell_1 \ell_2} - B_{\ell_1}(I_{rd'} - L)^{-1}A_{\ell_2}$ is nonzero for some indices $\ell_1,\ell_2$, we can find (partially commutative) matrix substitutions to the variables in $Z$ such that ${T}_{d'}(Z+\ubar{M})$ evaluated on such substitutions (let say of dimension $\hat{d}$) will be of rank more than $rd'\hat{d}$. Then 
$\pcrank({T}_{d'}(Z+\ubar{M}))>rd'$ implying that $\pcrank({T})>r$ by Corollary \ref{cor:pc-blow-up}. 
%$\tilde{p}^{(k)}_{\ell_1\ell_2}$ of dimension $d'$ for the variables $\{{z}^{(k)}_{\ell_1\ell_2}\}_{1\leq \ell_1, \ell_2\leq d, 1\leq k\leq n}$, such that the rank of $T_d(Z_1+p_1, \ldots, Z_n+p_n)$ on that substitution is more than $rdd'$. Therefore, $\ncrank(T_d(Z_1+p_1, \ldots, Z_n+p_n)) > rd$.
%More precisely, let $\tilde{\ubar{p}}=\left(\tilde{p}^{\{k\}}_{\ell_1,\ell_2}\right)_{1\leq \ell_1, \ell_2\leq d, 1\leq k\leq n}$. 
%$\ubar{q}$ from sufficiently large dimension $dd'$ such that 
%Let $\ubar{q}=\iota_{d'}(\tilde{\ubar{p}})$ and $\ubar{p}\otimes I_{d'}=(p_1\otimes I_{d'}, \ldots, p_n\otimes I_{d'})$, then the rank of $T_d(\ubar{q}+\ubar{p}\otimes I_{d'})$ is more than $rdd'$. 
%This is same as saying the rank of $T_d(Z)$ is more than $rdd'$ when the variables $z^{(k)}_{\ell_1,\ell_2}$ is substituted by $\tilde{p}^{k}_{\ell_1,\ell_2} + p_k(\ell_1,\ell_2) \otimes I_{d'}$ for $1\leq k\leq n, 1\leq \ell_1, \ell_2\leq d$.
%Hence $\ncrank(T_d)>rd$. 
%By Lemma \ref{lem:scaling-blowup}, we get that $\ncrank(T)>r$. 
\end{proof}

\subsubsection{A partially commutative ABP identity testing reduction step}\label{sec:pcabp-step}
Now we vary over all choices for $\ell_1,\ell_2$ and apply Lemma \ref{lem:bound-zero-finding} to find a nonzero of the series represented by $S_{\ell_1\ell_2}$ for some choice of $\ell_1,\ell_2$.\footnote{Notice that $C_{\ell_1 \ell_2}$ is a linear term and the  degree of the other terms in the series is at least $2$.} Next, we describe how to update the matrix tuple 
$\ubar{M}$ to a new tuple that will be the assignment for the $X_{[k]}$ variables. 

Suppose, by Lemma \ref{lem:bound-zero-finding}, we obtain matrix assignments $\{M'_{x,i,j}\}_{x\in X_1, 1\leq i,j\leq d_\ell}$
 to the $Z_\ell$ variables, $1\le \ell\le k$. Consider $x\in X_{\ell}$ ($1\leq \ell\leq k$). The matrix assignment for $Z_x$ will be the matrix $M'_x$ obtained by replacing the variable $z_{x,i,j} : 1\leq i,j\leq \ell$  by the matrix $M'_{x,i,j}$ of dimension $\p_{\ell}$. 
 Now, let
 %we ensure each $p_{\ell}$ is a distinct prime number larger than all $d_j$ for each $j\in [k]$ as mentioned in Remark~\ref{rem:pc-pit-dimension}). %By padding sufficient number of zero rows and columns, we can assume that the dimension of such matrices are $2rd'$. Let $d''_1=2rd'd_1, \ldots, d''_k=2rd'd_k$. Define, 
\[
M''_x = M_x\otimes I_{\p_{\ell}}  + M'_x. 
\]
For $x\in X_{\ell}$, we substitute $x$ by type-$\ell$ $k$-fold tensor 
\[
I_{d''_1}\otimes\cdots\otimes I_{d''_{\ell-1}}\otimes M''_x\otimes I_{d''_{\ell+1}}\otimes\cdots\otimes I_{d''_k},
\]
where we note that each $M''_x, x\in X_\ell$ is of dimension $d''_\ell=d_\ell \p_\ell$.
We denote the resulting tuple of matrices by $\widetilde{\ubar{M}}$. Proof of the next claim is analogous to the proof of 
Corollary \ref{cor:immediate}. 
\begin{claim}\label{claim:extend}
 $ \rank(T(\widetilde{\ubar{M}})) > r d''_1 d''_2\cdots d''_k.$
\end{claim}
\begin{remark}
We observe the following additional properties of our construction:
\begin{enumerate}
    \item W.l.o.g, we can ensure that $\p_1,\p_2,\ldots,\p_k$ are distinct odd primes such that each $\p_\ell>d_j$ %, 1\le j\le k$, and $d''_{\ell}=d_\ell \p_\ell > 2r^3$ for each $\ell \in [k]$. This point is also 
    as discussed in Remark~\ref{rem:pc-pit-dimension}.

    \item The above choice of the primes $\p_\ell$ ensures that the dimensions $d''_\ell, 1\le \ell \le k$ are pairwise relatively prime since the $d_\ell$ are distinct prime numbers.
\end{enumerate}
\end{remark}

\subsubsection{Rounding step}\label{pc-rounding}

Recall from the last section, we have already computed a matrix tuple $\widetilde{\ubar{M}}$ of shape $(d''_1, \ldots, d''_k)$ such that $\rank(T(\widetilde{\ubar{M}})) > rd''_1\cdots d''_k$, where the $d''_\ell$ are
all pairwise relatively prime. 
%Moreover, each $d''_j$ is a product of two primes. 
We now describe the algorithm to obtain a witness of $\pcrank$ $r+1$ if
$\rank(T)\ge r+1$.

\begin{lemma}\label{lem:pc-rounding}
Given a linear matrix $T$ over $X_{[k]}$ of size $s$ and matrix tuple $\widetilde{\ubar{M}}$ of shape $(d''_1, \ldots, d''_k)$ such that $\rank(T(\widetilde{\ubar{M}})) > r d''_1 d''_2\cdots d''_k$ and the $d''_i$ are pairwise relatively prime, we can compute another matrix tuple $\widehat{\ubar{M}}$ in deterministic $\poly(s, d''_1, \ldots, d''_k)$ time such that $\rank(T(\widehat{\ubar{M}})) \geq (r+1)\cdot d''_1 d''_2\cdots d''_k$.
\end{lemma}

\begin{proof}
If $\rank(T(\widetilde{\ubar{M}}))$ is a multiple of each $d''_i$ then the hypothesis already implies
$\rank(T(\widetilde{\ubar{M}}))\ge (r+1)\cdot d''_1 d''_2\cdots d''_k$, and there is nothing to prove. 
Now, suppose $\rank(T(\widetilde{\ubar{M}}))$ is not a multiple of $d''_i$ for some $i\in [k]$. The idea is to find a $d''_i\times d''_i$ matrix substitution $M''_x$ for each $x\in X_i$ and update the $i^{th}$ component of the matrix tuple $\widetilde{\ubar{M}}$ such that $\rank(T({\ubar{M''}}))$ is a multiple of $d''_i$ where $\ubar{M''}$ is the updated matrix tuple. To do so, we first substitute each $x\in X_{[k]\setminus \{i\}}$ by the restriction of the matrix tuple $\widetilde{\ubar{M}}$ of shape $(d''_1, \ldots, d''_{i-1}, d''_{i+1}, \ldots, d''_{k})$ by dropping the $i^{th}$ component and obtain a linear matrix $T_{[k]\setminus \{i\}}(X_i)$. 
    
    Now, we are left with an instance of the noncommutative rank computation over $X_i$ variables. 
    By Lemma \ref{lem:const-reg}, we can find matrix substitutions $M''_x : x\in X_i$ such that $\rank(T_{[k]\setminus \{i\}}(\{M''_x\}))$ is a multiple of $d''_i$. It also updates the matrix tuple $\widetilde{\ubar{M}}$ to $\ubar{M''}$ by updating only the $i^{th}$ component of $\widetilde{\ubar{M}}$ to $M''_x$. Now $\rank(T(\ubar{M''})) > rd''_1\cdots d''_k$ and $\rank(T(\ubar{M''}))$
    is a multiple of $d''_i$.
    
    We now do this for each $j\in [k]\setminus \{i\}$, to find a matrix substitution $\widehat{\ubar{M}}$ such that $\rank(T(\widehat{\ubar{M}}))$ is a multiple of $d''_j$ for each $j\in [k]$. As the $d''_j$ are pairwise relatively prime, $\rank(T(\widehat{\ubar{M}}))$ is also a multiple of $d''_1d''_2 \cdots d''_k$. Moreover, $\rank(T(\widehat{\ubar{M}})) > rd''_1\cdots d''_k$. Therefore, $\rank(T(\widehat{\ubar{M}})) \geq (r+1)d''_1\cdots d''_k$.
\end{proof}

Next, we describe the blow-up control step. 

\subsubsection{Blow-up and shape control step}\label{pc-blow-up-control}
Now the plan is to find another rank $r+1$ witness such that the dimension of the $i^{th}$ component is bounded by $s^3$. Moreover, the witness is of \emph{prime shape} $(\p_1, \ldots, \p_k)$ where the $\p_i$ are distinct prime numbers.
%for the variables in $X_{1}$ of dimension $d_{1}$ controlled appropriately by the parameter $s$ such that the rank will be $> (r+1) d_{1}$. 
%Then we can use the rank rounding step to push the rank further to at least $(r+1) d_{1}$.   
We need the following result about primes in short intervals, along with a nontrivial generalization of Lemma~\ref{lem:blow-up-control}, to prove the next lemma.

\begin{theorem}[prime number theorem in short interval \cite{LY92}]\label{thm:prime-number-thm}
    Let $n$ be a sufficiently large number, and $\pi(n)$ be the number of primes $\leq n$.  Moreover, let $n'=n^{\theta}$ for $1/2\leq \theta\leq 7/12$. Then, 
    \[
    1.01\frac{n'}{\log n}\geq \pi(n)-\pi(n-n') \geq 0.99\frac{n'}{\log n}
    \]
\end{theorem}

We are now ready to present the blow-up control step. For our purpose, we will choose $\theta=0.6$. 

\begin{comment}
\begin{lemma}\label{lem:pc-blow-up-control}
Suppose $T$ is a linear matrix of size $s$ over $X_{[k]}$ and $\widehat{\ubar{M}}$ is a matrix tuple of shape $(d''_1, \ldots, d''_k)$ such that $\rank(T(\widehat{\ubar{M}})) \geq (r+1) d''_1 d''_2\cdots d''_k$. Moreover, for $i\neq j$ the dimensions $d''_i$ and $d''_j$ are pairwise relatively prime, each $d''_i > 2s^3$, and it is a product of two distinct primes. Then we can compute another matrix tuple $\widehat{\ubar{N}}$ of shape $(\p_1, \ldots, \p_k)$ in deterministic $\poly(s, d''_1, \ldots, d''_k)$ time such that $\rank(T(\widehat{\ubar{N}})) \geq (r+1) \p_1 \cdots \p_k$ and for each $i$, $s^3 < \p_i \leq 2s^3$ is a prime number for sufficiently large $s$.
\end{lemma}
\end{comment}

\begin{lemma}\label{lem:pc-blow-up-control}
Suppose $T$ is a linear matrix of size $s$ over $X_{[k]}$ and $\widehat{\ubar{M}}$ is a matrix tuple of shape $(d''_1, \ldots, d''_k)$ such that 
\begin{itemize}
\item $\rank(T(\widehat{\ubar{M}})) \geq (r+1) d''_1 d''_2\cdots d''_k$. 
\item The dimensions $d''_i, 1\le i\le k$ are pairwise relatively prime. Moreover,
each $d''_i$ is a product of two distinct odd primes.
\end{itemize}
Then for all but finitely many $s$, in deterministic $\poly(s, d''_1, \ldots, d''_k)$ time, we can compute another matrix tuple $\widehat{\ubar{N}}$ of prime shape $(\p_1, \ldots, \p_k)$  such that $\rank(T(\widehat{\ubar{N}})) \geq (r+1) \p_1 \cdots \p_k$ and for each $i$, $\p_i \leq s^3$ is a prime number.
\end{lemma}

\begin{proof}
We will prove the statement by induction, replacing $d''_i$ by prime $\p_i$ for increasing indices
$i$. Inductively assume that we have computed a matrix tuple $\widetilde{\ubar{M}}$ of shape 
$(\p_1, \ldots,\p_\ell,d''_{\ell+1},\ldots, d''_k)$ such that 
\begin{itemize}
\item Each $\p_j\le s^3$ is an odd prime.
\item $\rank(T(\widetilde{\ubar{M}})) \geq (r+1) \p_1 \p_2\cdots \p_\ell d''_{\ell+1}\cdots d''_k$. 
\item The dimensions $\p_1,\p_2,\ldots,\p_\ell$ are distinct primes that are also relatively prime
to each $d''_i, i>\ell$.
\end{itemize}
Notice that the base case is $\ell=0$. In the inductive step, our goal is to replace $d''_{\ell+1}$
by a prime $\p_{\ell+1}$ satisfying the above. That will complete the proof.

Consider an invertible sub-matrix $A$ in $T$ of size $r+1$. We can find such $A$ since the rank of $T(\widetilde{\ubar{M}})$ is  $\geq (r+1)\p_1 \p_2\cdots\p_\ell d''_{\ell+1}\cdots d''_k$.

%such that $\pcrank(A)$ is at least $(r+1) d''_{1}$. We call procedure $\rankincrement_{k-1}$ over  $T({{\ubar{N}}_{1}}, X_2, \ldots, X_k)$ to compute matrix substitutions $X_i\leftarrow \ubar{M}'_i, 2\le i\le k$ that witness $\pcrank(T({{\ubar{N}}_{1}}, X_2, \ldots, X_k))$ is $(r+1)d''_1$. Let $\ubar{M}'_i$ be $d''_i$ dimensional, $2\le i\le k$. The resulting
%scalar matrix, call it $\hat{M}$, is of dimension $sd''_1d''_2\cdots d''_k$. Rank of $\hat{M}$ is $(r+1)d''_1\cdots d''_k$. Let $\hat{A}$ be the corresponding submatrix of $A$ witnessing that.
%Let $d''_i > r^3$ for some $i\in [k]$. %Let $d_0 = d''^{0.6}_i$. 
The following claim summarize how we will be applying the number-theoretic 
Theorem~\ref{thm:prime-number-thm} to find the prime $\p_{\ell+1}$.

\begin{claim}\label{claim:prime-dist}
For all but finitely many $d$ (depending on $k$) there are at least $2k + 1$ many prime numbers in the 
interval $(d - d^{0.6}, d]$. 
\end{claim}

We will apply the claim to $d=d''_{\ell+1}$. We can assume without loss of generality that 
$d''_{\ell+1}>s^3$. This is because we can always double the dimension $d''_{\ell+1}$ by
making the matrix components corresponding to $X_{\ell+1}$ in $\widetilde{\ubar{M}}$ 
block diagonal with two blocks. Notice that the resulting matrix tuple is still a witness of rank
at least $r+1$. Furthermore, notice that all the primes $\p_j$ and the dimensions $d''_\iota, \iota>\ell$
are all still pairwise relatively prime as $d''_{\ell+1}$ only changed by factors of $2$. 

By abuse of notation, let $\widetilde{\ubar{M}}$ still denote the modified matrix tuple
with $d''_{\ell+1}>s^3$. By the above Claim~\ref{claim:prime-dist}, for all but finitely many
$s$, we can find a prime $\p$ in the interval $(d''_{\ell+1} - d''^{0.6}_{\ell+1}, d''_{\ell+1}]$
that is relatively prime to all the $\p_j$ and to each $d''_\iota, \iota > \ell+1$.

%Therefore, we can always choose a prime $\p$ avoiding all the prime factors of $d''_1, \ldots, d''_k$ (there %are at most $2k$ such primes since each $d''_j$ is a product of \emph{two} distinct primes). 

Now from the matrices in $\widetilde{\ubar{M}}$, remove the last $d''_{\ell+1} - \p$ many rows and columns of the first component matrices (namely, the substitutions for variables in $X_{\ell+1}$) and keep other substitutions as they are. This yields another tuple ${\ubar{M'}}$ of matrix substitutions for the $X_{[k]}$ variables. 

Let $\Delta= \prod_{j=1}^\ell\p_j \prod_{\iota > \ell+1}d''_\iota$. We claim that $\rank(A({\ubar{M'}})) > r \p \Delta$. Suppose not.
Then, we have:
\begin{align*}
\rank(A(\widehat{\ubar{M}})) &\leq \rank(A({\ubar{M'}})) + 2(r+1)(d''_{\ell+1} - \p) \Delta\\
&\leq r \p \Delta + 2(r+1) (d''_1 - \p) \Delta \\
&= \Delta (r\p + 2(r+1) (d''_{\ell+1} - \p))\\
&= \Delta (2(r+1) d''_{\ell+1} - (r + 2)\p)\\
& < (r+1) d''_{\ell+1} \Delta
\label{eqn:new-blow-up}
%&< (r+1) d''_1 d''_2 \cdots d''_k.
\end{align*}
which contradicts the inductive assumption. To see the last strict inequality we observe that
$(r+1)d''_{\ell+1}<(r+2)\p$. This follows from the following:
\begin{align*}
(r+1)(d''_{\ell+1}-\p) &\le (r+1) d''^{0.6}_{\ell+1} \le (s+1)d''^{0.6}_{\ell+1} \le d''^{0.95}_{\ell+1} < \p
\end{align*}
because $(s+1)< d''^{0.34}_{\ell+1}$ and $\p> d''_{\ell+1} - d''^{0.6}_{\ell+1}$. 

We can now apply Lemma \ref{lem:pc-rounding}, with the matrix tuple ${\ubar{M'}}$ as input, to find a new matrix tuple on which 
$T$ will evaluate to a matrix of rank $\geq (r+1)\p \Delta$. Now, if $\p >s^3$ we can repeat
the above process with $\p$ instead of $d''_{\ell+1}$ until we finally get
$\p\le s^3$. Then we set $\p_{\ell+1}=\p$ completing the inductive step of the proof.

%Therefore repeating this process, we can find a $(r+1)$-rank witness of $T$ of shape $(\p_1, d''_2, \ldots, %d''_k)$ such that $s^3 < \p_1 \leq 2s^3$.    
%Therefore, we can transform the given input tuple to ensure each $d''_i \leq r^3$. 
%We will now replace each $d''_i$ by a suitable prime $\p_i$ for $i=1,2,\ldots,k$. At the $i^{th}$ step,
%inductively assume that we have 

%We now prove it inductively for each 
%We can now proceed inductively on each $i\in [k]$. 
%Suppose, at the $i^{th}$ stage, we choose a prime number such that $2^{i-1} r^3 < \p_i \leq 2^i r^3$. We need the additional slackness to ensure that the primes $\p_1, \ldots \p_i$ are all distinct.  W.l.o.g, we assume that $d''_i \geq 2^i r^3$ as explained in Remark~\ref{rem:pc-pit-dimension}. We now use Lemma \ref{lem:pc-rounding} to find a matrix tuple $\ubar{M_{\int}}$ of shape $(\p_1, \ldots, \p_i, d''_{i+1}, \ldots, d''_k)$ such that 
%\[
%\rank(T({\ubar{M}_{\int}})) \geq (r+1) \p_1 \cdots \p_i d''_{i+1}\cdots  d''_k. 
%\]

To summarize, we will finally obtain the claimed matrix substitution $\widehat{\ubar{N}}$ of prime shape $(\p_1, \ldots, \p_k)$ where each $\p_i \leq  s^3$. The runtime bound is easy to verify. \qedhere
%To see this, note that the rank of submatrix $A'$ is at least $(r+1)d''_1 d''_2\cdots d''_k -(r+1)d''_2\cdots d''_k$ because from $\hat{A}$ we have dropped that many rows and columns. As 
%\[
%(r+1)d''_1 d''_2\cdots d''_k -(r+1)d''_2\cdots d''_k > r(d''_1-1)d''_2 \cdots d''_k.
%\]

%Hence, as witnessed above, for this substitution $\ubar{N}'_1$ for $X_1$
%the $\pcrank$ of $T(\ubar{N}'_{1}, X_{2}, \ldots, X_k)$ is strictly more that $r(d''_1 -1)$. 

%Otherwise, $\pcrank(A)\leq \pcrank(A') + 2(r+1) \leq r(d''_{1}-1) + 2(r+1) < (r+1) d''_{1}$, which is a %contradiction. Now apply $\rankincrement_{k-1}$ over 
%$T({{\ubar{N}}_{1}}, X_2, \ldots, X_k)$ to compute matrix substitutions for the variables in $X_2, \ldots, %X_k$ that witness $\pcrank(T({{\ubar{N}}_{1}}, X_2, \ldots, X_k))$ more than $r(d''_1-1)$. 

%Now, we can apply the rounding procedure in Lemma \ref{lem:round-rankincrement} to compute another matrix tuple $\ubar{N}'_1$ inside a division algebra $D_1$ 
%such that $\pcrank(T(\ubar{N}'_1, X_2, \ldots, X_k))$ is at least $(r+1)(d''_1-1)$. 
%This step involves polynomial-time reduction to $\rankincrement_{k-1}$ subroutine.

%By repeating this process we can ensure that the dimension of the witness matrices is bounded by $r+2\leq s + 1$. Clearly, we can choose $\delta_2$ to be suitably large so that the lemma follows. 
%However, to the implement the rounding step again, we choose the new dimension $d_{1}$ to be a prime between $s+1$ and $2(s + 1)$. 
%Now, we apply the rounding step described above to obtain another tuple $\ubar{p}''_{1}$ such that the $\pcrank$ becomes at least $r+1$. 
\end{proof}

\begin{remark}
    %We can now bound the dimension of a witness of a matrix $T\in \F\angle{X_{[k]}}^{s\times s}$. It 
    The dimension of the final $(r+1)$-rank witness $\widehat{\ubar{N}}$ is bounded by $s^{3k}$.
\end{remark}

\subsubsection{Pseudo-code for rank increment}\label{sec:pcpseudocode} 
Given an input matrix $T$ over $X_{[k]}$ and type-$j$ $k$-fold tensor product matrix assignments 
for the variables in $X_j$ ($1\leq j\leq k$) such that $\rank(T(\ubar{M}))$ is at least $rd'$ where $\ubar{d}=(d_1,d_2,\ldots,d_k)$ is the shape of the tensor and $d'=d_1 d_2\cdots d_k$, we describe the pseudo-code of the rank increment procedure described above that finds another set of assignments to the variables in $X_j$ ($1\leq j\leq k$) that witness the $\pcrank(T(X_{[k]}))$ is at least $r+1$, if such a rank increment is possible. Moreover, for each $1\leq j\leq k : d_j\leq s^3$ %for a constant $c>0$ 
and $\ubar{M}$ represents the entire tuple of matrix assignment.   
%of the form 
%$I_{d_1}\otimes$

%$\{M_x\}_{x\in X_1}$ of dimension $d_1\leq s+1$ such that $\pcrank(T(\{M_x\}_{x\in X_1}, X_2, \ldots, X_k))$ is at least $r$, we describe the pseudo-code of the rank increment subroutine that finds another set of assignments to the variables in $X_1$ that witness the $\pcrank(T(X_{[k]}))$ is at least $r+1$, if such a rank increment is possible. 
%The pseudo-code given below implements the subroutine $\rankincrement$. 

\begin{description}
\item[Algorithm for $\RANKINCREMENT$ $(T_{[k]}, \ubar{M}, r)$]\
\item[Input :] A linear matrix $T$ over $X_{[k]}$ and the matrix tuple $\ubar{M}$ such that $\rank(T(\ubar{M}))$ is at least $rd'$ where the shape of the matrix tuples in $\ubar{M}$ are given by $\ubar{d}=(d_1, d_2, \ldots, d_k)$ such that the $d_j\leq s^3 $ ($1\leq j\leq k$) are distinct prime numbers and $d' = d_1 d_2 \cdots d_k$.  
%and $\{M_x\}_{x\in X_1}$ such that $\pcrank(T(\{M_x\}_{x\in X_1}, X_2, \ldots, X_k))$ is at least $r$. Also the dimension is bounded by $\dim(M_x)\leq s+1$.\
\item[Output :] Find another set of matrix assignments of shape $\ubar{d} = (d_1, \ldots, d_k)$ for $x\in X_{[k]}$ %$x\in \bigsqcup_{i=1}^k X_i$ 
that witness the $\pcrank(T)\geq r+1$, if such a rank increment is possible and the $d_j \leq s^3$ are distinct prime numbers. 
%Moreover the shape of the new tensor product matrices is the same as that of the matrices in $M$. 

%\item $\rankincrement(T_{0}=T, X_{1}, \ldots, X_k)$\\
%\begin{enumerate}
%\item 
%Base case: find a witness to the $X_{1}$ variables for rank $r=1$.
\item [Steps :]

 \begin{enumerate}
    \item Using the $Z$ variables and the matrix shift, construct the linear matrix $T_{d'}(Z+\ubar{M})$ as shown in Equation \ref{eqn:new6}\label{step3}. 
    \item Using Gaussian elimination, convert the matrix $T_{d'}(Z+\ubar{M})$ to the block diagonal shape shown in Equation \ref{eqn:new7}\label{step4}. 
    \item Use Lemma \ref{lem:bound-zero-finding}, to find the nonzero of a series originating from the bottom right of the block. If it fails to find a nonzero STOP the procedure\label{step5}. 
    \item Use Claim \ref{claim:extend} to compute a new set of matrix assignments of dimension $d''_1, d''_2, \ldots, d''_k$ (the $d''_j$ are pairwise relatively prime) to the variables in $X_1, X_2, \ldots, X_k$, such that after the evaluation, the rank of the resulting matrix is strictly more than $r\cdot d''_1 d''_2 \cdots d''_k$\label{step6}. 
    \item Use Lemma \ref{lem:pc-rounding} and Lemma \ref{lem:pc-blow-up-control} to implement the rounding and the blow-up control steps and compute the matrix assignments that witness $\pcrank(T)\geq r+1$. Moreover, the dimension of each component of the witness is bounded by $s^3$\label{step7}. 
\end{enumerate}
  \end{description}

%We now describe the construction of the ABP $\mathcal{A}$ which is the step $(d)$ of the subroutine $\rankincrement$ above. 

We complete the section with the proof of the main theorems. For the convenience of the reader, we restate the theorems.   

\begin{theorem}[Restate of Theorem \ref{thm:main-theorem}]\label{thm:main-theorem-restate}
Given an $s\times s$ matrix $T$ whose entries are $\Q$-linear forms over the partially commutative set of variables $X_{[k]}$ (where $|X_i|\leq n$ for $1\leq i\leq k$ and w.l.o.g $n\leq s$), the rank of $T$ over $\U_{[k]}$ can be computed in deterministic $s^{2^{O(k \log k)}}$ time. The bit complexity of the algorithm is also bounded by 
$s^{2^{O(k \log k)}}$.
%The field $\F$ is the field of rational numbers $\Q$. 
%The field $\F$ is either $\Q$ or any sufficiently large finite field. 
\end{theorem}

\begin{proof}
Firstly note that, due the blow-up control step, the shape of the matrix tuples is always determined by the size of the input matrix thus it remains as $\ubar{d}=(d_1, d_2, \ldots, d_k)$ where each $d_i\leq s^3$. 
Also, since the $\pcrank(T)$ is bounded by $s$, the subroutine $\RANKINCREMENT$ can be called for at most $s$ times. Let $t_k(s)$ be the time taken by the procedure $\RANKINCREMENT$ from rank $r$ to rank $r+1$. 
The size of the matrix 
$T_{d'}(Z+\ubar{M})$ is at most $sd' = s^{O(k)}$ since $d'=d_1 d_2 \cdots d_k$. Hence Step \ref{step3} and Step \ref{step4} can be performed in $s^{O(k)}$ time. 
%We can choose a constant $\beta>1$ suitably large.
%The theorem is proved by simply inspecting the pseudo-code above. 
%Let $t_k(s)$ be the time taken by the procedure. 
%In Step \ref{step3}, the size of the matrix $T_{d'}(Z+M)$ is at most $sd'$ which is bounded by $s^{O(c^k)}$. The Gaussian elimination in Step \ref{step4} takes time $(sd')^{O(1)}$. 
%We can choose a constant $\beta>1$ suitably large. 
In Step \ref{step5}, the application of Lemma \ref{lem:bound-zero-finding} calls $\pitsearch_{k}$ on a linear matrix of size $s^{O(k)}$ and additional $s^{O(k)}$ time for linear algebraic computation. 
As shown in Lemma \ref{lem:pc-rounding} and Lemma \ref{lem:pc-blow-up-control} that Step \ref{step7} takes at most $s^{O(k)}$ time. 
%many calls to $\rankincrement_{k-1}$ for linear matrices of size at most $s^{2^{O(k)}}$. 
%Step \ref{step2} recursively solves $\rankincrement_{k-1}$ on a linear matrix of size $O(s^2)$.  
%at most $s^{c^{\beta k}}t_{k-1}(s^{c^{\beta k}})$ in Step \ref{step5}. Step \ref{step7} again deals with reducing the the problem to $\rankincrement_{k-1}$ eventually and takes time $s^{c^{\beta k}}t_{k-1}(s^{c^{\beta k}})$. Finally Step \ref{step2} 
%recursively solve $\rankincrement_{k-1}$ for a linear matrix of size $O(s^{2})$. Combining everything, we get the following recurrence:
%\[
%t_k(s)\leq s^{c^{\tau k }} t_{k-1}(s^{c^{\tau k}}) + s^{c^{\tau k}}, 
%\]
%for a sufficiently large constant $\tau >1$. Assuming as induction hypothesis $t_{k-1}(s)\le s^{c_2^{(k-1)^2}}$,
%where $c_1=c^\tau$ and $c_2=3c_1$, notice that for almost all $s$
%\[
%t_k(s)\le s^{c_1^k}(s^{c_1^k})^{c_2^{(k-1)^2}}+s^{c_1^k} = s^{c_1^k}(s^{c_1^k\cdot c_2^{(k-1)^2}} +1)
%\le s^{3c_1^k\cdot c_2^{(k-1)^2}} \le s^{c_2^{k^2}}.
%\]
%\ACnote{I am adding my analysis here.}

Let $T_1(s,k)$ be the running time of the $\pitsearch_{k}$ subroutine on an ABP of size $s$ over $X_{[k]}$, and $T_2(s,k)$ be the running time of the $\rankincrement_{k}$ subroutine on a linear matrix of size $s$ over $X_{[k]}$. Then, for a suitable constant $\beta >0$ we can bound 
\[
t_k(s)\leq s^ {\beta k} T_1(s^{\beta k}, k) + s^{\beta k}. 
%t_k(s)\leq T_1(s^{2^{O(k)}}, k) + s^{2^{O(k)}} T_2(s^{2^{O(k)}}, k-1) + s^{2^{O(k)}}
\]
Now, we simultaneously analyze the recurrences for 
$T_1(s,k)$ and $T_2(s,k)$. Notice that, $T_1(s,k)\leq T_2(O(s),k)$, since size $s$ ABPs have linear pencils of size $O(s)$ (Proposition \ref{prop:abp-pencil}). From Theorem \ref{thm:pittorank} and from the time analysis of $\RANKINCREMENT$ subroutine as shown above,
as $T_2(s,k)\le s t_k(s)$ we have:
%From the recursive analysis, we have,
\begin{align*}
    T_1(s,k) &\leq s T_1(s^4, k-1) + s^6 T_2(s^6, k-1) + s^{O(1)}\\
    T_2(s,k) &\leq s T_1(s^{\gamma k}, k) + s^{\gamma k}. 
\end{align*}
for some constant $\gamma > 0$. 
From the first inequality above, $T_1(s,k) \leq 2^ks^{O(1)}T_2(s^6, k-1)$ for all but finitely many $s$. Combined with the second inequality above, we have $T_2(s, k) \leq s^{\tau k} T_2(s^{\tau k}, k-1)$ for a suitable constant $\tau>\beta$.
\[
\text{Therefore,}\quad T_2(s, k) \leq s^{\tau k} T_2(s^{\tau k}, k-1) \leq s^{\tau k}\cdot s^{\tau {k}}\cdots s^{\tau {k}}\cdot T_{\nsing}(s^{(\tau {k})^k}), %\leq s^{\sum_{i=1}^k \tau^{ik}}\cdot T_{\nsing}(s^{\tau^{k^2}}),
\]
where $T_2(s,1)=T_{\nsing}(s)=\poly(s)$ is the running time of the $\nsing$ algorithm on a linear matrix of size $s$.
Therefore, we have 
\[T_2(s,k) \leq (s^{(\tau k)^2}) \poly(s^{(\tau k)^k}) \leq s^{2^{O(k \log k)}}.\]
%As $\sum_{i=1}\tau^{ik}\le 2\tau^{k^2}$, we have $T_2(s, k)\le s^{2\tau^{k^2}}\cdot T_{\nsing}(s^{\tau^{k^2}})\le s^{\eta^{k^2}}$, for a sufficiently large constant $\eta>0$. Also, the bit complexity of the procedure can also be bounded by $s^{\eta^{k^2}}$, as the bit complexity of $\nsing$ is polynomial in input size. 
We can bound the bit complexity of the algorithm along the same line and noting the fact that the bit complexity of the $\nsing$ algorithm is polynomially bounded.
\end{proof}

Next, we prove Theorem \ref{thm:main-theorem-2}. 
\begin{theorem}[Restate of Theorem \ref{thm:main-theorem-2}]\label{thm:main-theorem-2-restate}
Given an ABP of size $s$ whose edges are labeled by $\Q$-linear forms over the partially commutative set of variables $X_{[k]}$ (where $|X_i|\leq n\leq s$ (w.l.o.g) for $1\leq i\leq k$), there is a deterministic $s^{2^{O(k \log k)}}$ time algorithm %$s^{\gamma {k^k}}$ time algorithm (for a constant $\gamma >0$)
to check whether the ABP computes the zero polynomial. As a corollary, the equivalence testing of $k$-tape weighted automata can be solved in deterministic polynomial time for $k=O(1)$. The bit complexity of the algorithm is also bounded by $s^{2^{O(k \log k)}}$.%$s^{\gamma {k^k}}$. 
\end{theorem}

\begin{proof}
The proof follows directly from the analysis of the recurrence for $T_2(s,k)$ in the proof of Theorem \ref{thm:main-theorem-restate} above. 
\end{proof}

\section{Discussion}\label{sec:discussion}
We find the interplay between symbolic determinant identity testing, concepts from formal language theory, and noncommutative algebra very fascinating. Apart from yielding a deterministic polynomial-time algorithm for
the $k$-tape weighted automata equivalence problem, 
the most interesting aspect of the $\pcsing$ problem is that it provides a common framework spanning both  $\sing$ and $\nsing$. We state a few questions for further study.  
\begin{enumerate}
\item It would be satisfactory to obtain a deterministic algorithm for $\pcsing$ over a $k$-partitioned set of $n$ variables such that setting $k=1$ captures the best-known algorithm for $\nsing$ and setting $k=n$ yields the best-known algorithm for $\sing$. For $k=1$, we obtain a deterministic polynomial-time algorithm for $\nsing$. In contrast, as the runtime of our algorithm is  
doubly exponential in $k$, applied to the $\sing$ problem (where $k=n$) the time bound becomes even worse than an exhaustive search. Of course, finding an $(nsk)^{O(1)}$ algorithm for $\pcsing$ would be a breakthrough as it would imply a circuit lower bound \cite{KI04}.  

\item It is to be noted that the running time of the randomized algorithm for equivalence testing of $k$-tape weighted automata by Worrell \cite{Worrell13} is  indeed $(ns)^{O(k)}$. Thus, it would be plausible and interesting to obtain a \emph{deterministic} algorithm for equivalence testing of $k$-tape weighted automata  
 with runtime closer to $(ns)^{O(k)}$. 
 
 \item Another interesting problem is to understand the complexity of the equivalence testing of multi-tape weighted automata for unbounded number of tapes.
\end{enumerate}
%\newpage
\bibliographystyle{alpha}
\bibliography{ref2}

\appendix

%\section{The Proof of Lemma \ref{lemma-pcrank-connection}}\label{sec:rank-computation}
\section{Appendix}

The idea is to reduce the computation of $\pcrank$ of a matrix with $\U_{[k]}$ entries to $\pcrank$ computation of a linear matrix incurring a small blow-up in the size. To show the reduction, we need the following lemma.

\begin{lemma}\label{lemma-nc-rank2by2}
Let $X=X_{[k]}$ and $\U_{[k]}$ be the universal skew field over $\F\angle{X_{[k]}}$. 
Let $P\in {\U_{[k]}}^{{m\times m}}$ such that,
\[
P = \begin{bmatrix}
A &B\\
C &D
\end{bmatrix},
\]
where $A\in \U_{[k]}^{r\times r}$ is invertible. Then,
\[
\pcrank(P) = r+ \pcrank(D - CA^{-1}B),
\]
\end{lemma}

\begin{proof}
If $Q$ is an $m\times m$ invertible matrix over $\U$
then
\[
\pcrank(QP)=\pcrank(PQ)=\pcrank(P).
\]
For if $P=MN$ then $QP=(QM)N$ and if $QP=MN$ then $P=(Q^{-1}M)N$.
Similarly for $PQ$.

The matrix
\[
\begin{bmatrix}
A^{-1} &0\\
0 &I_{m-r}
\end{bmatrix}
\]
is full rank. Similarly, the matrix 
\[
\begin{bmatrix}
I_r &0\\
-C &I_{m-r}
\end{bmatrix}
\]
is full rank because
\[
\begin{bmatrix}
I_r &0\\
-C &I_{m-r}
\end{bmatrix}
\begin{bmatrix}
I_r &0\\
C &I_{m-r}
\end{bmatrix} =
\begin{bmatrix}
I_r &0\\
0 &I_{m-r}
\end{bmatrix}.
\]
Hence, $\pcrank(P)$ equals $\pcrank(R)$ where
\[
R= \begin{bmatrix}
I_r &0\\
-C &I_{m-r}
\end{bmatrix}\cdot
\begin{bmatrix}
A^{-1} &0\\
0 &I_{m-r}
\end{bmatrix}\cdot 
\begin{bmatrix}
A &B\\
C &D
\end{bmatrix} =
\begin{bmatrix}
I_r &A^{-1}B\\
0 & D- CA^{-1}B
\end{bmatrix}
\]

Post-multiplying by the invertible matrix
$\begin{bmatrix}
I_r &-A^{-1}B\\
0 &I_{m-r}
\end{bmatrix}$ we obtain
$\begin{bmatrix}
I_r &0\\
0 &D-CA^{-1}B
\end{bmatrix}$.

It is easy to see that its inner rank is $r+\pcrank(D-CA^{-1}B)$.
\end{proof}

%We now prove Theorem~\ref{thm-nc-rank} assuming this lemma to be true.
For the sake of reading, we restate Lemma \ref{lemma-pcrank-connection}. 

\begin{lemma}[Restate of Lemma \ref{lemma-pcrank-connection}]\label{lemma-ncrank-connection-appendix}
Let $X=X_{[k]}$ be a set of partially commutative variables. Let  $M \in \F\angle{X_{[k]}}^{m\times m}$ be a matrix where each $(i,j)^{th}$ entry $M_{ij}$ is computed as the $(1,s)^{th}$ entry of the inverse of a linear pencil $L_{ij}$ of size $s$. Then, one can construct a linear pencil $L$ of size $m^2s + m$ such that,
\[
\pcrank(L) = m^2s + \pcrank(M).
\]
\end{lemma}

\begin{proof}
We first describe the construction of the linear pencil $L$ and then argue the correctness.
%W.l.o.g. we may assume that each linear matrix $L_{ij}$ is $s\times s$ (by padding it, if required, with an identity matrix of suitable size). 

\begin{equation}\label{eq:ncrank-gen}
\text{Let}\quad
L = \left[\begin{array}{ c c c c | c}
L_{11} &0  &\cdots &0  &B_{11}\\
0   &L_{12} &\cdots &0  &B_{12}\\
\vdots   &\vdots &\ddots  &\vdots &\vdots\\
0   &0  &\cdots &L_{mm} &B_{mm}\\
\hline
-C_{11} &-C_{12} &\cdots &-C_{mm} &0
\end{array}
\right], 
\end{equation}
where each $C_{ij}$ is an $m\times s$ and $B_{ij}$ is an $s\times m$ rectangular matrix defined below. Let $e_i$ denote the column vector with 1 in the $i^{th}$ entry and the remaining entries are zero. We define
\[
C_{ij} = \left[\begin{array}{ c| c| c| c}
   & &  &\\
   & &  &\\
   e_i  &0 &\cdots  &0\\
   & &  &\\
   & &  &
\end{array}
\right]
\quad\text{and, }\quad
B_{ij} = \left[\begin{array}{ c c c c c}
   & &0 & &\\
   \hline
& &0 & &\\
   \hline
   & &\vdots & &\\
   \hline
& &e_j & &
\end{array}
\right],
\]
where $e_j$ is a row vector in $B_{ij}$. 
To argue the correctness of the construction, we write $L$ as a $2\times 2$ block matrix. As each $L_{ij}$ is invertible (otherwise $M_{ij}$ would not be defined), the top-left block entry is invertible. Therefore, we can find two invertible matrices $U,V$ implementing the required row and column operations such that,

\[
L = 
U \left[
\begin{array}{c c c c | c}
L_{11} &0  &\cdots &0  &0\\
0   &L_{12} &\cdots &0  &0\\
\vdots   &\vdots &\ddots  &\vdots &\vdots\\
0   &0  &\cdots &L_{mm} &0\\
\hline
0 &0 &\cdots &0 &\widetilde{D}
\end{array}
\right]V, 
\]
for some $m\times m$ matrix $\widetilde{D}$.

\begin{claim}
The matrix $\widetilde{D}$ is exactly the input matrix $M$.
\end{claim}

\claimproof
From the $2\times 2$ block decomposition we can write,
\[
\widetilde{D} = [C_{11}  C_{12}  \cdots  C_{mm}] 
\left[
\begin{array}{ c c c c}
L^{-1}_{11} &0  &\cdots &0  \\
0   &L^{-1}_{12} &\cdots &0 \\
\vdots   &\vdots &\ddots  &\vdots \\
0   &0  &\cdots &L^{-1}_{mm}
\end{array}
\right]
\left[
\begin{array}{c}
B_{11}\\
B_{12}\\
\vdots\\
B_{mm}
\end{array}
\right]
 = 
 \sum_{i,j} C_{ij}L^{-1}_{ij}B_{ij}.
 \]
Observe that, for each $i,j$, $C_{ij}L^{-1}_{ij}B_{ij}$ is an $m\times m$ matrix with $M_{ij}$ as the $(i,j)^{th}$ entry and remaining entries are 0. Hence, $\widetilde{D} = M$. \qed

Notice that the top-left block of $L$ in Equation~\ref{eq:ncrank-gen} is invertible as for each $i,j\in [m]$, $L_{ij}$ is invertible. Now the proof follows from Lemma~\ref{lemma-nc-rank2by2}.
\end{proof}

\end{document}